\documentclass[twocolumn]{aastex631}

\usepackage{natbib}
\usepackage{epsfig}
\usepackage{graphicx}
\usepackage{amsthm}
\usepackage{amssymb}
\usepackage{rotating}
\usepackage{multirow}
\usepackage{longtable}
\usepackage{savesym}
\savesymbol{tablenum}
\restoresymbol{SIX}{tablenum}
\usepackage{booktabs}
\usepackage{xcolor}

\defcitealias{2019ApJ...873...51L}{Paper~I}
\defcitealias{2020ApJ...888...98L}{Paper~II}
\defcitealias{2021ApJ...923..198D}{Paper~III}
\defcitealias{2022ApJ...933...60D}{Paper~IV}
\defcitealias{2023ApJ...949...82D}{Paper~V}
\defcitealias{2024ApJ...963...55D}{Paper~VI}
\defcitealias{2025ApJD}{Paper~VII}

\makeatletter
\renewcommand{\paragraph}{\@startsection{paragraph}{4}{0ex}%
   {-3.25ex plus -1ex minus -0.2ex}%
   {1.5ex plus 0.2ex}%
   {\normalfont\small\itshape\centering}}
\makeatother

\stepcounter{secnumdepth}
\stepcounter{tocdepth}

\accepted{by ApJ February 13, 2026}

\shorttitle{Surveying G\ion{H}{2} Regions: VIII. W43 Main}
\shortauthors{De Buizer et al.}

\begin{document}

\title{Surveying the Giant \ion{H}{2} Regions of the Milky Way with SOFIA: VIII. W43 Main}

\email{jdebuizer@seti.org}

\author[0000-0001-7378-4430]{James M. De Buizer}
\affil{Carl Sagan Center for Research, SETI Institute, 339 Bernardo Avenue, Suite 200, Mountain View, CA 94043, USA}

\author[0000-0003-4243-6809]{Wanggi Lim}
\affil{IPAC, Mail Code 100-22, Caltech, 1200 East California Boulevard, Pasadena, CA 91125, USA}

\author[0000-0003-0740-2259]{James T. Radomski}
\affil{Independent Researcher, Sunnyvale, CA 94085, USA}

\author[0000-0003-3682-854X]{Nicole Karnath}
\affil{Space Science Institute, 4765 Walnut Street, Suite B, Boulder, CO 80301, USA}

\begin{abstract}
In this eighth paper of the SOFIA-FORCAST series on Milky Way G\ion{H}{2} regions, we present an analysis of the massive star-forming complex W43~Main. We compared our $11\text{--}37\mu$m maps with multi-wavelength observations from the near-infrared to radio, and investigated the physical nature of compact sources and dust substructures. We applied SED fitting to constrain properties of the compact infrared objects, and examined the evolutionary states of the extended subregions. We identified 20 compact infrared objects, 16 (80\%) of which we classify as MYSOs or candidate MYSOs. W43~Main resides at the junction of the Scutum spiral arm and the Galactic Bar, a location where enhanced turbulence is anticipated and has been proposed as a potential influence on star-formation activity. Nevertheless, our analysis shows that its Lyman continuum photon production rate, the mass of its most massive MYSO, and its MYSO density are all consistent with the survey-wide median values. We therefore conclude that, despite W43~Main's unique Galactic environment, its present star formation activity appears broadly consistent with that of an average Galactic G\ion{H}{2} region.
\end{abstract}


\section{Introduction} 

Giant \ion{H}{2} (G\ion{H}{2}) regions are extreme sites of massive star formation, playing host to the largest young and forming OB stellar clusters within a galaxy. They are characterized by their intense Lyman continuum emission ($\gtrsim$10$^{50}$ LyC photons s$^{-1}$), which heats and excites their surrounding molecular clouds, creating vast, bright ionized zones \citep{1970IAUS...38..107M, 2004MNRAS.355..899C}. Their infrared and radio continuum emissions, often spanning several to tens of parsecs, make them dominant contributors to a normal galaxy’s bolometric luminosity \citep{1980A&A....90..246I}. These regions provide insights into the earliest stages of life of massive stars, which are powerful enough to influence the chemistry and dynamics of the interstellar medium, and can trigger or suppress new generations of star formation. Studying G\ion{H}{2} regions helps improve our understanding of stellar populations across different galactic environments and serve as templates for studying similar star-forming environments in distant galaxies. In particular, they are ideal laboratories for studying clustered high-mass star formation, particularly in starburst-like galactic environments. Therefore, understanding G\ion{H}{2} regions is key to unraveling the larger cosmic processes driving star and galaxy evolution.

To better comprehend G\ion{H}{2} regions in our Milky Way, we have been conducting a survey using data from the Stratospheric Observatory For Infrared Astronomy (SOFIA) and its mid-infrared instrument, FORCAST. These observations provide the highest spatial resolution ($\lesssim 3\arcsec$) mid-infrared images of the entire infrared-emitting areas of G\ion{H}{2} regions without saturation, essential for detecting the earliest stages of massive star formation, which are often obscured in optical and near-infrared wavelengths due to high extinction. We combine these data with archival near-infrared Spitzer and far-infrared Herschel data to quantify the the physical properties of G\ion{H}{2} regions and their presently-forming massive stellar populations. 

In this paper we will concentrate on the G\ion{H}{2} region W43~Main. This G\ion{H}{2} region exists within the much larger W43 cloud complex, an expansive giant molecular cloud structure in the Milky Way's first Galactic quadrant, spanning 1.6$\arcdeg$ of sky, or $>$150\,pc. With a Lyman continuum photon rate of $logN_{LyC}=50.76^{+0.18}_{-0.22}$ photons s$^{-1}$ \citep{2022ApJ...933...60D}, the W43~Main G\ion{H}{2} region ranks as the 14th most-powerful G\ion{H}{2} region in the Milky Way. Although a Wolf-Rayet/OB stellar cluster has been identified near the center of the region behind $\sim$30 magnitudes of extinction \citep{1999AJ....117.1392B}, its known stellar population alone cannot account for the measured total value of $logN_{LyC}$. This mismatch implies that the protostellar population likely contributes substantially to the ionizing photon budget measured. Indeed, W43~Main has many hallmarks of prolific and efficient present star formation, and has sufficient dense molecular reservoirs for future, perhaps intense episodes of massive star formation \citep{2013ApJ...775...88N}. The distance to W43 has been determined through trigonometric parallax observations of masers to be $5.49^{+0.39}_{-0.34}$~kpc, placing it close to the meeting point of the Galactic bar and the Scutum spiral arm \citep{2014ApJ...781...89Z}.

Our first papers\footnote{Specifically, Lim \& De Buizer 2019 (\citetalias{2019ApJ...873...51L}); Lim \& De Buizer 2020 (\citetalias{2020ApJ...888...98L}); De Buizer et al. 2021 (\citetalias{2021ApJ...923..198D}); De Buizer et al. 2023 (\citetalias{2023ApJ...949...82D}); and De Buizer et al. 2024 (\citetalias{2024ApJ...963...55D}).} provide in-depth analyses of individual G\ion{H}{2} regions at relatively large Galactic radii ($R_{GC} = 6.3-10.2$~kpc), whereas our most recent paper (De Buizer et al. 2025; hereafter \citetalias{2025ApJD}), covered three G\ion{H}{2} regions near the Galactic Center ($R_{GC} \sim0.1$~kpc) in the Milky Way's central molecular zone (CMZ). It was found that there are significant differences in the properties of these two groups of G\ion{H}{2} regions, due to the far more extreme environmental conditions experienced by the G\ion{H}{2} regions near the Galactic Center, including tidal forces, shearing forces, crossing orbits from material streams, and much higher overall levels of heat and turbulence. At the meeting point of the Scutum arm and the Galactic bar, W43 represents a source in a Galactic environment distinct from our previously studied G\ion{H}{2} regions. It lies far enough from the Galactic Center that it does not experience the same disruptive forces as the CMZ G\ion{H}{2} regions. But at a position near where a spiral arm and the Galactic bar interact, its environment may not be as quiescent as the other G\ion{H}{2} regions we have studied at large Galactic radii. Indeed, W43's location is thought to be subject to increased turbulence and complexity due to converging flows of material \citep{2011A&A...529A..41N}, making the source an interesting case study distinct from our previously studied G\ion{H}{2} regions.

\begin{figure*}[tb!]
\epsscale{0.90}
\plotone{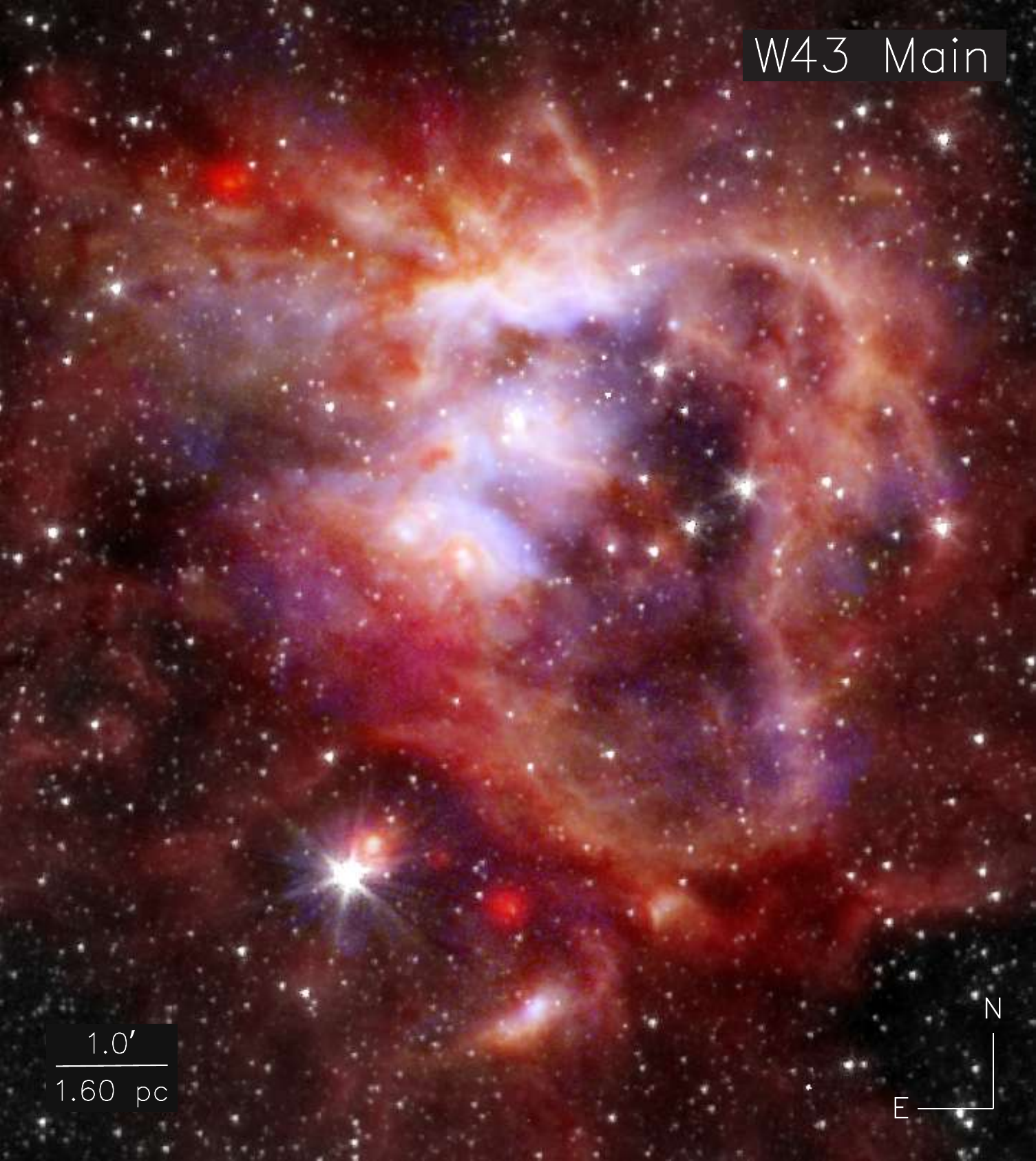}
\caption{A 4-color image of the central $9\farcm2\times10\farcm3$ (14.7$\times$16.5\,pc) region of W43 Main. Blue is the SOFIA-FORCAST 20\,$\mu$m image, green is the SOFIA-FORCAST 37\,$\mu$m image, and red is the Herschel-PACS 70\,$\mu$m image. Overlaid in white is the Spitzer-IRAC 3.6\,$\mu$m
image, which traces the revealed stars within W43 Main, field stars, and hot dust.\label{fig:fig1}}
\end{figure*}

\begin{figure*}[tb!]
\epsscale{0.90}
\plotone{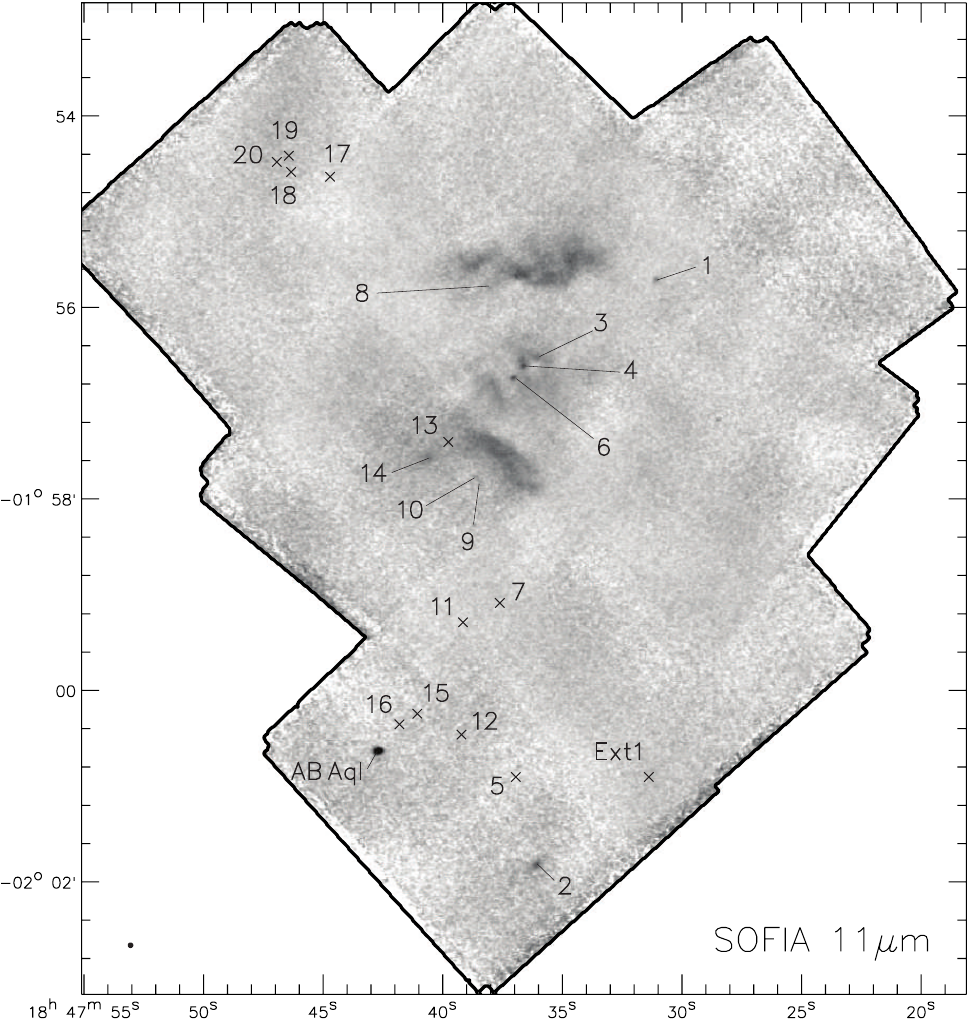}
\caption{W43~Main image mosaic taken at 11\,$\mu$m by SOFIA shown in inverse color (i.e. brighter features are darker in color). Sources labeled with numbers are the mid-infrared compact sources identified in this work. The positions of the mid-infrared sources marked with an `x' are not seen at this wavelength. The small black dot in the lower left indicates the resolution of the image at this wavelength.\label{fig:fig2}}
\end{figure*}

\begin{figure*}[tb!]
\epsscale{0.90}
\plotone{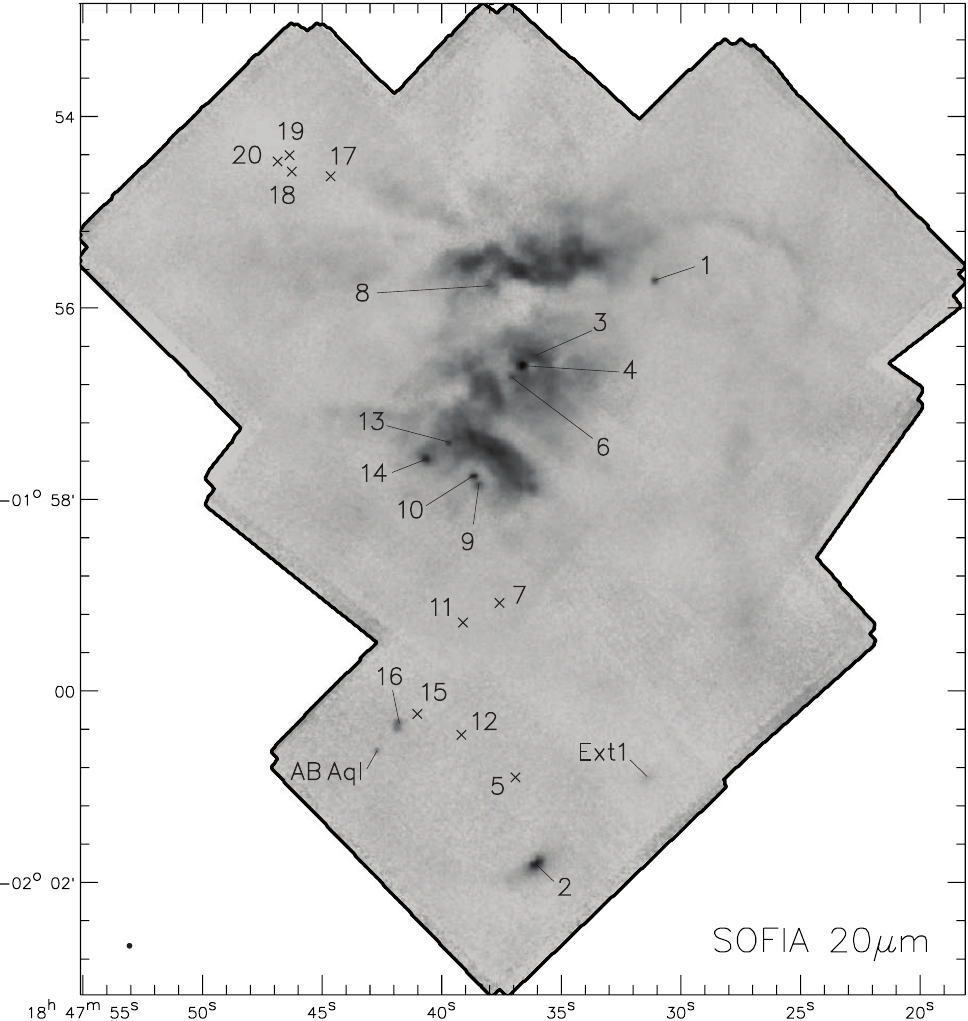}
\caption{W43~Main image mosaic taken at 20\,$\mu$m by SOFIA shown in inverse color (i.e. brighter features are darker in color). See Figure~\ref{fig:fig2} caption for further information.\label{fig:fig3}}
\end{figure*}

\begin{figure*}[tb!]
\epsscale{0.90}
\plotone{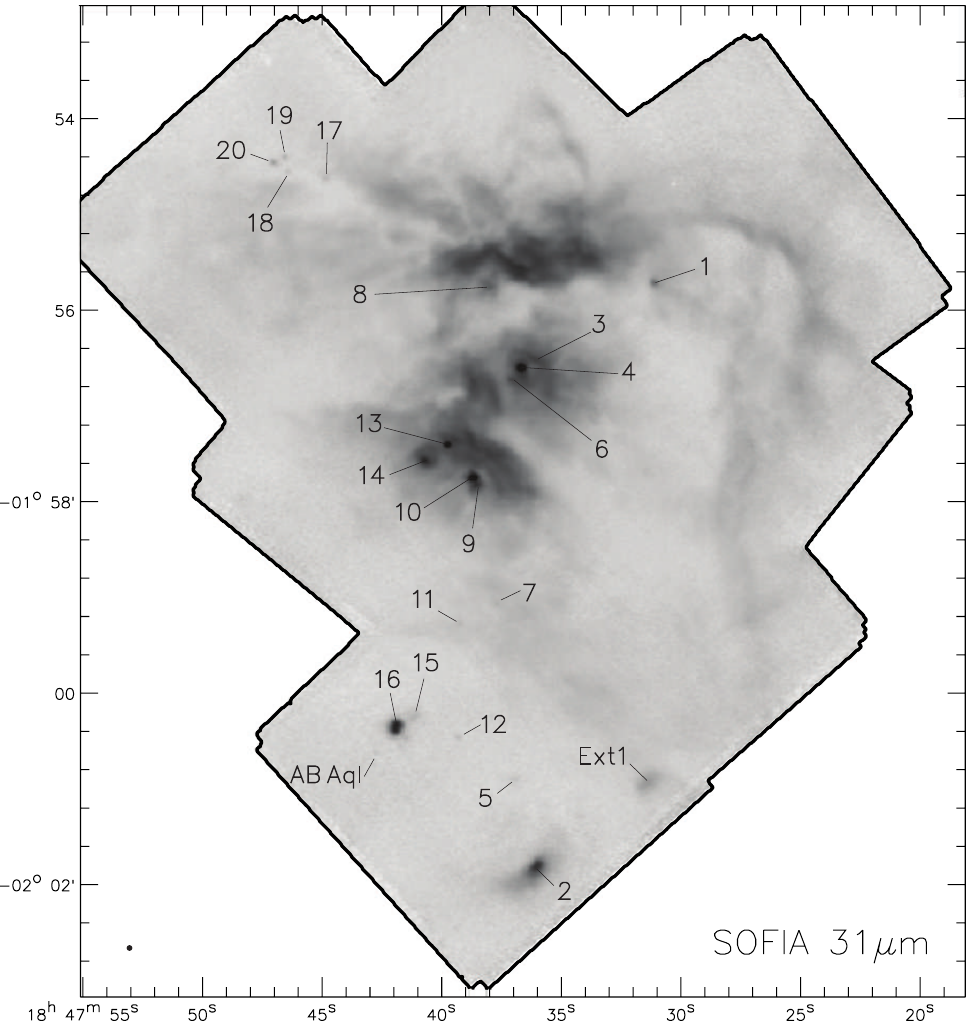}
\caption{W43~Main image mosaic taken at 31\,$\mu$m by SOFIA shown in inverse color (i.e. brighter features are darker in color). See Figure~\ref{fig:fig2} caption for further information. \label{fig:fig4}}
\end{figure*}

\begin{figure*}[tb!]
\epsscale{0.90}
\plotone{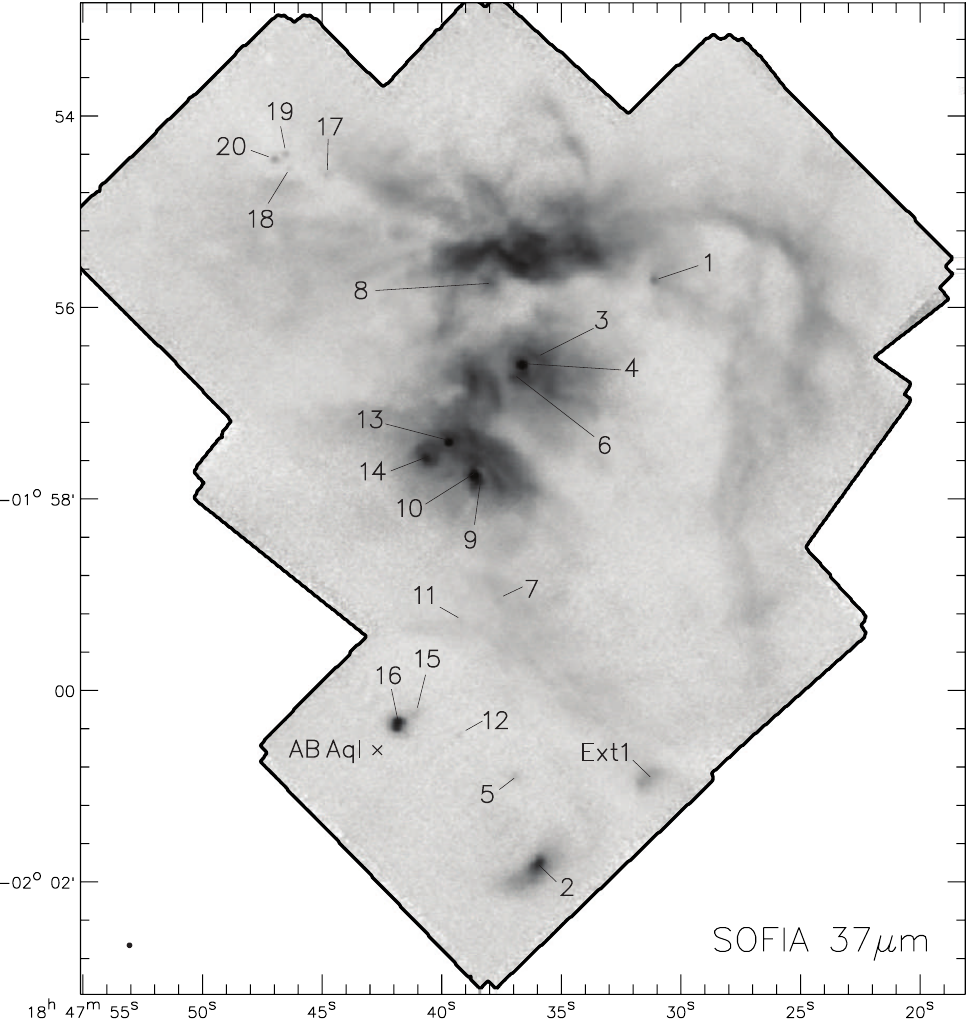}
\caption{W43~Main image mosaic taken at 37\,$\mu$m by SOFIA shown in inverse color (i.e. brighter features are darker in color). See Figure~\ref{fig:fig2} caption for further information.\label{fig:fig5}}
\end{figure*}

\begin{figure*}[tb!]
\epsscale{1.15}
\plotone{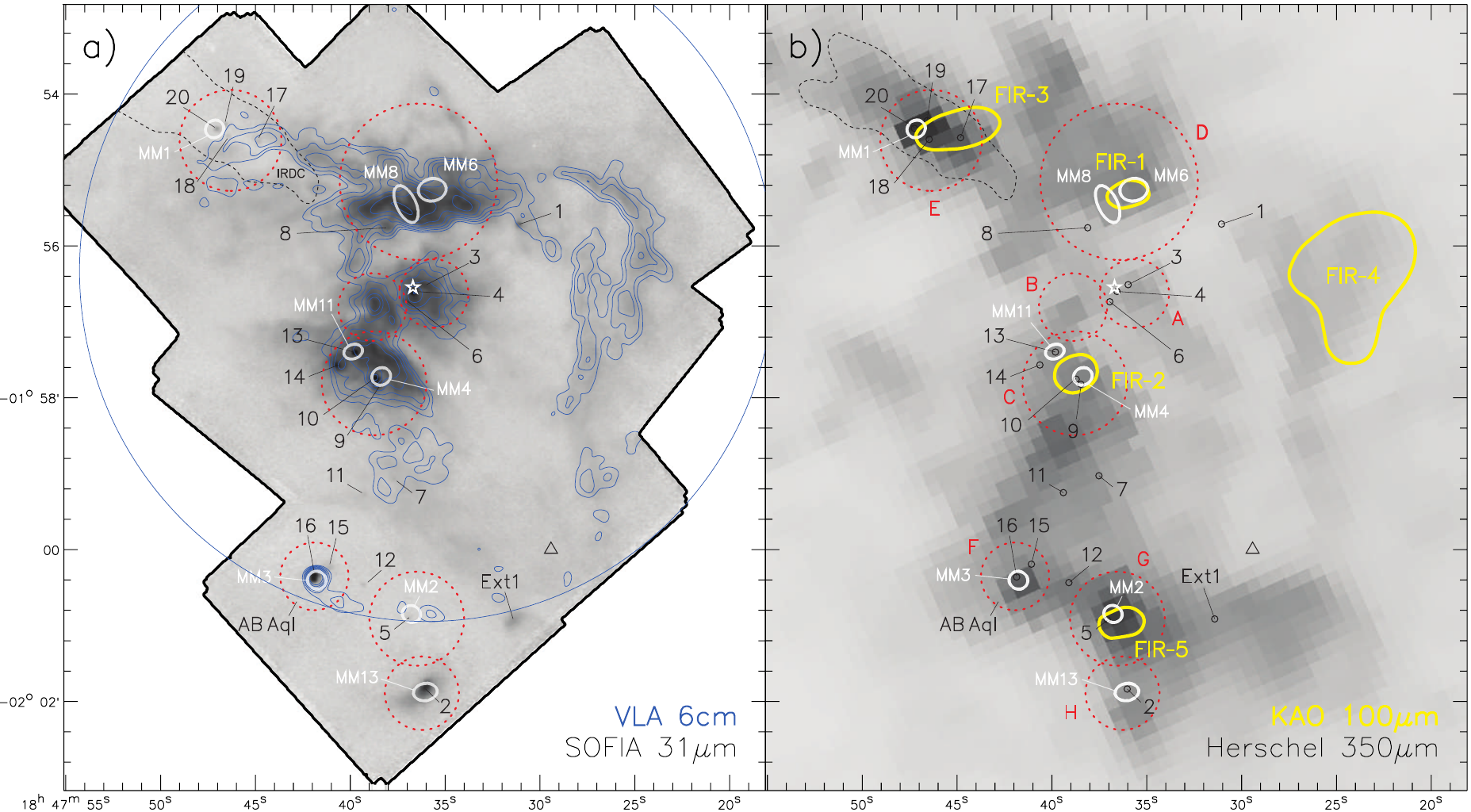}
\caption{Multi-wavelength comparisons of W43~Main. a) The SOFIA 31\,$\mu$m image overlaid with VLA 6\,cm archival data contours (with 4.6$\arcsec$ resolution) in dark blue showing the spatial similarities in wide-spread ionized gas and warm dust emission. The large dark blue circle shows the field of view of this radio continuum image, which does not extend as far south as the SOFIA data. b) The Herschel 350\,$\mu$m image showing the spatial distribution of the cold dust. Overlaid in yellow is the brightest contour for each source from the KAO 100\,$\mu$m image from \citet{1985ApJ...296..565L}, with 100\,$\mu$m source names also labeled in yellow. In both panels of this figure, we give the locations of the SOFIA compact sources in black, as well as the locations and 2-dimensional FWHM sizes of several millimeter cloud fragments from \citet{2003ApJ...582..277M}, which are shown as white ellipses with with their names labeled in white. Also shown in both panels is a white star marking the center of the WR/OB stellar cluster. The area contained within the dashed black lines in the upper left of both panels is an infrared dark cloud (IRDC). The triangle symbol shows the location of the MYSO candidate of \citet{2017ApJ...839..108S}, though we do not detect this source at any SOFIA wavelength. The red dotted circles and red labels mark the identified mid-infrared sub-regions from this work used in the evolutionary analyses, with the circles representing the size of the apertures used for photometry (see Sections~\ref{sec:esm} and \ref{sec:es}). \label{fig:fig6}}
\end{figure*}

\section{Observations and Data Reduction} \label{sec:obs}

The observational techniques and reduction processes employed on the data were similar to all previously studied sources in this survey. Below we will highlight some of observation and reduction details specific to the data for W43~Main. For a more in-depth discussion of these details and techniques we refer the reader to \citetalias{2019ApJ...873...51L}.

Observations were made with the airborne astronomical observatory, SOFIA \citep{2012ApJ...749L..17Y}, utilizing the FORCAST instrument \citep{2013PASP..125.1393H}. Data were taken of W43~Main across two nights during Cycle 3: May 30, 2015 (Flight 212, Cycle OC3C) and June 4, 2015 (Flight 215, Cycle OC3C). All observations were made on flights departing from Palmdale, CA (Latitude $35\arcdeg$) and observed between latitudes of $40\arcdeg$ and $43\arcdeg$. Observations of W43~Main on Flights 212 and 215 were taken at an altitude of 43000\,ft, which has a typical precipitable water vapor overburden of $\sim$6$\,\mu$m (the water vapor monitor was unavailable on these flights). Further observations of the outer areas of W43~Main were taken September 7, 2018 (Flight 501, Cycle OC6J), but were not incorporated into our final data products or analyses due to high noise and distortion from a damaged entrance window on FORCAST.

FORCAST is a facility imager and  spectrograph that employs a Si:As 256$\times$256 blocked-impurity band (BIB) detector  array to cover a wavelength range of 5 to 25\,$\mu$m and a Si:Sb 256$\times$256 BIB array  to cover the range from 25 to 40\,$\mu$m. As with most of the SOFIA data taken for this survey, imaging observations were obtained in the 20\,$\mu$m ($\lambda_{eff}$=19.7\,$\mu$m; $\Delta\lambda$=5.5\,$\mu$m) and 37\,$\mu$m  ($\lambda_{eff}$ = 37.0\,$\mu$m; $\Delta\lambda$ = 3.3\,$\mu$m) filters simultaneously using an internal dichroic. However, unlike our previous papers, we additionally have observations of W43 Main obtained in  the 11\,$\mu$m ($\lambda_{eff}$=11.0\,$\mu$m; $\Delta\lambda$=1.0\,$\mu$m) and 31\,$\mu$m  ($\lambda_{eff}$ = 31.4\,$\mu$m; $\Delta\lambda$ = 5.7\,$\mu$m) filters (also simultaneously taken using the dichroic). In imaging mode the arrays cover a  3$\farcm$40$\times$3$\farcm$20 instantaneous field-of-view with a pixel scale of  0$\farcs$768 pixel$^{-1}$ after distortion correction.

All images were obtained by employing the standard chop-nod observing technique used in ground-based thermal infrared observing, with chop throws ranging from 4$\farcm$3 to 4$\farcm$8 and nod throws from 4$\farcm$8 to 11$\farcm$6 in order to be sufficiently large enough to sample clear off-source sky regions uncontaminated by the extended infrared emission of W43~Main. Since the mid-infrared emitting area of W43~Main was much larger than the FORCAST field of view, it had to be mapped using seven pointings with each field stitched together to create a final image mosaic. The final mosaics at each wavelength extend across an area of approximately $9\farcm2\times10\farcm3$, but do not fill in that entire area, only covering regions of significant mid-infrared emission (with a fill factor of $\sim$56$\%$; see Figures~\ref{fig:fig1}-\ref{fig:fig5}). Each of the seven pointings for the mosaics ranged in total integration times (i.e., SOFIA header value TOTINT) from 85 to 240\,s (not including overlap regions) with the four combined mosaics having average total integration times of $\sim$1000\,s each at 11, 20, 31, and 37\,$\mu$m. 

Due to the data for this project being taken prior to the creation of the Level 4 processing package for the SOFIA Data Cycle System pipeline, the SOFIA Data Archive does not contain any Cycle 3 mosaicked images (i.e., Level 4) products. Therefore, the data mosaics for W43~Main presented here were combined using the final released version of SOFIA Redux software \citep{https://doi.org/10.5281/zenodo.8219569}. Flux calibration was provided by the SOFIA Data Cycle System pipeline and the final total photometric errors in the images were derived using the same process described in \citetalias{2019ApJ...873...51L}. The estimated total photometric errors are 15\% for 20\,$\mu$m and 10\% for 37\,$\mu$m images (which are the same as our previous papers), and 10\% for both the 11 and 31\,$\mu$m images (which are used here in this project for the first time).

Image quality for SOFIA was fairly stable throughout its history, and is quoted as being $\sim$2.6$\arcsec$ (FWHM) resolution at 11.1\,$\mu$m, $\sim$2.5$\arcsec$ at 20\,$\mu$m, $\sim$2.9$\arcsec$ at 31\,$\mu$m, and $\sim$3.2$\arcsec$ at 37\,$\mu$m in the FORCAST Handbook for Archive Users\footnote{https://irsa.ipac.caltech.edu/data/SOFIA/docs/instruments/
handbooks/FORCAST\_Handbook\_for\_Archive\_Users\_Ver1.0.pdf}. 

As was the case with a significant portion of the Cycle 3 FORCAST data, the raw data for W43~Main had highly erroneous World Coordinate System (WCS) astrometry values in the headers. Therefore, the final mosaics presented here had their astrometry absolutely calibrated using Spitzer (24\,$\mu$m if possible, otherwise 8\,$\mu$m) imaging data by matching up the centroids of point sources in common between the Spitzer and SOFIA data using \textit{Aladin Sky Atlas}\footnote{https://aladin.cds.unistra.fr/} \citep{2000A&AS..143...33B}. Absolute astrometry of the final SOFIA images is assumed to be better than $1\farcs5$.

\hspace{2cm}

\section{Comparing SOFIA Images to Previous Imaging Observations} \label{sec:overview}

W43\footnote{a.k.a. G30.8+0.0 or G30.8-0.02, and often mislabeled as G30.8-0.2} was originally discovered as a large radio continuum source by \citet{1958BAN....14..215W} while performing a 21\,cm survey of the Galactic plane. Other coarse resolution ($\gtrsim$3$\arcmin$) observations followed in the radio (e.g., 6\,cm by \citealt{1970AuJPA..14....1G}; 73\,cm by \citealt{1970AuJPA..14...77S}) as well as cold dust \citep[e.g., 100\,$\mu$m continuum by][]{1971ApJ...170L..89H} and molecular material \citep[e.g., $^{12}$CO $J=1-0$ by][]{1986ApJS...60..297C,1986ApJ...305..892D} showed that this was a contiguous and expansive ($>$150\,pc, or 1.6$\arcdeg$) giant molecular cloud complex. The W43 complex has two prominent peaks, the brightest peak seen in the aforementioned surveys is associated with the G\ion{H}{2} region of this study, and is referred to as W43~Main (though, confusingly, just this G\ion{H}{2} region itself is often called W43). The second peak, which lies 0.79$\arcdeg$ away from W43~Main (and therefore not covered by our observations) is often referred to as W43~South (or G29.944-0.042), and is also considered to be a G\ion{H}{2} region in its own right \citep[$logN_{LyC}=50.31^{+0.16}_{-0.19}$ photons s$^{-1}$;][]{2022ApJ...933...60D}, with its radio and infrared emission dominated by the bright compact \ion{H}{2} region G29.96-0.02 \citep{2013A&A...552A.123B, 2002ApJ...564L.101D}. 

The overall shape of the W43~Main G\ion{H}{2} region itself was first resolved by \citet{1969ApJ...156..269S} at 2\,cm, but at a nominal resolution of 2$\arcmin$, they failed to resolve any internal structures. \citet{1974ApJ...194..279T} were the first to resolve W43~Main into multiple peaks in the radio with a resolution $<$20$\arcsec$, however limited baseline coverage led to a complex beam function and the images were hard to interpret. \citet{1978ApJ...221..816R} were able to resolve two main emitting areas at 1.3\,cm with 80$\arcsec$ resolution. The first sub-arcminute resolution view of W43~Main came from \citet{1974ApJ...193..283P} who generated an infrared continuum image of the area at 12.6\,$\mu$m with 12$\arcsec$ resolution (and reported integrated fluxes for the region at 4.9, 8.4, 11.1, 12.6 and 19\,$\mu$m), detecting two large ($\gtrsim$1$\arcmin$ in length) ridges of mid-infrared emission. In our SOFIA data, these two ridges correspond to the infrared extended emission region running east-to-west towards the top of our field (i.e., the sub-region labeled D in Figure~\ref{fig:fig6}), and the extended emission region running northeast-to-southwest region west toward the center of our field (i.e., the sub-region labeled C in Figure~\ref{fig:fig6}).

However, it was the multi-wavelength study of \citet{1985ApJ...296..565L} that really began to bring the nature of the W43~Main region into clearer focus. This study not only produced 6\,cm VLA radio continuum images that resolved internal detail with $\sim$8$\arcsec$ resolution, but also included 40, 50, 100, and 175\,$\mu$m observations with Kuiper Airborne Observatory at an effective 40$\arcsec$ resolution. Despite the difference in spatial resolutions, \citet{1985ApJ...296..565L} were able to see that the ionized gas distribution in W43~Main was well-matched to the far-infrared morphology. They identified 5 main far-infrared emitting regions at 100\,$\mu$m, with the two most prominent peaks (i.e., FIR-1 and FIR-2) coincident with those seen by \citet{1974ApJ...193..283P} at 12.6\,$\mu$m (Figure~\ref{fig:fig6}b). At 50\,$\mu$m, they detected an additional peak in between FIR-1 and FIR-2 (i.e., the sub-region labeled A in Figure~\ref{fig:fig6}), and found that the far-infrared color temperature peaks there. Using a single channel near-infrared bolometer on the NASA Infrared Telescope Facility (IRTF), they detected an unresolved ($<$6$\arcsec$) K-band (2.2\,$\mu$m) source at this location, and speculated that it was the location of the source or sources ionizing the W43~Main G\ion{H}{2} region. 

\citet{1999AJ....117.1392B} were the first to resolve the 2\,$\mu$m source seen by \citet{1985ApJ...296..565L} into a tight stellar cluster at J (1.2\,$\mu$m), H (1.7\,$\mu$m), and K that is highly reddened ($A_V=34$ mags), with the brightest three stars identified spectroscopically as one Wolf-Rayet (W43 \#1, or WR\,121a) and two O giant or supergiant stars (named W43 \#2 and \#3). \citet{2011A&A...532A..92L} were able to resolve both W43 \#1 and W43 \#3 into binaries (in JHK filters with slightly higher resolution). The location of the center of this $\sim3\times10^6 L_{\sun}$ cluster \citep{2013ApJ...775...88N} is marked by the star in Figure~\ref{fig:fig6}.

Observing at 1.3\,mm with $\sim$11$\arcsec$ resolution, \citet{2003ApJ...582..277M} resolved the cold dust structure of the W43~Main region into $\sim$50 fragments, and identified the four brightest fragments (labeling them MM1-MM4) which they characterized as massive protoclusters. The locations and 2-dimensional FWHM sizes of several of these 1.3\,mm fragments are shown as white ellipses in Figure~\ref{fig:fig6}. These sources have become the subjects of much of the recent studies of W43~Main, as we will describe in the discussion of individual and extended sources below (Section~\ref{sec:data}). 

The highest resolution mid to far-infrared images to-date of W43~Main come from the Spitzer Space Telescope and Herschel Space Telescope. Spitzer 8\,$\mu$m and Herschel 70, 160, 250, 350, and 500\,$\mu$m images of the W43~Main G\ion{H}{2} region (e.g., see Figure~\ref{fig:fig6}b) were presented by \citet{2010A&A...518L..90B}, and they identified an infrared dark cloud (IRDC) associated with W43-MM1, the western part of which is traced by the dearth of extended mid-infrared emission surrounding infrared compact sources 17, 18, 19, and 20 (shown as the black dashed area in Figure~\ref{fig:fig6}). Later \citet{2017ApJ...839..108S} presented Spitzer 3.6, 4.5, 5.8, and 8.0\,$\mu$m data of the entire W43 complex identifying 14 massive young stellar object candidates (MYSOs) dispersed throughout. However, only one lies in the W43~Main area covered by our SOFIA data (see triangle symbol in Figure~\ref{fig:fig6}), but we detect no emission from this source at any SOFIA wavelength (and it is therefore highly unlikely to be a MYSO). While the 3.6-8.0\,$\mu$m Spitzer data are fine, W43~Main was too bright for the Spitzer 24 and 70\,$\mu$m channels, and the resultant images are saturated over a large majority of the interesting star-forming areas. The images from the Wide-field Infrared Survey Explorer (WISE) are also saturated at 12 and 22\,$\mu$m. Thus, the best unsaturated mid-infrared images (i.e., 8\,$\mu$m $ < \lambda <$ 70\,$\mu$m) that cover all of W43~Main are from the 12.1 and 21.3\,$\mu$m Midcourse Space Experiment (MSX) images which have a resolution of $\sim$18$\arcsec$ at 21.3\,$\mu$m \citep[see][]{2004MNRAS.355..899C}.

That brings us to the SOFIA mid-infrared data we now present in this work. Our data have an image quality of $\sim$3$\arcsec$ from 11 to 37\,$\mu$m, allowing us to see much finer detail (Figures \ref{fig:fig1}, \ref{fig:fig2}, \ref{fig:fig3}, \ref{fig:fig4}, and \ref{fig:fig5}) than ever before of the entirety of W43~Main at these wavelengths. We see that the large-scale shape in the mid-infrared is a cavity, blowing out to the west (classifying it as a ``cavity-type'' G\ion{H}{2}; see \citealt{2024ApJ...963...55D} for description). \citet{2003ApJ...582..277M} speculate that all (or at least the vast majority) of the free-free emission seen at 3.5\,cm in W43~Main is optically thin and due to the WR/OB cluster (see Figure~\ref{fig:fig6}), and further that the ionization and heating of the cluster dominates the cm and infrared characteristics. However, the combined ionizing output of the confirmed stellar members of the WR/OB cluster is insufficient to account for the total Lyman-continuum photon rate of the entire G\ion{H}{2} region. The required ionizing flux ($logN_{LyC}\sim50.8$) corresponds to approximately 40 O7-type stars, or, expressed in terms of the spectral types of the three confirmed cluster members, roughly 15 O5 or WN7 stars \citep{1996ApJ...460..914V, 2007ARA&A..45..177C}, and thus there is likely a significant contribution from the protostellar population, or from as yet unidentified evolved cluster members. Nevertheless, the ridges and extended structures located to the north and south of the cluster, as revealed in the SOFIA mid-infrared images, coincide spatially with cm-wavelength radio continuum emission and may therefore indeed be tracing ionization fronts, including those at the far western periphery (Figure~\ref{fig:fig6}a). In Figure~\ref{fig:fig1}, there is a bright blue haze around the location of the WR-OB cluster and the ionized ridges just north and south of it, showing the location of the hotter dust. Figure~\ref{fig:fig1} also shows that the stellar cluster has photoablated some of these extended structures into trunks or protuberances, which we will discuss in more detail in Section~\ref{sec:trunks}.  

In Figure~\ref{fig:fig6}, it can be seen that there are bright regions at the longer Herschel wavelengths (e.g., at 350\,$\mu$m) that show little to no extended mid-infrared emission at the SOFIA wavelengths. In these cases, these regions are likely the coldest and densest regions of W43~Main. For instance, the IRDC containing MM1 can be seen as the brightest feature at 350\,$\mu$m and this is seen as a void of mid-infrared extended emission, though it is populated with highly extinguished compact mid-infrared sources (i.e., sources only seen at the longest SOFIA wavelengths). Similarly, the whole cold dust ridge extending from MM3 to MM2/FIR-5 shows no mid-infrared extended emission, but also contains heavily extinguished compact mid-infrared sources.

In the SOFIA images presented in Figures \ref{fig:fig2}, \ref{fig:fig3}, \ref{fig:fig4}, and \ref{fig:fig5}, we see that W43~Main looks similar across the span of mid-infrared wavelengths covered, with some notable exceptions. The 11\,$\mu$m image appears to have far less extended emission than all other wavelengths, and also has the lowest signal. This is the case for two reasons: 1) the 11\,$\mu$m filter has by far the narrowest bandpass ($\Delta\lambda$ = 0.95\,$\mu$m) of the four filters, and thus is indeed effectively less sensitive by comparison, and 2) the whole of W43~Main lies behind a relatively high extinction, as we are viewing it through the tangent of a spiral arm, and thus even the central, evolved, WR/OB cluster has a high measured extinction ($A_V\approx34$ mag; \citealt{1999AJ....117.1392B}). That said, at 11\,$\mu$m the region looks similar to the 12.6\,$\mu$m observations of \citet{1974ApJ...193..283P} (at least for the areas with overlapping coverage). Though the large-scale emission at 20\,$\mu$m in Figure \ref{fig:fig3} appears similar to that at 31 and 37\,$\mu$m, the 20\,$\mu$m image is most noticeably missing the compact sources to the northeast (our sources labeled 17--20), and several to the south (our sources labeled 5, 7, 11, 12, and 15), due to high extinction in those cold, dense regions (more on this in Section~\ref{sec:sources}). The 31 and 37\,$\mu$m images (Figures \ref{fig:fig4} and \ref{fig:fig5}) look almost identical, with the exception of the marginal detection of AB Aql at 31 but not 37\,$\mu$m.

\section{The distance to W43~Main}\label{sec:dist}

\citet{2014ApJ...781...89Z} determined the distance to W43 through trigonometric parallax observations of masers, finding a value of $5.49^{+0.39}_{-0.34}$~kpc. This value was determined via a variance-weighted average of the distances determined from four water and methanol maser locations within the larger W43 complex, none of which were located directly in the W43~Main G\ion{H}{2} region. Given the contiguous emission seen throughout the complex at far-infrared wavelengths, and in molecular data \citep[including being enveloped in an \ion{H}{1} envelope $\sim$270\,pc in diameter;][]{2014A&A...571A..32M}, the whole W43 complex is believed by many to be a single entity \citep[][and references therein]{2013A&A...560A..24C}. This distance would place W43 close to the meeting point of the Galactic bar and the Scutum spiral arm. This distance is also consistent to within the errors with the kinematically derived near distances quoted for this region, for example, the H91$\alpha$ transition velocity of 92.02$\pm$0.04 km/s from \citet{2006ApJ...653.1226Q}, which yields a near kinematic distance of 5.57$_{-0.73}^{+0.38}$ kpc \citepalias{{2022ApJ...933...60D}}. Evidence that the near kinematic distance is preferred over the far distance comes from \citet{2011MNRAS.411..705M} who spectrophotometrically derived a distance of 4.90$\pm$1.91\,kpc by directly observing the spectral properties of several stars in the central WR/OB cluster of the W43~Main G\ion{H}{2} region. This spectrophotometrically derived distance agrees with the one found via maser parallaxes to within the quoted errors.  

In the more recent work of \citet{2019ApJ...885..131R}, parallax and proper motions are given for two methanol maser sites directly within the W43~Main G\ion{H}{2} region. The first, called G30.70-0.06, is coincident with MM2 (source 5) and using their published proper motions and inputting them into the \textit{Parallax-Based Distance Calculator v2}\footnote{From the Bar and Spiral Structure Legacy (BeSSeL) Survey website, http://bessel.vlbi-astrometry.org.}, the distance to this maser site is derived to be 5.47$\pm$0.74\,kpc. The second maser site, called G30.74-0.04, is coincident with MM11 (source 13), and we find a distance of 5.27$\pm$0.72\,kpc using the same methodology as used for the first maser site.

Since these derived parallactic distance values for masers directly within the W43~Main G\ion{H}{2} region have slightly larger errors but are consistent with the more precise distance derived by \citep{2014ApJ...781...89Z}, we adopt their distance of $5.49^{+0.39}_{-0.34}$~kpc in this work.

\begin{deluxetable*}{lccl}
\tabletypesize{\scriptsize}
\tablecolumns{4}
\tablewidth{0pt}
\tablecaption{SOFIA-Detected Infrared Sources in W43~Main}
\tablehead{\colhead{    }&
           \multicolumn{2}{c}{Source Centroid}&          
           \colhead{    } \\ [-4pt]
           \colhead{Source }&
           \colhead{ R.A.}&
           \colhead{ Decl. }&  
           \colhead{ Coincidences } 
}
\startdata
Ext1 &	18 47 31.03	&	-02 00 55.6	 &	\\
1	&	18 47 31.07	&	-01 55 42.5	 &	G030.7688-0.0078	    \\
2$^b$	&   18 47 35.99 &   -02 01 48.5  & Gaia DR3 4259130986533128448 \\ 
3	&	18 47 36.06	&	-01 56 30.2	 &   G030.7660-0.0322		\\
4	&	18 47 36.59	&	-01 56 35.8	 &   W43\#3$^d$, G030.7662-0.0348		\\
5$^a$	&	18 47 36.91	&	-02 00 53.8	 & Gaia DR3 4259131222754720512, MM2		\\ 
6	&	18 47 37.01	&	-01 56 43.2	&   G030.7652-0.0375, Gaia DR3 4259134864887880064 		\\ 
7$^a$	&	18 47 37.59	&	-01 59 04.1	&	G030.7314-0.0576, Gaia DR3 4259131433209784192	\\ 
8	&	18 47 37.94	&	-01 55 45.8	&   G030.7808-0.0333		\\
9$^b$	&	18 47 38.44  &  -01 57 50.7	&	G030.7514-0.0511	\\ 
10	&	18 47 38.61	&	-01 57 45.5	&   G030.7529-0.0512		\\
11$^{a,b}$	&	18 47 39.17	&	-01 59 17.0	&	G030.7315-0.0647\\
12$^a$	&	18 47 39.23	&	-02 00 28.9	& Gaia DR3 4259131325835056000, MM10 \\ 
13	&	18 47 39.67	&	-01 57 24.7	&   G030.7598-0.0525, MM11		\\
14	&	18 47 40.61	&	-01 57 34.6	&	G030.7594-0.0571	\\
15$^{a,b,c}$	&18 47 41.12	&	-02 00 15.9	& G030.7202-0.0797		\\ 
16$^{b,c}$	&	18 47 41.78	&	-02 00 20.0 &   G030.7206-0.0825, MM3		\\ 
AB Aql	&	18 47 42.65	&	-02 00 37.6	&   Gaia DR3 4259131158334377472		\\ 
17$^a$	&	18 47 44.79	&	-01 54 36.7	&   G030.8109-0.0501		\\
18$^a$	&	18 47 46.34	&	-01 54 32.9	& MM5		\\
19	&	18 47 46.45	&	-01 54 23.7	&		\\
20$^a$	&	18 47 46.92	&	-01 54 26.9	& MM1
\enddata
\tablecomments{Coordinates are from the SOFIA 20\,$\mu$m data, unless otherwise noted. Potentially coincident source names beginning with ``G'' are from \citet{2017ApJ...839..108S}, and are coincident with the SOFIA source coordinates to within 1.5$\arcsec$ unless otherwise noted. The ``MM'' sources are the millimeter fragment names from \citet{2003ApJ...582..277M}, and are given for sources where the mid-infrared peak lies within the 1\,mm size ellipse. GAIA sources coincident to within the FORCAST astrometric accuracy (1.5$\arcsec$) are also given.}
\tablenotetext{a}{Due to not being detected at 20\,$\mu$m, the coordinates for this source are based upon the 37\,$\mu$m data.} 
\tablenotetext{b}{Due to source asymmetry, the aperture centroid was different than the source centroid for these sources. Aperture centroids are: 2 (18 47 36.05,	-02 01 49.8), 9 (18 47 38.40, -01 57 52.5), 11 (18 47 39.22, -01 59 16.2), 15 (18 47 41.07, -02 00 14.6), and 16 (18 47 41.78, -02 00 21.5). }
\tablenotetext{c}{Has two peaks marginally resolved from each other so source centroid is given for brightest peak in mid-infrared.} 
\tablenotetext{d}{From \citet{1999AJ....117.1392B}.}
\label{tb:SOFIA_compactA}
\end{deluxetable*}

\begin{deluxetable*}{lccrrrrrrrrrrrr}
\tabletypesize{\scriptsize}
\tablecolumns{9}
\tablewidth{0pt}
\tablecaption{SOFIA Photometry of Mid-Infrared Sources in W43~Main}
\tablehead{\colhead{  }&
           \multicolumn{3}{c}{${\rm 11\mu{m}}$}&
           \multicolumn{3}{c}{${\rm 20\mu{m}}$}&
           \multicolumn{3}{c}{${\rm 31\mu{m}}$}&
           \multicolumn{3}{c}{${\rm 37\mu{m}}$}\\ [-4pt]
           \cmidrule(lr){2-4} \cmidrule(lr){5-7} \cmidrule(lr){8-10} \cmidrule(lr){11-13}\\ [-12pt]
           \colhead{ Source }&       
           \colhead{ $R_{\rm int}$ } &
           \colhead{ $F_{\rm int}$ } &
           \colhead{ $F_{\rm int-bg}$ } &
           \colhead{ $R_{\rm int}$ } &
           \colhead{ $F_{\rm int}$ } &
           \colhead{ $F_{\rm int-bg}$ } &
           \colhead{ $R_{\rm int}$ } &
           \colhead{ $F_{\rm int}$ } &
           \colhead{ $F_{\rm int-bg}$ } &
           \colhead{ $R_{\rm int}$ } &
           \colhead{ $F_{\rm int}$ } &
           \colhead{ $F_{\rm int-bg}$ } \\ [-6pt]
	   \colhead{  } &	   
	   \colhead{ ($\arcsec$) } &
	   \colhead{ (Jy) } &
	  \colhead{ (Jy) } &
	   \colhead{ ($\arcsec$) } &
	   \colhead{ (Jy) } &
	   \colhead{ (Jy) } &
	   \colhead{ ($\arcsec$) } &
	   \colhead{ (Jy) } &
	  \colhead{ (Jy) } &
	   \colhead{ ($\arcsec$) } &
	   \colhead{ (Jy) } &
	   \colhead{ (Jy) } \\ [-12pt]
}
\startdata
Ext1	&	9.2	&	UD	&	0.108	&	9.2	&	7.51	&	3.93	&	12.3	&	52.4	&	36.3	&	12.3	&	70.6	&	55.5  \\
1	&	6.1	&	1.16	&	0.592	&	6.1	&	12.3	&	5.26	&	6.9	&	34.7	&	12.6	&	7.7	&	52.4	&	15.2       \\
2	&	10.7	&	6.24	&	3.01	&	10.7	&	52.2	&	41	&	10.7	&	164	&	137	&	10.7	&	208	&	164  \\
3   &   5.4	&	2.76	&	1.17	&	6.9	&	53.1	&	17.6	&	6.9	&	150	&	43	&	6.9	&	177	&	48.3 \\
4	&	6.1	&	3.99	&	1.32	&	6.1	&	54.2	&	26.7	&	6.9	&	222	&	126	&	6.9	&	276	&	161  \\
5	&	3.8	&	UD	&	0.108	&	3.8	&	UD	&	0.262	&	8.4	&	3.41	&	3.37	&	8.4	&	10.2	&	8.53  \\
6	&	4.6	&	1.96	&	0.584	&	4.6	&	17.8	&	4.63	&	4.6	&	64.2	&	14.8	&	4.6	&	82.1	&	21.7  \\
7	&	3.8	&	UD	&	0.108	&	3.8	&	UD	&	0.262	&	6.1	&	12	&	0.776	&	6.1	&	19.5	&	1.77 \\
8	&	7.7	&	3.3	&	0.836	&	8.4	&	38.7	&	12.5	&	8.4	&	120	&	38.9	&	8.4	&	150	&	42.8 \\
9	&	4.6	&	1.06	&	0.117	&	5.4	&	16.1	&	6.02	&	6.1	&	97.4	&	35.9	&	6.1	&	145	&	56.6  \\
10	&	5.4	&	1.56	&	0.192	&	4.6	&	16.8	&	8.76	&	6.1	&	159	&	73.7	&	6.1	&	236	&	104 \\
11	&	3.8	&	UD	&	0.108	&	3.8	&	UD	&	0.262	&	5.4	&	6.43	&	1.22	&	5.4	&	7.78	&	1.59 \\
12	&	3.8	&	UD	&	0.108	&	3.8	&	UD	&	0.262	&	4.6	&	1.82	&	1.06	&	4.6	&	1.34	&	1.3  \\
13	&	3.8	&	1.16	&	UD	&	3.8	&	14.2	&	3.16	&	4.6	&	100	&	39.1	&	4.6	&	140	&	62.2  \\
14	&	8.4	&	6.29	&	2.04	&	8.4	&	48.6	&	26.2	&	10.7	&	307	&	143	&	10.7	&	415	&	206  \\
15	&	3.8	&	UD	&	0.108	&	3.8	&	UD	&	0.262	&	5.4	&	7.63	&	6.07	&	5.4	&	11	&	8.86 \\
16	&	3.8	&	UD	&	0.108	&	7.7	&	14.5	&	7.72	&	10.0	&	116	&	110	&	10.7	&	150	&	142 \\
AB Aql	&	6.1	&	6.1	&	5.45	&	5.4	&	5.05	&	1.95	&	3.8	&	1.15	&	0.362	&	5.4	&	UD	&	0.608  \\
17	&	3.8	&	UD	&	0.108	&	3.8	&	UD	&	0.262	&	5.4	&	4.09	&	2.91	&	5.4	&	8.03	&	4.12	\\
18	&	3.8	&	UD	&	0.108	&	3.8	&	UD	&	0.262	&	3.8	&	1.2	&	0.586	&	3.8	&	1.6	&	0.696	\\
19	&	3.8	&	UD	&	0.108	&	3.8	&	UD	&	0.262	&	3.8	&	2.21	&	1.4	&	3.8	&	3.59	&	2.87	\\
20	&	3.8	&	UD	&	0.108	&	3.8	&	UD	&	0.262	&	5.4	&	4.88	&	4.27	&	5.4	&	7.71	&	7.04	
\enddata
\tablecomments{If there is no $F_{\rm int-bg}$ value for a source, then the source is not well resolved from other nearby sources and/or extended emission. For these sources, the $F_{\rm int}$ value is used as the upper limit in the SED modeling. `UD' means the source is undetected. For all undetected sources (as well as unresolved sources), if there is significant background emission or contamination from nearby sources, then the flux upper limit given is the measured flux in the indicated aperture, otherwise the 3-sigma upper limit for a point source detection is used.}
\label{tb:SOFIA_compactB}
\end{deluxetable*}

\section{Data Analysis and Results}\label{sec:data}

As we have done in our other papers in this survey, we further subdivide the infrared sources for each region into whether they are compact or extended sources. The two categories denote which objects we believe are star-forming cores (compact; $\lesssim$0.3 pc in size) versus the larger star-forming molecular clumps (extended). As we have done previously for the compact sources, we will fit SED models to their multi-wavelength photometry to estimate their physical properties and identify which compact infrared sources potentially house MYSOs. For the regions identified as extended, we will employ the same evolutionary analyses that we have in our previous papers to try to learn more about the evolutionary properties of the clumps and the entire G\ion{H}{2} region as a whole.  

\begin{figure*}[tb!]
\epsscale{0.6}
\plotone{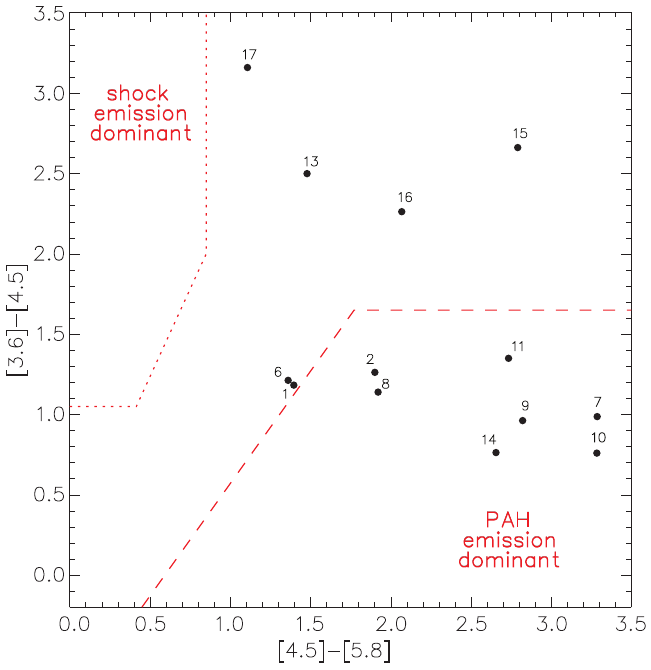}
\caption{A color-color diagram for compact sources in W43~Main utilizing our background-subtracted Spitzer-IRAC 3.6, 4.5, and 5.8\,$\mu$m source photometry to distinguish ``shocked emission dominant'' and ``PAH emission dominant'' YSO candidates from our list of compact sub-components. Above (up-left) the dotted line indicates shock emission dominant regime. Below (bottom-right) the dashed line indicates PAH dominant regime. We adopt this metric from \citet{2009ApJS..184...18G}. Some sources are not included in this diagram due to non-detection or saturation in the Spitzer-IRAC bands.}\label{fig:ccd}
\end{figure*}

\begin{figure*}[t]
\epsscale{0.98}
\plotone{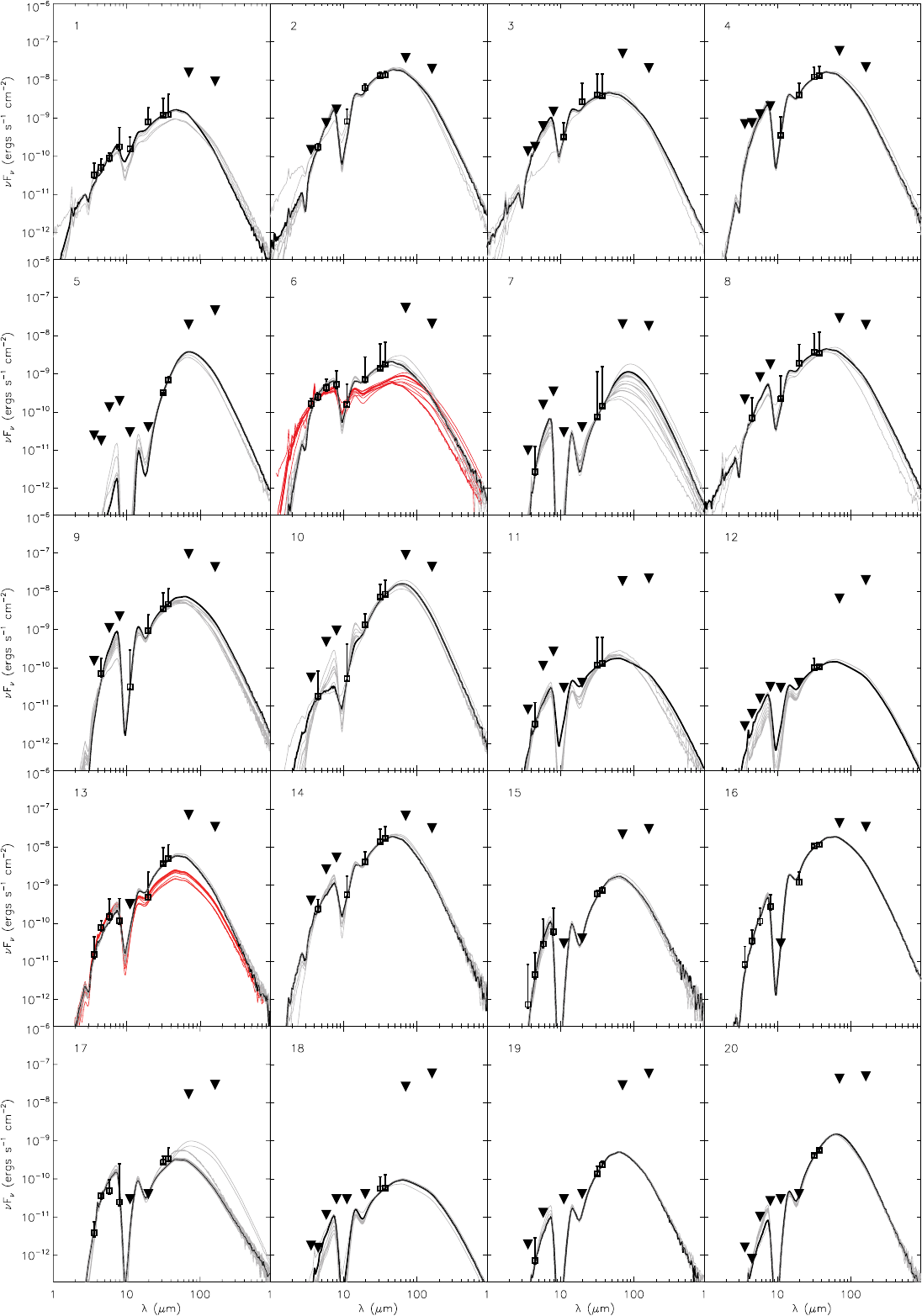}
\caption{SED fitting with ZT MYSO SED models for compact sources in W43~Main. Black lines are the best fit model to the SEDs, and the system of gray lines are the remaining fits in the group of best fits (from Table \ref{tb:seds}). Upside-down triangles are data that are used as upper limits in the SED fits. For Sources 6 and 13 only, the red lines are fits using the nominal upper error bars, and black/gray lines are fits use upper errors matched to the lower errors.\label{fig:SED}}
\end{figure*}

\subsection{Compact Infrared Sources: SED Model Fitting and Determining MYSO Candidates}\label{sec:cps}

A compact source is defined as one that exhibits a distinct peak, maintaining a consistent location across multiple Spitzer and SOFIA wavelengths and is detectable at more than one wavelength. To identify compact source candidates, we first look for resolved sources or peaks in the SOFIA data and then cross-check these with Spitzer-IRAC, Herschel-PACS, and cm radio data to confirm spatial coincidences. We define compact sources as having physical sizes smaller than approximately 0.3~pc, which aligns with the typical size of molecular cores, around 0.1 pc \citep[e.g.,][]{2007ARA&A..45..481Z}. 

We identified 20 compact infrared sources in the SOFIA-mapped area of W43~Main, and  Table~\ref{tb:SOFIA_compactA} provides details on the compact source positions. Often the compact sources seen in the SOFIA data are situated to within our astrometric accuracy of being coincident with known sources detected at other wavelengths, and we tabulate these potential source aliases in Table~\ref{tb:SOFIA_compactA}. However, given the differences in wavelength (and often spatial resolution), and given the density of sources seen in projection toward W43~Main, it is not certain that in all cases the emission of the sources given by these aliases are actually tracing the same sources as we are seeing in the SOFIA-FORCAST data. Since there exists no previously labeled sources at the wavelengths and resolution of our FORCAST data, we will give our compact infrared sources unique numerical labels to avoid assuming any potentially erroneous coincidences.

Table~\ref{tb:SOFIA_compactB} provides the radii used for aperture photometry of each source given in Table~\ref{tb:SOFIA_compactA}, flux densities within those apertures, and background-subtracted flux densities measured at all SOFIA wavelengths. The aperture sizes were found at each wavelength via a optimal extraction technique as was used in \citetalias{2019ApJ...873...51L}, and background subtraction was also performed in the same way as \citetalias{2019ApJ...873...51L} (i.e. using background statistics from an annulus outside the optimal extraction radius which had the least environmental contamination).

For each SOFIA-detected source, we combine the SOFIA photometry values with the photometry values we derive from data from the Spitzer and Herschel missions (provided in Appendix B) in order to create spectral energy distributions (SEDs). Specifically, we perform aperture photometry on the Spitzer-IRAC 3.6, 4.5, 5.8, 8.0\,$\mu$m data and Herschel-PACS 70 and 160\,$\mu$m data, employing the same optimal extraction technique and background subtraction technique as was used for the SOFIA photometry.

As we did in \citetalias{2019ApJ...873...51L}, we used a color-color ($3.6-4.5$\,$\mu$m vs. $4.5-5.8$\,$\mu$m) plot analysis developed by \citet{2009ApJS..184...18G} to determine which Spitzer-IRAC photometric data may be contaminated with excess flux at 3.6, 5.8, and 8.0\,$\mu$m from polycyclic aeromatic hydrocarbons (PAHs) or at 4.5\,$\mu$m from shocked-excited H$_2$ emission. Figure~\ref{fig:ccd} shows the color-color plot for compact sources in W43~Main. This plot reveals that seven sources show contamination from PAH emission, and as a consequence all such sources have their Spitzer-IRAC 3.6, 5.8, and 8.0\,$\mu$m photometric values set as upper limits in the SED diagrams (i.e., only the 4.5\,$\mu$m data will be treated as a valid nominal data point). No sources are found to be shock emission dominant. Therefore, for all other sources plotted on the diagram in Figure~\ref{fig:ccd}, we use the photometric values measured in all four Spitzer bands as nominal data points in the SED diagrams and fitting. There are some sources missing from the color-color diagram (sources 3-5, 12, and 18-20) due to non-detection or confusion with nearby sources in the Spitzer-IRAC bands. We also do not perform color-color analysis or SED fitting for AB Aql, as it is not a YSO (see Section~\ref{sec:mm2mm3}). As we did in \citetalias{2019ApJ...873...51L}, we will conservatively assume that the sources missing from Figure~\ref{fig:ccd} are PAH emission dominant, and therefore only the 4.5\,$\mu$m data will be treated as a nominal data point, as our previous papers have found that PAH emission dominant sources are most common in the G\ion{H}{2} regions.

In constructing the spectral energy distributions (SEDs) for our compact sources, we treat the Herschel-PACS measurements as upper limits. This approach accounts for the considerable and uncertain contamination from extended background emission, as well as source blending caused by the lower angular resolution at 70 and 160\,$\mu$m (i.e., $\sim$5.6$\arcsec$ and $\sim$11.4$\arcsec$, respectively), both of which hinder reliable isolation of the flux of the individual compact flux components intrinsic to the source.

Regarding photometric uncertainties, we follow the procedure established in our earlier work. Specifically, we assign the upper error bars based on the value of the subtracted background flux, recognizing that while background subtraction can fluctuate, it does not exceed the subtracted amount. Consistent with our previous analyses, the lower error bars are determined using the mean total photometric uncertainty at each wavelength: 20\%, 15\%, and 10\% for the 4.5, 20, and 37\,$\mu$m bands, respectively. For the new SOFIA-FORCAST wavelengths in this work, we use 10\% for both the 11\,$\mu$m and 31\,$\mu$m photometric uncertainties. For the Spitzer-IRAC bands at 3.6, 5.8, and 8.0\,$\mu$m, we assume a 20\% uncertainty, provided the source is free from contamination by polycyclic aromatic hydrocarbon (PAH) emission features (as determined from Figure~\ref{fig:ccd}). In line with \citetalias{2019ApJ...873...51L}, we adopt photometric error estimates of 40\% and 30\% for the Herschel 70 and 160\,$\mu$m bands, respectively.

We constructed near- to far-infrared spectral energy distributions (SEDs) for all compact sources using photometric data and associated uncertainties from SOFIA, Spitzer, and Herschel. These SEDs were then analyzed using the theoretical model SEDs for MYSOs and fitting algorithm developed by \citet{2011ApJ...733...55Z}. In that work, radiative transfer is simulated under the assumption of a massive protostar, surrounded by an active accretion disk, which is encompassed in a hydrostatic core with dust-free outflow cavities, at multiple orientations, yielding model SEDs. The fitter from that work compares the observed data to a grid of these discretely parameterized theoretical models (which we refer to here as the ``ZT MYSO SED models''). Each fit yields a normalized minimum chi-squared value ($\chi_{\rm nonlimit}^2$), which serves as a metric for assessing the quality of the fit. Consistent with the methodology adopted in our previous studies, we selected the group of models whose $\chi_{\rm nonlimit}^2$ values are comparable to that of the best-fit model and clearly separated from the next group of models with significantly higher chi-squared values. In cases where the top one or several fits yield substantially lower $\chi_{\rm nonlimit}^2$ values than subsequent models, we include these additional fits in the best-fit group to ensure a minimum of five representative models is retained for further analysis. Figure~\ref{fig:SED} shows the observed SEDs, as well as the best and group of best fits by the ZT MYSO SED models for the 20 sources in W43~Main. The physical parameters of these best-fit models are summarized in Table~\ref{tb:seds}. 

For two compact infrared sources, 6 and 13, we show two sets of best fits in Figure~\ref{fig:SED}. The red fits are the nominal fits, but as can be seen for both of these sources, the SED model fitter does a poor job of fitting the longer wavelength SOFIA data points. The black sets of fits were achieved by decreasing the upper errors to match the lower errors for all wavelengths. This shows that the data can indeed be well fit by MYSO models, and for these two sources the parameters we report in Table~\ref{tb:seds} are derived from these better fits, which we believe to be the best estimates of these parameters. Further testing of the ZT SED model fitting algorithm shows that in some instances it deals with large errors poorly, especially when data density is increased. In these instances, it can overweigh the significance of the data points with lower errors at the expense of not fitting the data points with very large errors at all. This issue was not encountered in our previous papers, but we have additional SOFIA data points for the first time in this work (and hence, higher data density than before). 

Table~\ref{tb:seds} lists the physical properties of the MYSO SED model fits for each source. The observed bolometric luminosities, $L_{\rm obs}$, of the best fit models are presented in column~2 and are found via a simple model assuming isotropic radiation that is fit to the data with a correction for foreground extinction (using the distance to the source). The true total bolometric luminosities, $L_{\rm tot}$, are in column~3 and are calculated assuming the presence of the core geometry described above and correcting for foreground extinction and system viewing angles. The extinction and the stellar mass of the single best models are listed in column~4 and 5, respectively. Columns~6 and 7 present the ranges of the foreground extinction and stellar masses derived from the models in the group of best model fits for which we provide the number of the models in those groups in column~8. 

In the far right column of Table~\ref{tb:seds} we give our assessment of the nature of each compact source. Using the results of the group of best-fit model parameters we identify sources as massive young stellar objects (MYSOs) or potential MYSO candidates (pMYSOs). Following our previously established criteria, a source is classified as a MYSO candidate if it meets the following criteria: (1) the observed SED is reasonably well-fit by the MYSO models (column 9 of Table~\ref{tb:seds}), (2) the best-fit model yields a stellar mass $M_{\rm star} \geq 8M_{\sun}$ (column 5), and (3) all models within the selected group of best-fit solutions have $M_{\rm star} \geq 8M_{\sun}$ (column 7). Sources that satisfied only the first two criteria were designated as “potential MYSOs” (pMYSOs), as in our earlier papers. Additional information used in the assessment of the nature of each sources comes from Column 10 of Table\,\ref{tb:seds}, which lists other auxiliary indicators of sources being YSOs, like the presence of water or hydroxyl masers, or if the source has a known outflow (which is an indicating of ongoing accretion). Beside ionized emission in the form of cm radio continuum detections, a source is very likely to be a massive YSO if there are methanol masers present (a signpost of MYSO activity).

We must add the caveat that, the ZT MYSO SED model fitting assumes a single central star. Given the likely high multiplicity fraction among massive stars in general \citep[e.g.,][]{2007ARA&A..45..481Z}, several of the infrared compact sources considered here may host protobinaries (or even small protoclusters). As we discussed in \citetalias{2021ApJ...923..198D}, given the high total luminosity and internal mass estimates from the fittings, at least one member of a potential protobinary would have to be massive, and determining which compact infrared objects host MYSOs is the main goal here. Furthermore, the use of the axillary indicators described can provide further evidence that a particular source hosts a MYSO.

We have identified three compact infrared sources (specifically, 3, 4, and 6) that appear very close to the ionizing cluster in projection. Given the caustic nature of immediate environments around the WR/OB cluster, it is likely that these sources lie farther away but along the line of sight. However, this does beg the question of how much of an influence the stellar cluster may have on the measured luminosities we report in Table~\ref{tb:seds}. The next farthest source from the cluster is Source 8. Using it as an illustrative case, we measure that it lies $\sim50\arcsec$ from the cluster, corresponding to a projected separation of 1.33 pc, which represents a lower limit to the true three-dimensional distance. Owing to the limited spatial resolution and contamination from extended emission at Herschel wavelengths, we adopt the core size measured at 37\,$\mu$m, where the source has $FWHM=5.6\arcsec$. Under these assumptions, the core would intercept approximately $\sim2.4\times10^3 L_{\sun}$ from the $\sim3\times10^6 L_{\sun}$ cluster, equivalent to roughly one quarter of the $\sim1\times10^4 L_{\sun}$ luminosity inferred for this source from our model fits (Table~\ref{tb:seds}). However, this externally-generated luminosity represents an extreme upper limit, since the true separation is likely larger than the projected distance, the calculation assumes unity absorption and heating efficiency by dust grains, and neglects any attenuation by intervening gas or dust between the cluster and the core.

More importantly, this external heating is unlikely to significantly influence our fitted source properties. At a projected distance of 1.33 pc, the cluster-facing surface of the core would reach temperatures of order $\sim90$ K; however, such temperatures are confined to a thin externally heated layer. The observed contribution of this warm material depends strongly on geometry: it is maximized only when the heated surface is oriented toward the observer, whereas for other orientations the emission is partially absorbed and reprocessed through cooler material within the core. Consequently, although the outer surface may approach $\sim90$ K, the emission integrated over the source is expected to correspond to a characteristic dust temperature of $\sim20-40$ K. At these temperatures, the primary effect is an enhancement of the far-infrared emission ($\gtrsim$70\,$\mu$m). Since our fitting procedure already treats the Herschel fluxes as upper limits, any FIR excess arising from external heating is expected to have only a minor impact on the derived model parameters, including the inferred luminosities and central stellar masses. In summary, while external heating may be non-negligible for sources located near the cluster, its principal effect is to elevate the far-infrared flux, and it does not significantly alter the overall results of our SED fitting.

Furthermore, our compact-source criteria (described at the beginning of this section) inherently selects against objects that are predominantly externally heated if they are resolved (as most of our sources are). Sources whose emission peaks exhibit significant wavelength-dependent positional shifts are interpreted as being externally heated and are excluded from the compact sample. For example, source P1 (which will be discussed in Section~\ref{sec:trunks}) was initially identified as a compact infrared object candidate in the SOFIA data; however, the observed displacement of its emission peak with wavelength (i.e., shorter infrared wavelengths peaking closer to the stellar cluster) led to its reclassification as a predominantly externally heated source.

\begin{deluxetable*}{lcccccccccc}[tb!]
\tabletypesize{\scriptsize}
\tablecolumns{1}
\tablewidth{0pt}
\tablecaption{SED Fitting Parameters of Compact Infrared Sources in W43~Main \label{tb:seds}}
\tablehead{\colhead{Source   }  &
           \colhead{  $L_{\rm obs}$   } &
           \colhead{  $L_{\rm tot}$   } &
           \colhead{ $A_v$ } &
           \colhead{  $M_{\rm star}$  } &
           \colhead{$A_v$ Range}&
           \colhead{$M_{\rm star}$ Range}&
           \colhead{ Best }&
           \colhead{ Well }&
           \colhead{  }&
           \colhead{  }\\[-6pt]
	   \colhead{        } &
	   \colhead{ ($\times 10^3 L_{\sun}$) } &
	   \colhead{ ($\times 10^3 L_{\sun}$) } &
	   \colhead{ (mag.) } &
	   \colhead{ ($M_{\sun}$) } &
       \colhead{(mag.)}&
       \colhead{($M_{\sun}$)}&
       \colhead{  Models   }&
           \colhead{ Fit? }&
           \colhead{ Features }&
           \colhead{ MYSO? }
}
\startdata
1 &      2.22 &     11.20 &     16.8 &      8.0 &   2.6 - 16.8 &   8.0 - 16.0 &  6  &Y &  &MYSO \\ 
2 &     25.95 &     36.53 &      0.8 &     16.0 &   0.8 - 33.5 &  12.0 - 24.0 &  9 &Y &cm  &MYSO\\ 
3 &      7.27 &      9.95 &      4.2 &      8.0 &   1.7 - 26.5 &   8.0 - 12.0 &  7 &Y &cm  &MYSO\\
4 &     21.71 &     80.58 &     47.8 &     24.0 &  31.0 - 55.3 &  16.0 - 32.0 &  7 &Y &cm &MYSO \\
5$^a$ &      3.69 &      9.99 &    172.2 &      8.0 & 159.3 - 241.1 &   8.0 - 12.0 &  5 &Y &m,h,MM2 &MYSO \\
6  &      3.13 &     71.35 &     41.1 &     24.0 &  22.6 - 83.0 &   8.0 - 24.0 &  8 &Y$^b$ &cm  & MYSO \\
7 &      1.13 &      1.98 &    276.7 &      2.0 & 164.3 - 285.1 &   2.0 - 8.0 & 14  &Y & & \\ 
8 &      6.48 &      9.95 &      1.7 &      8.0 &   1.7 - 41.9 &   8.0 - 12.0 &  5  &Y &cm  &MYSO\\
9 &      9.54 &     19.87 &     81.3 &     12.0 &  36.9 - 98.0 &   8.0 - 32.0 & 15 &Y  &   &MYSO \\
10 &     17.24 &     50.38 &     53.0 &     12.0 &  26.5 - 132.5 &  12.0 - 16.0 &  9 &Y &  &MYSO \\
11 &      0.28 &      0.31 &     11.7 &      2.0 &  11.7 - 156.3 &   2.0 - 32.0 &  8  &Y &  &  \\ 
12$^a$ &      0.23 &      0.27 &      2.5 &      2.0 &   0.8 - 32.7 &   2.0 - 2.0 &  8 &Y  &MM10   & \\ 
13 &      6.90 &    158.30 &     42.4 &     32.0 &  39.7 - 55.6 &  24.0 - 32.0 &  8 &Y$^b$  &m,MM11   &MYSO \\
14 &     22.26 &    616.72 &     34.4 &     64.0 &  34.4 - 106.0 &  32.0 - 64.0 &  8 &Y  &cm  &MYSO \\
15 &      1.82 &    343.52 &    193.4 &     48.0 & 166.9 - 234.8 &   8.0 - 48.0 & 10 &Y  &  &MYSO \\
16 &     23.36 &     38.58 &     38.6 &     16.0 &  38.6 - 43.6 &  16.0 - 16.0 &  9 &Y &cm,m,MM3  &MYSO \\  
17 &      0.47 &      9.90 &    148.4 &      8.0 & 137.8 - 201.2 &   4.0 - 32.0 & 15  &Y  &  &pMYSO \\
18$^a$ &      0.14 &      0.17 &     32.7 &      1.0 &   7.5 - 32.7 &   1.0 - 1.0 &  8 &Y  &o,MM5   &  \\ 
19 &      0.54 &     31.36 &    145.7 &     16.0 & 143.1 - 166.9 &  16.0 - 16.0 &  5 &Y  &m,h,o  &MYSO\\
20$^a$ &      1.50 &     13.30 &    159.0 &      8.0 & 132.5 - 159.3 &   8.0 - 16.0 &  8 &Y  &m,h,o,MM1 & MYSO   
\enddata
\tablecomments{\footnotesize In the Features column: `cm' means the source is coincident with a radio continuum source or peak, `m' means it is associated with methanol maser emission, `h' means it is associated with hydroxyl maser emission, and `o' means it is driving an outflow. The “MM” designations correspond to the spatially coincident millimeter fragments (see Table~\ref{tb:SOFIA_compactA}).}
\tablenotetext{a}{These sources are well-fit by the SED fitter, however only two nominal data points used in the SED fitting, thus the results of the modeling are less reliable.} 
\tablenotetext{b}{These sources are well-fit by the SED fitter when using the same upper limit photometric errors as the lower limit errors. The fitter does not fit these sources well using the nominal upper error limits, which are much larger.} 
\end{deluxetable*}

\subsection{Discussion of Individual Sources and Their Physical Properties} \label{sec:sources}

Here we will discuss the individual infrared sources detected by SOFIA within W43~Main and compare them qualitatively and quantitatively to previous multi-wavelength observations. We will also discuss the physical nature of the individual sources. This discussion will include our interpretations based upon our findings from the SED modeling that we described in Section \ref{sec:cps}. 

\subsubsection{The MM1 Region: Sources 17--20}

\begin{figure*}[tb!]
\epsscale{0.9}
\plotone{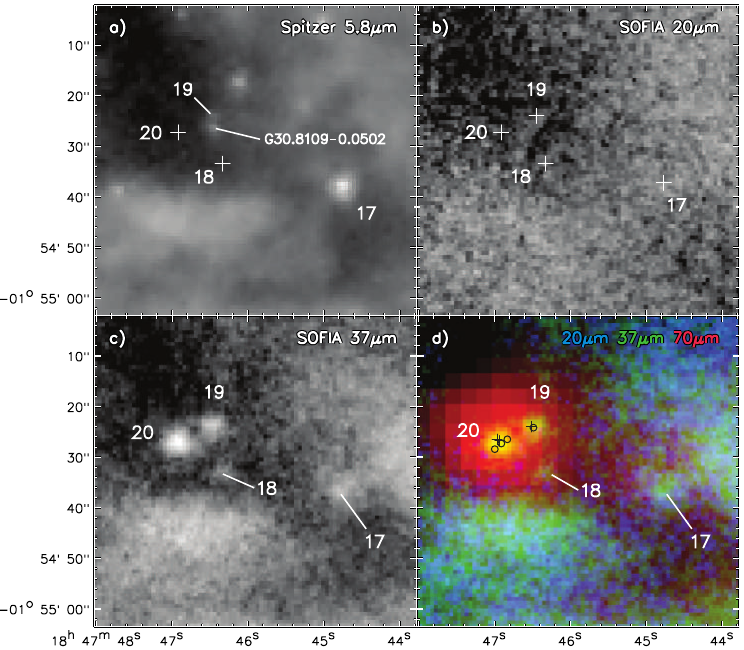}
\caption{Images of the MM1 region containing infrared sources 17-20, with (a) Spitzer 5.8\,$\mu$m, (b) SOFIA 20\,$\mu$m, and (c) SOFIA 37\,$\mu$m. In panel (d), the region is shown as a three-color infrared composite and black crosses mark the positions of 6.7 GHz methanol masers from \citet{2022A&A...666A..59N}, and black circles mark the locations of OH masers from \citet{2019A&A...628A..90B}. In all panels the infrared sources are labeled numerically, but source positions marked by a cross are undetected at that wavelength. \label{fig:mm1}}
\end{figure*}

Mid-infrared sources 18, 19, and 20 (Figure~\ref{fig:mm1}) lie inside an infrared dark cloud identified as IRDC 18454 \citep[see][and references therein]{2012A&A...538A..11B}. This IRDC can be seen as the dark region in the upper left quadrant of Figure~\ref{fig:mm1} and as a cold dust structure in Figure~\ref{fig:fig6}b. First identified as a separate subregion by \citet{1985ApJ...296..565L}, this part of W43~Main was labeled as FIR-3 in their 100\,$\mu$m images (Figure~\ref{fig:fig6}b). In that same paper, the 50\,$\mu$m images do not show prominent emission at this location, hinting to its nature as a colder and/or denser region. 

\begin{figure*}[tb!]
\epsscale{1.18}
\plotone{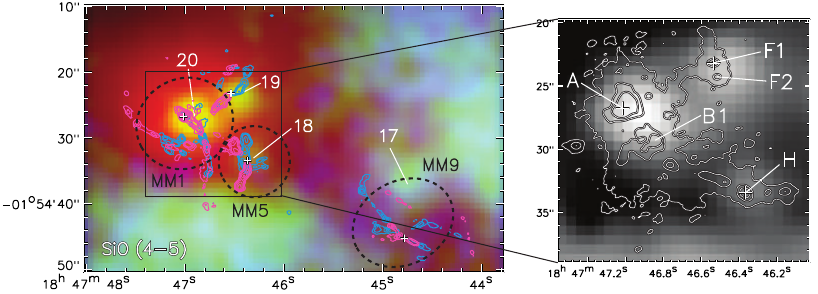}
\caption{(Left) The same 3-color image of the MM1 region as shown in Figure~\ref{fig:mm1} overlaid with the SiO (4--5) outflow contours from \citet{2020A&A...636A..38N}. Light blue contours denote blue-shifted outflow velocities, and light red contours denote red-shifted velocities. The dashed black ellipses denote the 2D Gaussian sizes of the millimeter fragments with names given by \citet{2003ApJ...582..277M}. The white crosses (in both images) denote the mm core locations thought to be at the center of the outflows. (Right) A zoomed in area on the region around mid-infrared sources 18, 19, and 20 at 31\,$\mu$m. White contours are the high-resolution (0.44$\arcsec$) 1.3\,mm emission showing the individual cores from \citet{2018NatAs...2..478M}. Alphabetical labels of the mm cores from \citet{2016ApJ...825L..15C}. \label{fig:mm1flow}}
\end{figure*}

Our mid-infrared source 20 lies near the center of the 1.3\,mm fragment identified as MM1 by \citet{2003ApJ...582..277M}, and source 18 lies near the center of the the position of the mm fragment MM5 (Figure~\ref{fig:mm1flow}). Though originally identified as separate millimeter fragments, MM1 and MM5 (and often also MM9) have taken on the more recent moniker of being collectively called the ``W43-MM1 region'' or ``W43-MM1 ridge''\citep[e.g.,][]{2019ApJ...884...48C,2013ApJ...775...88N}, though some older work refer to the region as ``G30.79 FIR 10'' \citep[e.g.,][]{2010A&A...519A..35C}. \citet{2003ApJ...582..277M} find  MM1 to be the most massive dust fragment in all of W43 (3590\,$M_{\sun}$)\footnote{However, multiple observations over the years at different wavelengths and using different techniques have derived different mass values for the fragments, with some \citep[e.g.,][]{2010A&A...518L..90B} claiming MM2 is more massive.}. \citet{2003ApJ...582..277M} claimed that the MM1 dust fragment likely houses a massive protocluster that is gravitationally bound. Correspondingly, \citet{2010A&A...519A..35C} measure infalling motions and determine that this region seems to be in a state of collapse. \citet{2003ApJ...582..277M} claim that since MM1 is not coincident with any near-infrared or mid-infrared sources (or cm radio continuum emission) that it is in a very early evolutionary phase, prior to the UC\ion{H}{2} phase, but perhaps even prior to the MYSO phase. Of course, our SOFIA observations at 31 and 37\,$\mu$m show that both MM1 and MM5 house mid-infrared sources.  

Source 20 is not detected in any of our data at any wavelength below 31\,$\mu$m. It is the brightest object at 31 and 37\,$\mu$m of the three sources (i.e., 18, 19, and 20) that lie in the IRDC here, and is the only one of the three with strong compact emission at 70\,$\mu$m (though it would be hard to resolve fainter 70\,$\mu$m components from sources 18 or 19 from the 70\,$\mu$m emission of source 20). Though the MM1 fragment coincident with mid-infrared source 20 breaks up into several millimeter cores at higher angular resolution (Figure \ref{fig:mm1flow}), it appears to be dominated by core A (as designated by \citealt{2016ApJ...825L..15C}). This mm core is the closest to our mid-infrared source 20 peak (offset 1.2$\arcsec$, which is coincident to within our astrometric accuracy), and appears to house a YSO driving an outflow (\citealt{2020A&A...636A..38N}; see Figure \ref{fig:mm1flow}) and display hot core chemistry \citep{2022A&A...665A.140B}. As shown in Figure~\ref{fig:mm1}d, methanol masers, which are signposts of MYSOs, are also found here \citep{2022A&A...666A..59N} in addition to hydroxyl masers \citep{2019A&A...628A..90B}. Our SED fitting of this source shows that the data are best fit by a $8M_{\sun}$ MYSO model, however, because the fitting only utilizes the 31 and 37\,$\mu$m data points as nominal data (with all other wavelengths constraining the fits via upper limits) our results are not as reliable as for other sources. However, given it's relatively large and steeply increasing fluxes from 31 to 37\,$\mu$m (as well as the bright component seen at 70\,$\mu$m) and the presence of methanol masers, we are fairly confident that mid-infrared source 20 is tracing a MYSO.

At 4.5 and 5.8\,$\mu$m, source 19 lies just 2.2$\arcsec$ north of another near-infrared point source (see Figure~\ref{fig:mm1}a) identified as G30.8109-0.0502 in Spitzer-IRAC data by \citet{2017ApJ...839..108S}. However, G30.8109-0.0502 is not present in any of the SOFIA data, so we will not discuss it further here. Source 19 is too faint to be detected in the Spitzer 3.6\,$\mu$m image, and the 8.0\,$\mu$m Spitzer image is plagued with artifacts at this location (and thus we do not attempt to calculate a flux at this wavelength). While source 19 is readily seen at 31 and 37\,$\mu$m, it is not detected in either our 11 nor 20\,$\mu$m images. At 0.44$\arcsec$ resolution, the 1\,mm ALMA observations of \citet{2018NatAs...2..478M} show there are two cores very close to each other (1$\arcsec$ separation in declination) which \citet{2016ApJ...825L..15C} label as F1 and F2. Though both are coincident with the peak of mid-infrared source 20 to within our astrometric errors (Figure~\ref{fig:mm1flow}), F2 is situated slightly closer to the peak ($\sim$0.8$\arcsec$) than F1 ($\sim$1.0$\arcsec$). Millimeter core F2 is fainter in 1\,mm continuum emission than core F1, but core F2 was found to be much hotter, and consistent with this, \citet{2022A&A...665A.140B} finds core F2 to have relatively bright emission in several hot core indicators, whereas core F1 does not. According to the mass estimates of \citet{2018NatAs...2..478M}, core F2 is only $2M_{\sun}$, and thus this is a lower mass hot core. However, the peak of mid-infrared source 19 is also coincident with methanol maser emission \citep{2022A&A...666A..59N} which is only thought to come from massive YSOs (Figure~\ref{fig:mm1}d). Therefore, these masers are likely associated with millimeter core F1, which \citet{2018NatAs...2..478M} estimates is $14M_{\sun}$. Our SED model fitting shows the photometry for source 19 to be best fit by a $16M_{\sun}$ YSO, and thus consistent with perhaps being associated with F1 and being an MYSO (though good fits are achieved down to $4M_{\sun}$). Furthermore, all fits require considerable extinction ($A_V=50-167$), consistent with the source's location within an IRDC. Additional support for millimeter core F1 being the MYSO comes from the observations of \citet{2020A&A...636A..38N} that show it to be the center of a relatively collimated outflow in CO\,$(2-1)$ (see Figure \ref{fig:mm1flow}).

Like source 20, source 18 is not detected in any of our data at any wavelength below 31\,$\mu$m. It is present at 37\,$\mu$m but is not much brighter than at 31\,$\mu$m, and the far-infrared component of the source is not bright enough to stand out as an elongation or tongue of emission off of the very bright compact emission seen coincident with source 20. Of the three sources in the center of the mid-infrared IRDC (i.e., 18, 19 and 20), source 18 is by far the faintest at 31 and 37\,$\mu$m. Given the low fluxes, our SED fitting yields only low-mass YSO fits, with only modest extinctions (see Table~\ref{tb:seds}). We do caution that, because the fitting only utilizes the 31 and 37\,$\mu$m data points as nominal data (with all other wavelengths constraining the fits via upper limits) our results are not as reliable as for other sources. However, even if we impose SED fitting results with only high extinctions similar to those we find for sources 19 and 20, we still get groups of best fits dominated by low-mass YSOs (though we do get some high-mass YSO fits).

Source 18 lies at the center of the millimeter fragment MM5 (Figure~\ref{fig:mm1flow}). However, \citet{2018NatAs...2..478M} find that MM5 breaks up into smaller cores with the dominant one being the millimeter core named H by \citet{2016ApJ...825L..15C}, with an estimated mass of $59M_{\sun}$ \citep{2018NatAs...2..478M}. Core H is coincident with the peak of our mid-infrared source 18 to within $\sim$0.5$\arcsec$ (Figure~\ref{fig:mm1flow}). It is therefore interesting that our modeling only finds a low-mass YSO (best fit $1M_{\sun}$) associated with this relatively massive core, when we found our mid-infrared source 19 to be a MYSO associated with a millimeter core of only $14M_{\sun}$. While mid-infrared source 18 does not have any known methanol masers indicative of it being a MYSO, it does lie at the center of a collimated outflow (\citealt{2020A&A...636A..38N}; Figure \ref{fig:mm1flow}), and displays a hot core gas chemistry \citep{2018A&A...618L...5N}, so must at a minimum house a low-to-intermediate mass YSO.

Due to the fact that the MM1 region is found to break into so many dust core sub-components at 1 and 3\,mm with in the high spatial resolution ALMA observations \citep[i.e.,][]{2016ApJ...825L..15C, 2018NatAs...2..478M, 2022A&A...662A...9G, 2022A&A...662A...8M} and evidence of present star formation activity (i.e., masers and outflows), W43-MM1 is sometimes referred to as a mini-starburst protocluster \citep[e.g.,][]{2020A&A...636A..38N, 2022A&A...664A..26P}.

As can be seen in Figures~\ref{fig:mm1} and \ref{fig:mm1flow}, approximately 30$\arcsec$ to the southwest of sources 18-20 lies source 17. This source lies on the periphery of the MM9 millimeter fragment of \citet{2003ApJ...582..277M}. Like MM1 and MM5, MM9 houses at least one outflowing YSO at its center (Figure~\ref{fig:mm1flow}), but unlike those millimeter fragments there is no detected mid-infrared emission coming from the location of the center of the outflows. The peak of the mid-infrared emission from source 17 lies $\sim$6$\arcsec$ to the north of the millimeter peak of MM9, so it is unclear if they are physically related or not. Also, in the Spitzer-IRAC data there is a peak here that \citet{2017ApJ...839..108S} names G30.8109-0.0501, however this peak is offset 1.4$\arcsec$ to the south of the mid-infrared peak, which is just within our margin of astrometric error, but may mean that the NIR source is not the same as the MIR source that we are detecting at wavelengths $\ge$31\,$\mu$m. Assuming that the flux in the NIR is the same source as that in the MIR, our SED modeling finds that source 17 is best fit by a $8M_{\sun}$ MYSO, however we get good fits with models as low as $4M_{\sun}$ as well, so it is not clear if this YSO is genuinely massive or not (and the range of best SED fits with the near-infrared fluxes set as upper limits yield similar results).

\subsubsection{The MM2 \& MM3 region: Sources 5, 12, 15, \& 16 }\label{sec:mm2mm3}

\begin{figure*}[tb!]
\epsscale{1.05}
\plotone{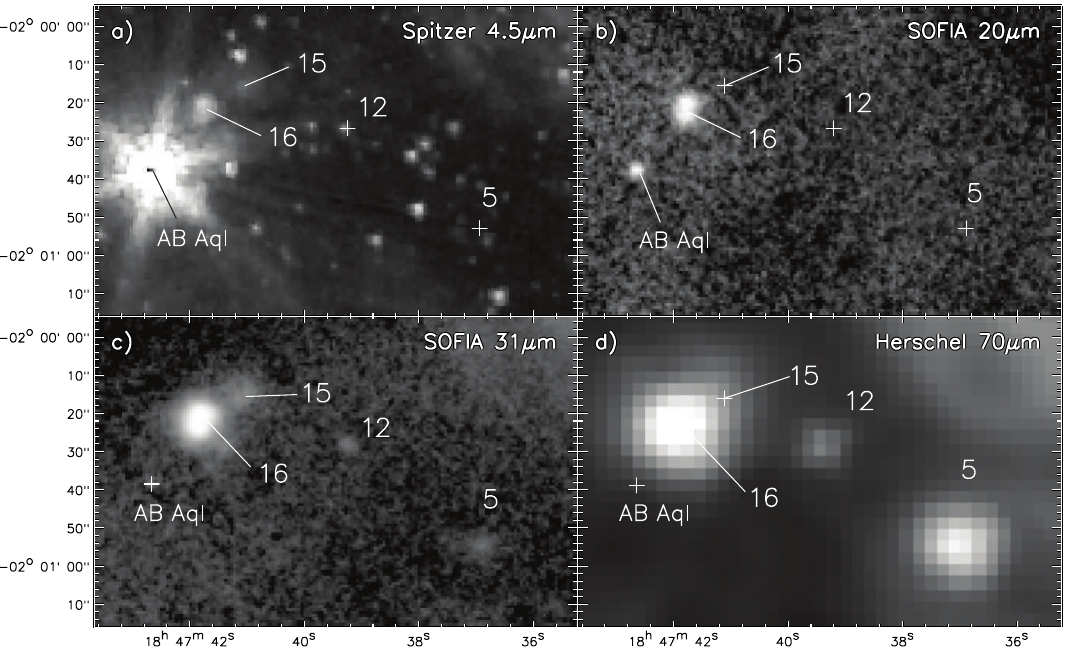}
\caption{Images of the MM2 \& MM3 region containing infrared sources 5, 12, 15 and 16, with (a) Spitzer 4.5\,$\mu$m, (b) SOFIA 20\,$\mu$m, (c) SOFIA 31\,$\mu$m, and (d) Herschel 70\,$\mu$m. In all panels the infrared sources are labeled numerically, but source positions marked by a cross are undetected at that wavelength. \label{fig:mm2}}
\end{figure*}

\begin{figure*}[tb!]
\epsscale{0.95}
\plotone{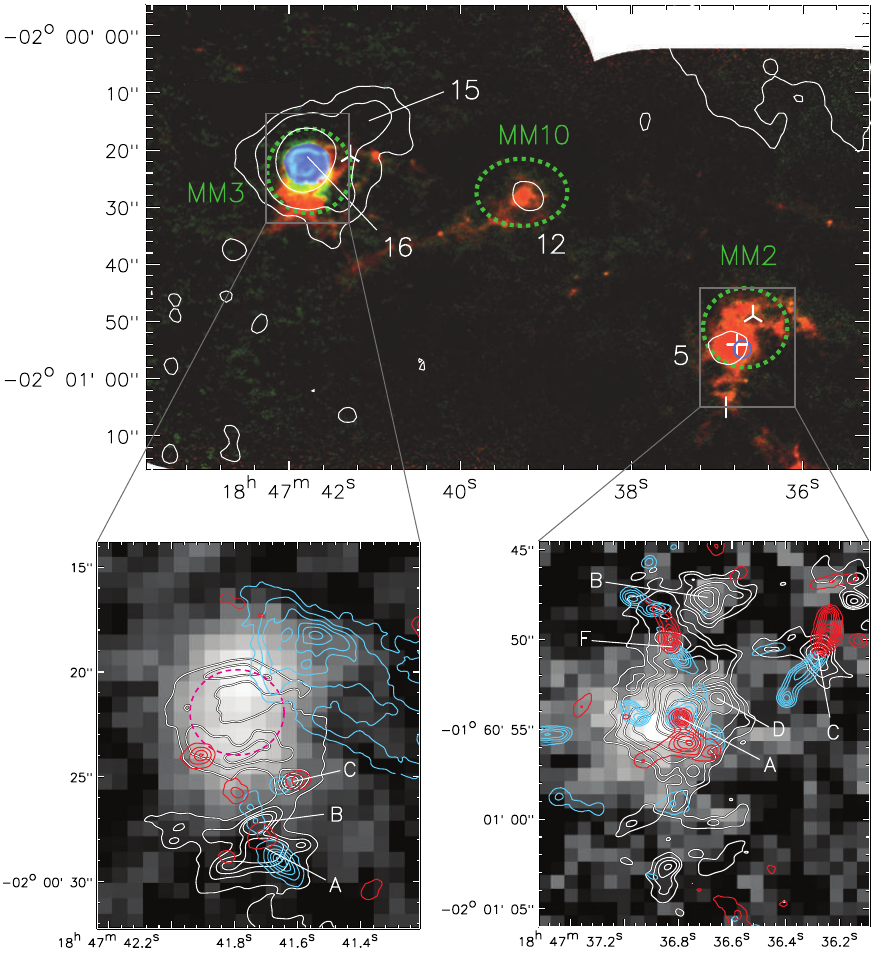}
\caption{(Top) Three-color RGB image of the MM2 \& MM3 region from \citet{2022A&A...664A..26P} showing cold dust continuum at ALMA 1.3\,mm (green) and 3\,mm (red). Free-free emission at H41$\alpha$ recombination line (blue) show the location of the UC\ion{H}{2} region. White contours are from the SOFIA 31\,$\mu$m emission. The dashed green ellipses denote the 2D Gaussian sizes and positions of the mm fragments MM2, MM3, and MM10 defined by \citet{2003ApJ...582..277M}. White numerical labels are given for the mid-infrared sources identified in this present work. White upside-down `Y' symbols mark the locations of the Class I methanol masers \citep[from][]{2017ApJS..231...20Y}, white crosses mark the locations of the center of the Class II methanol maser emission \citep[from][]{2022A&A...666A..59N}, and the blue circle symbol marks the location of the center of the hydroxyl maser emission \citep[from][]{2019A&A...628A..90B}. (Bottom-Left) The SOFIA 20\,$\mu$m image of mid-infrared source 16, with 1.3\,mm contours of \citet{2019ApJ...884...48C} with the brightest mm cores labeled. The magenta ellipse traces the donut-shaped mm shell discussed in the text. (Bottom-Right) The SOFIA 31\,$\mu$m image of mid-infrared source 5, with 1.3\,mm contours of \citet{2019ApJ...884...48C} with the brightest mm cores labeled. In both bottom panels the red and blue contours are the red- and blue-shifted CO outflow emission from \citet{2023A&A...674A..75N}.\label{fig:mm2mm}}
\end{figure*}

Located near the southern end of our map of W43~Main lies a region of star formation that was first identified as source FIR-5 in \citet{1985ApJ...296..565L}, and was considered in that work to be the third-most prominent peak at 100\,$\mu$m (Figure~\ref{fig:fig6}b). In the 6\,cm image presented in that same paper, only a single bright compact source was detected in the region, offset a little over an arcminute to the east of the 100\,$\mu$m peak.

In the millimeter observations of \citet{2003ApJ...582..277M}, this region is broken up into several dust fragments, with the brightest to being MM2 and MM3. MM2 lies near the peak in the 100\,$\mu$m emission of \citet{1985ApJ...296..565L}, and MM3 is coincident with the bright and compact 6\,cm radio continuum source from that same work. The MM2 and MM3 fragments lie in a ridge of mm emission along with several other identified fragments. Collectively, MM2, MM3 and the fragment between them, MM10, have taken on the more recent moniker of being collectively called the ``W43-MM2 \& MM3 region'' or ``W43-MM2 ridge''\citep[e.g.,][]{2022A&A...664A..26P,2013ApJ...775...88N}. More recent ALMA millimeter results \citep[i.e.,][]{2022A&A...662A...8M, 2022A&A...662A...9G, 2022A&A...664A..26P} show this whole region and the fragments it contains break up into many cores, with \citet{2022A&A...664A..26P} estimating that the entire MM2 \& MM3 ridge is home to over 200 cores with masses ranging between $\sim$0.1 and 70\,$M_{\sun}$. Like MM1, this region is consider by those authors to be a mini-starburst protocluster.

We detect five mid-infrared compact sources within this sub-region (Figure~\ref{fig:mm2}). Our source 5 peaks near the location of the mm peak of MM2, source 16 peaks at the location of the MM3 mm peak, and our source 12 peaks at the location of the MM10 mm peak (Figure~\ref{fig:mm2mm}). Sources 5 and 12 are not detected at 20\,$\mu$m or shorter wavelengths (Figure~\ref{fig:mm2}), nor at cm radio continuum wavelengths, but are detected in our SOFIA 31 and 37\,$\mu$m images as well as Herschel 70\,$\mu$m. MM3 houses a UC\ion{H}{2} region \citep[e.g.,][]{2022A&A...664A..26P} near its millimeter peak which is coincident with the peak of source 16 at SOFIA 20 and 37\,$\mu$m, and also at Spitzer-IRAC wavelengths (Figure~\ref{fig:mm2mm}).

Observations with ALMA \citep[e.g.,][]{2019ApJ...884...48C, 2023A&A...674A..76P} show that MM2 is dominated by a single bright core, which \citet{2019ApJ...884...48C} label core A and estimate is 426\,$M_{\sun}$ (see Figure~\ref{fig:mm2mm}, lower-right), the most massive that they found in their survey of W43~Main. Both \citet{2003ApJ...582..277M} and \citet{2019ApJ...884...48C}, claim that the fact that MM2 has no cm radio or 24\,$\mu$m emission means it is perhaps a star-forming clump at an early evolutionary stage. However, as stated earlier, we do indeed see mid-infrared emission at 31 and 37\,$\mu$m here which we call source 5, mimicking the extent of the extended emission around core A, though our mid-infaerd peak is $\sim$1.5$\arcsec$ to the southeast of the mm peak of core A (Figure~\ref{fig:mm2mm}, lower-right). Further evidence of present stellar formation activity is evident by the presence of methanol and OH masers \citep[Figure~\ref{fig:mm2mm};][]{2022A&A...666A..59N,2019A&A...628A..90B}, as well as the multiple outflow lobes found in the region \citep{2023A&A...674A..75N}. Several outflows are centered on the identified mm cores (Figure~\ref{fig:mm2mm}, lower-right), with core A hosting what appears to be several outflows alone. Our SED modeling of our source 5 also indicate that it is a MYSO, with a best fit mass of $8M_{\sun}$. Furthermore, in order to fit both the relatively bright fluxes at 31 and 37\,$\mu$m and the upper limits at 11 and 20\,$\mu$m requires high extinction values, leading to a best-fit group range of $A_V= 159-241$. However, with only two nominal data points used in the fits, the results for source 5 are a bit less robust than for other sources in our survey. Interestingly, at 31\,$\mu$m (only) we may be picking up some mid-infrared emission from mm core B (Figure~\ref{fig:mm2mm}, lower-right), though it may just be spurious noise.

For MM3, \citet{2019ApJ...884...48C} show that there is a round shell of millimeter emission (see Figure~\ref{fig:mm2mm}), likely the result of the expanding UC\ion{H}{2} region at the center of the MM3 fragment, and this lies just to the north of some of the millimeter cores resolved at higher resolutions. As can be seen in Figure~\ref{fig:mm2mm}, this shell of emission around the UC\ion{H}{2} region is donut-shaped but brighter to the north and south than east and west. Our SOFIA data show source 16 to be double peaked, barely resolved from each other at both wavelengths, but best seen at 20\,$\mu$m. These two peaks coincide with the brighter millimeter areas of the donut-shaped MM3 source. As can be seen in Figure~\ref{fig:mm2mm}, there is nearby Class I methanol maser emission \citep[][]{2017ApJS..231...20Y}, which is often associated with outflows from MYSOs \citep[e.g.,][]{2012ApJ...760L..20C}. Indeed, there is a large-scale blue shifted outflow lobe in the area of this maser (Figure~\ref{fig:mm2mm}, bottom-left) as well as several other outflows found in this region \citep{2023A&A...674A..75N}, though none appear to be directly associated with the UC\ion{H}{2} region (i.e., our source 16). Our SED fitting of the photometry for source 16 gives us a group of best fits that all indicate the source is a $16M_{\sun}$ MYSO. The UC\ion{H}{2} region here is 115\,mJy at 6\,cm \citep{2010A&A...518L..90B} which we calculate to be equivalent to $log(N_{LyC}) = 47.8$ photons/s, which is B0.5 star \citep{1996ApJ...460..914V}, and about a $16M_{\sun}$ \citep{2000AJ....119.1860B} which is consistent with our results from SED modeling of the infrared emission. 

\cite{2023A&A...674A..76P} find that, unlike the MM2 and MM3 millimeter fragments, MM10 breaks up into a collection of only lower mass cores, with none greater than $11.5M_{\sun}$. Consistent with this, we find from our SED model fitting of the mid-infrared photometry of source 12 that it is best fit with intermediate mass YSO models of $2M_{\sun}$. This is based upon only two nominal photometric data points, so is a less robust results than most of our SED fitting results. However, there is a peak here in the Herschel 70\,$\mu$m image. But even if we add in the 70\,$\mu$m data point as a nominal data point (even though it is likely contaminated by environmental emission), and insist on models with only $A_V>10$ (since we have non-detection at all wavelengths $\leq20$\,$\mu$m), we still do not get fits over $4M_{\sun}$.   

Located $\sim$12$\arcsec$ to the northwest of the peak of source 16, is source 15. This source can be seen at all Spitzer-IRAC wavelengths, as well as in the 31 and 37\,$\mu$m SOFIA data. At 70\,$\mu$m, there is also an elongation of emission off of source 16 toward the location of source 15 (Figure~\ref{fig:mm2}). However, there appears to be no mm fragments or cores in this area \citep[e.g.,][]{2023A&A...674A..75N}. Source 15 itself has a main peak in the infrared with its own extension of emission towards the northwest ($\sim$2.5$\arcsec$ in extent). The position of this mid-infrared tongue of emission seems to move slightly in position depending on wavelength, and thus it is likely not a separate YSO. Under the assumption of the mid-infrared peak and tongue coming from a single source, it is well fit by the MYSO SED fitter, with a best-fit mass of $48M_{\sun}$.  

Interestingly, source 15 does not appear in either the SOFIA 11 or 20\,$\mu$m images. However, on the larger scale, the environment surrounding sources 5, 12, 15, and 16 is severely depressed in extended near-infrared and mid-infrared emission compared to the majority of W43~Main, but this same large area glows brightly at far-infrared wavelengths (see Figure~\ref{fig:fig1}), signifying a overall dense and cold dust region that constitutes a reservoir for continued and future star formation. 

Also contained in this area is an unresolved mid-infrared point source that is coincident with AB Aql, a late-type M star \citep{1977PASP...89..829C}, which the GAIA Data Release 3 (DR3) catalog places at a distance of 685\,pc (see Figure~\ref{fig:mm2}). This source is saturated at all Spitzer-IRAC wavelengths, very bright at 11\,$\mu$m, readily seen at 20\,$\mu$m, barely detected at 31\,$\mu$m, and undetected at 37 and 70\,$\mu$m -- behavior typical of photospheric emission from a foreground late M-type field star. We therefore do not apply the MYSO modeling to this source, nor discuss it further here.

\begin{figure*}[tb!]
\epsscale{1.05}
\plotone{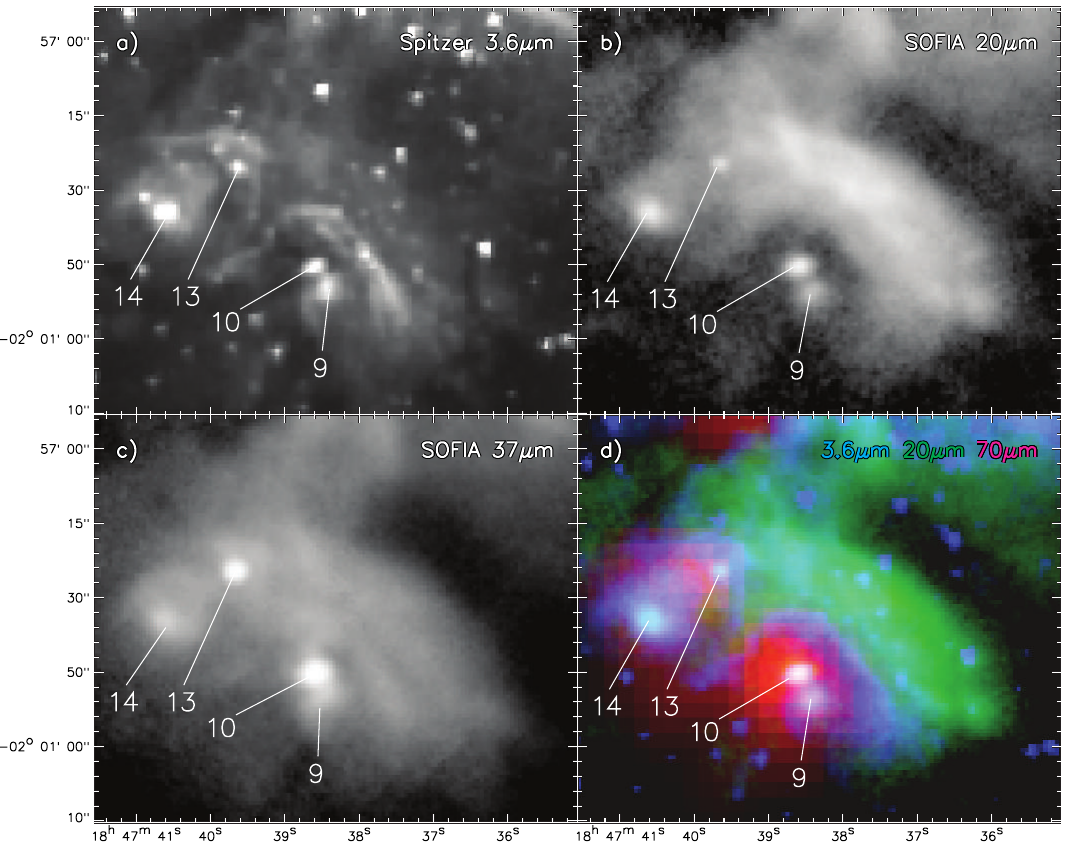}
\caption{Images of the MM4 and MM11 region containing infrared sources 9, 10, 13 and 14, with (a) Spitzer 3.6\,$\mu$m, (b) SOFIA 20\,$\mu$m, (c) SOFIA 37\,$\mu$m, and (d) a three-color RGB image showing cold dust continuum Herschel 70\,$\mu$m (red), warm dust at SOFIA 20\,$\mu$m (green), and hot dust and photospheric emission at Spitzer 3.6\,$\mu$m (blue). In all panels the infrared sources are labeled numerically.\label{fig:mm4a}}
\end{figure*}

\subsubsection{The MM4 \& MM11 Region: Sources 9, 10, 13, \& 14}

\begin{figure*}[tb!]
\epsscale{1.10}
\plotone{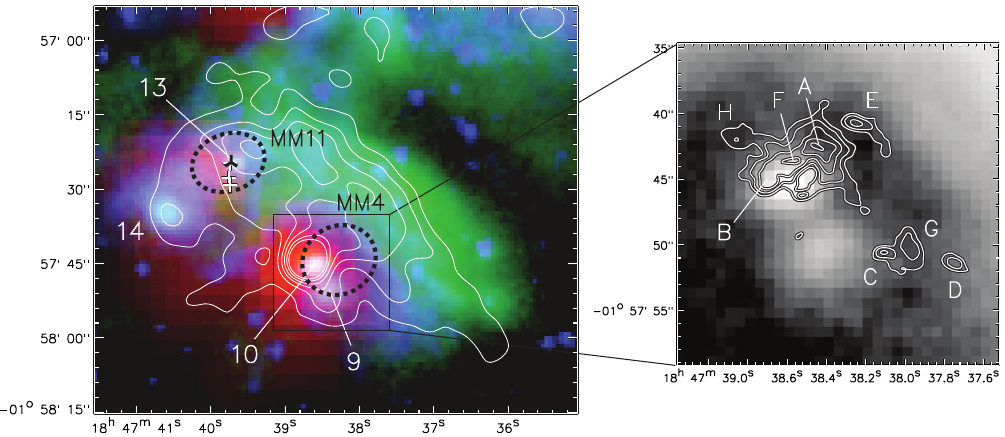}
\caption{(Left) Three-color RGB image of the MM4 and MM11 region (from Figure~\ref{fig:mm4a}), overlaid with contours from the 6\,cm data from the VLA archive with 4.5$\arcsec$ resolution. The dashed black ellipses denote the 2D Gaussian sizes and positions of the mm fragments MM4 and MM11 defined by \citet{2003ApJ...582..277M}. White numerical labels are given for the mid-infrared sources identified in this present work. White crosses mark the locations of the center of the Class II methanol maser emission \citep[from][]{1998MNRAS.301..640W, 2016ApJ...833...18H, 2019ApJ...885..131R}, and the black upside-down `Y' symbol marks the locations of the Class I methanol maser \citep[from][]{2022A&A...666A..59N}.  (Right) Closeup 20\,$\mu$m image of the region around sources 9 and 10, with the 1.3\,mm map of MM4 from \citet{2019ApJ...884...48C}, with individual mm cores labeled as letters.\label{fig:mm4b}}
\end{figure*}

This region is also known as FIR-2 as well as G30.8S from \citet{1985ApJ...296..565L} which, at the resolution of their 100\,$\mu$m map, encompassed the entire area and collection of mid-infrared sources described in this subsection (see Figure~\ref{fig:fig6}b). The emission in the 100\,$\mu$m image peaks very close to the position of millimeter fragment MM4, and FIR-2/G30.8S has the brightest peak (and integrated flux) at 100\,$\mu$m. This region appears as a large ($\sim$1$\arcmin$-diameter), contiguous but clumpy radio continuum region at modest resolutions ($\sim$4.5$\arcsec$) at 6\,cm, whose overall morphology appears similar at the SOFIA mid-infrared wavelengths at similar resolutions (Figures~\ref{fig:mm4a} and \ref{fig:mm4b}). In addition to millimeter fragment MM4, this region hosts nearby fragment MM11. At the SOFIA wavelengths we identify four mid-infrared sources or peaks which we label sources 9, 10, 13, and 14 (Figure~\ref{fig:mm4a}).

SOFIA sources 9 and 10 are situated at the southeastern periphery of the MM4 fragment (Figure~\ref{fig:mm4b}). At higher resolution in the millimeter with ALMA \citep{2019ApJ...884...48C}, MM4 breaks up further into cores, with several being aligned in a northwest to southeast fashion (Figure~\ref{fig:mm4b}), and one core (labeled A) dominates the integrated flux of the region. Though extended mid-infrared emission seems ubiquitous throughout this entire region, there is a marked absence of mid-infrared emission at the location of core A (Figure~\ref{fig:mm4b}). The string of mm cores appear to wrap around but avoid the mid-infrared emission peak, perhaps indicating that the infrared is leaking out of the less dense and less extinguished regions, rather than tracing the cores themselves in this case.  

Source 10 is coincident with near-infrared source G30.7529-0.0512 identified in the catalog of \citet{2017ApJ...839..108S}, which they classify as a PAH-dominated source (which we confirm; see Figure~\ref{fig:ccd}). The peak of source 10 is 1.7$\arcsec$ southwest of the very bright nearby radio peak, but this radio peak is embedded in extended radio emission (Figure~\ref{fig:mm4b}). \citet{2018A&A...615A.103K} claim the peak is a UC\ion{H}{2} region, and derive a source luminosity (70,800\,$L_{\sun}$) equivalent to a $\sim$18$M_{\sun}$ ZAMS star (based upon SED fitting of from 1 to 870\,$\mu$m and a measured 6\,cm flux of 302\,mJy). Though source 10 lies too far from the radio peak to be tracing emission directly from the UC\ion{H}{2} region, our SED modeling of source 10 finds a best fit for a $12M_{\sun}$ source, but it has good fits as high as $16M_{\sun}$, which is consistent with that derived for the UC\ion{H}{2} region. This may mean that source 10 is a younger MYSO that has yet to ionize its surroundings that lies near the UC\ion{H}{2} region, or it is somehow emission from the UC\ion{H}{2} region leaking out from a less dense part of the MM4 molecular fragment.   

Though hard to see in Figure~\ref{fig:mm4b}, there is fainter radio continuum source to the southwest of the peak of the UC\ion{H}{2} region in the direction of the source 9 (which can be seen as a tongue of radio emission coming from the main peak), however in actuality source 9 lies between these two radio peaks and does not appear to have radio continuum emission peak of its own. No millimeter cores appear to be associated with source 9 either (Figure~\ref{fig:mm4b}), though the Herschel 70\,$\mu$m emission is extended in this direction. Source 9 is prominent enough in the near-infrared that is was cataloged as G030.7514-0.0511 by \citet{2017ApJ...839..108S}, who claim it is a Class I YSO. Our SED fitting shows that this source is well-fit by MYSO models with a best fit of $12M_{\sun}$. However, we caution that this modestly extended source seems to have a size and peak location that changes with wavelength. While the 3.6, 5.8, 8.0, 31, and 37\,$\mu$m peaks are all coincident to within a FORCAST pixel (0.768$\arcsec$), the 4.5 and 20\,$\mu$m peaks are coincident with each other but lie 1.4$\arcsec$ south of the average position of the other wavelengths. These are all shifts within the astrometric accuracy of the FORCAST data, however, nearby source 10 does not display such shifts with wavelength, and appears similar in shape and extent at all infrared wavelengths, and thus these peak shifts seen for source 9 are likely real. However, the shifts don't follow the pattern expected for a single externally heated knot (where the shifts increase in a direction away from the heating source with wavelength). This could mean that this source may contain multiple unresolved internal heating sources creating different peaks at different wavelengths.   

In fact, the misalignment of emission peaks seen in high-resolution infrared, radio, and millimeter images in and around the immediate region of sources 9 and 10 may be evidence that W43-MM4 is indeed hosting a massive and embedded protocluster. This was first suggested by \citet{2003ApJ...582..277M} who stated that millimeter fragment MM4 is a massive ($M\sim500M_{\sun}$), and is probably gravitationally bound.  

Mid-infrared source 13 is coincident with the peak of the MM11 millimeter fragment. Source 13 is also coincident with \citet{2017ApJ...839..108S} near-infrared catalog source G30.7598-0.0525, which they classify as a Class I YSO. The presence of mid-infrared emission from 3 to 37\,$\mu$m hints that star formation is indeed occurring within MM11. In fact, MM11 is associated with several OH and water maser emission detections \citep[e.g., ][]{1997A&A...325..255B, 2004A&A...414..235S, 2001A&A...368..845V}, which are signs of star formation activity (however their positions are known only to $\sim$30$\arcsec$ precision). Moreover, the peak of MM11 and source 13 is coincident with class II methanol masers \citep[e.g., ][]{1998MNRAS.301..640W, 2016ApJ...833...18H, 2019ApJ...885..131R}, which have sub-arcsec positional certainties and are only associated with massive star formation activity (Figure~\ref{fig:mm4b}), and also displays Class I methanol maser emission \citep[][]{2022A&A...666A..59N}. Our SED modeling of source 13 indeed shows it to be a MYSO, with a best-fit mass of $24M_{\sun}$. Though the extended 6\,cm continuum emission region does have a weak peak near the location of MM11, the peak of source 13 is a little more than 4$\arcsec$ from this radio peak. Source 13, therefore, does not appear to host a UC\ion{H}{2} region at its center. Unlike most MYSOs prior to the onset of UC\ion{H}{2} regions, this infrared source does not seem to have a peak at $\ge$70um either, though it does lie close to ($\sim$3.5$\arcsec$ north west of) a 70\,$\mu$m peak. 

Source 14 has a broad peak at 37\,$\mu$m that seems to be slightly offset (1.7$\arcsec$) from a broad peak in the 6\,cm images (Figure~\ref{fig:mm4a} and \ref{fig:mm4b}). Source 14 is bright with a compact core in the Spitzer and SOFIA 11, 20, and 31\,$\mu$m data. At 70\,$\mu$m it is hard to resolve source 14 from the brighter source 13, but there exists a tongue of 70\,$\mu$m emission toward it's location. There is no corresponding 1.3\,mm core present at this location in the maps of \citet{2003ApJ...582..277M}. Source 14 is also known as G030.7594-0.0571 from \citet{2017ApJ...839..108S} who classified it as a Class II YSO based on the near-infrared spectral index. However, our photometry is best fit by extremely massive MYSO models ($32-64M_{\sun}$) with a best-fit mass of $64M_{\sun}$, making it the most-massive MYSO in all of W43~Main.

We speculate further on the nature of MM4 and MM11 and their relationship to the mid-infrared sources in Section~\ref{sec:trunks}.

\subsubsection{The WR/OB Cluster Region: Sources 3, 4, \& 6 }

\begin{figure*}[tb!]
\epsscale{1.05}
\plotone{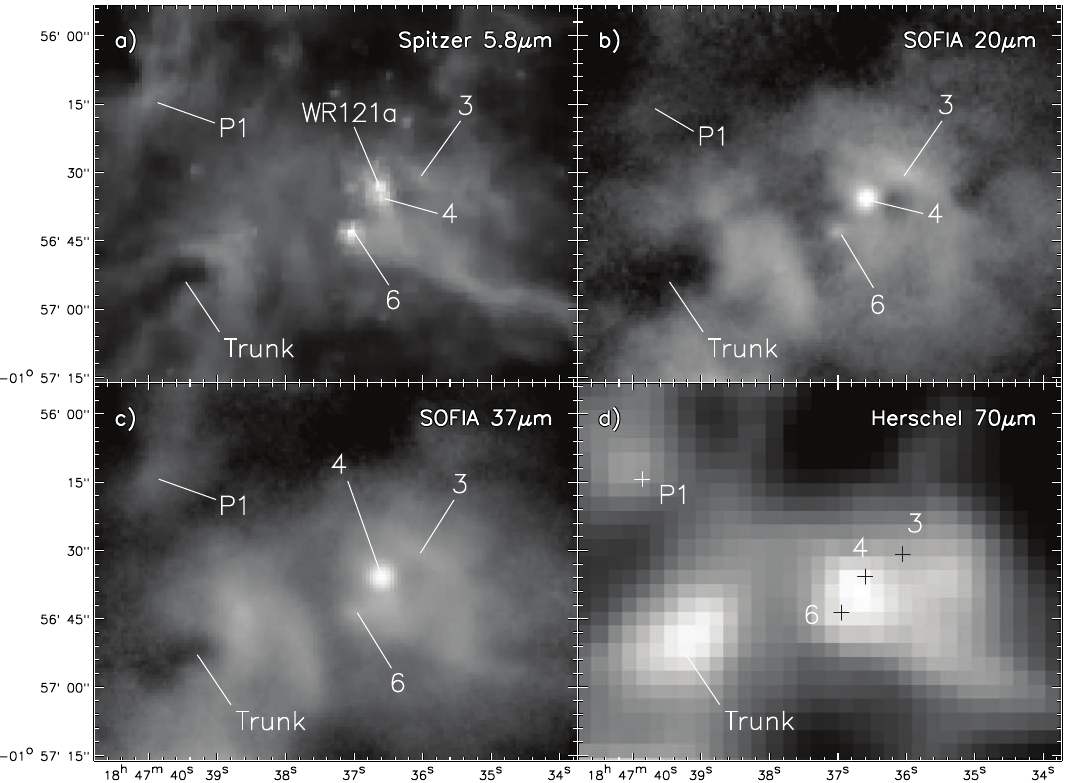}
\caption{Images of the central W43~Main region containing mid-infrared sources 3, 4, and 6, with (a) Spitzer 5.8\,$\mu$m, (b) SOFIA 20\,$\mu$m,(c) SOFIA 37\,$\mu$m, and (d) Herschel 70\,$\mu$m. Panel a) also shows the position of WR 121a (a.k.a. W43\#1a). Also labeled are the protuberance P1 and a photoablated trunk as identified in Figure~\ref{fig:trunks}. In all panels the infrared sources are labeled numerically, but source positions marked by a cross are undetected at that wavelength.\label{fig:mm15a}}
\end{figure*}

\begin{figure*}[tb!]
\epsscale{1.10}
\plotone{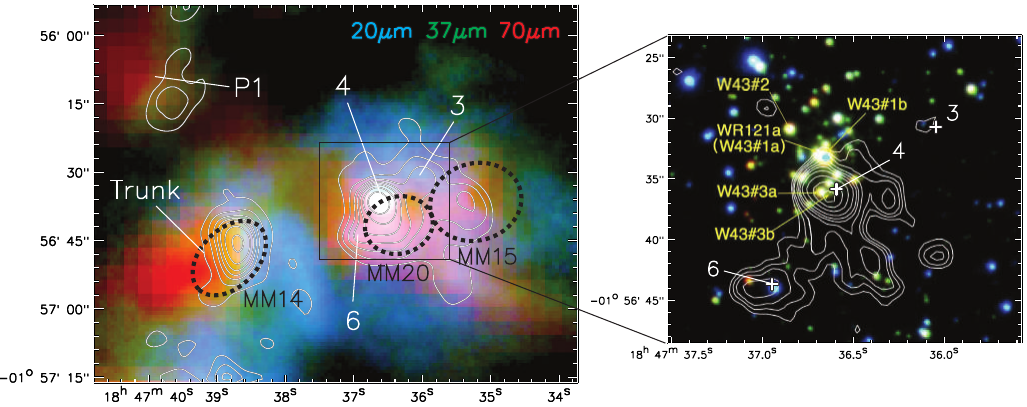}
\caption{(Left) Three-color RGB image of the central W43~Main region with Herschel 70\,$\mu$m (red), SOFIA 37\,$\mu$m (green), and SOFIA 20\,$\mu$m (blue), overlaid with contours from the 6\,cm data from the VLA archive with 4.5$\arcsec$ resolution. White numerical labels indicate the location and name of the mid-infrared sources. The dashed black ellipses denote the 2D Gaussian sizes and positions of the mm fragments MM14, MM15, and MM20 defined by \citet{2003ApJ...582..277M}. (Right) A 3-color image of the W43 stellar cluster made with ESO VLT J, H, and Ks bands in blue, green, and red color, respectively \citep[adapted from][]{2011A&A...532A..92L}. Contours are from VLA 6\,cm data taken in its extended B-configuration with $\sim$1.5$\arcsec$ resolution. Yellow labels show the WR/OB cluster members identified by \citet{1999AJ....117.1392B} and \citet{2011A&A...532A..92L} in the near-infrared. White crosses indicate the locations of the mid-infrared source peaks at 37\,$\mu$m.\label{fig:mm15b}}
\end{figure*}

Though not given a numerical designation by \citet{1985ApJ...296..565L} because it was less conspicuous at 100\,$\mu$m, this region shows up as a peak in their 50\,$\mu$m maps and is the hottest region in their dust color temperature maps. They also were the first to find 2\,$\mu$m emission coming from within this region, which was later resolved into the WR/OB stellar cluster now known to reside there by \citet{1999AJ....117.1392B}. This cluster is optically-obscured having an estimated $A_{V}\sim34$\,mag of extinction \citep{1999AJ....117.1392B}. 

We detect three compact mid-infrared sources in the $\sim$1$\arcmin$$\times$1$\arcmin$ area surrounding the location of this central WR/OB cluster (Figure~\ref{fig:mm15a}). The closest mid-infrared source (in projection, at least) to the center of this stellar cluster is source 4 (a.k.a. G30.7662-0.0348 from \citealt{2017ApJ...839..108S}), which has a peak coincident with (to with the SOFIA astrometric errors) a fairly bright (88 mJy at 6\,cm; \citealt{2015MNRAS.448.3572I}) and compact radio continuum source (Figure~\ref{fig:mm15b}). The mid-infrared/radio peak appears to lie on the northeastern periphery of the millimeter fragment MM20 \citep{2003ApJ...582..277M}, and given this positioning, it may not be directly associated with the fragment (Figure~\ref{fig:mm15b}). It may simply be an area tracing the hot dust and ionized surface of the mm fragment that faces the stellar cluster. That said, given it's compact size and the fact that the radio and MIR peaks are seemingly coincident, this is likely to contain a MYSO. Indeed, our SED fitting shows that the photometry for source 4 is best fit by a $24M_{\sun}$ MYSO. Of the stellar cluster members identified and studied by \citet{1999AJ....117.1392B}, it appears that the mid-infrared/radio peak lies close to the location of W43\#3. However, \citet{2011A&A...532A..92L} resolved W43\#3 into a double, which they denote as W43\#3a and W43\#3b, and which they find are an O-type supergiant and a main sequence O star, respectively. Though the mid-infrared/radio source we are seeing is close in projection to these stars, it is unlikely that either star is the optical component of source 4, as revealed main sequence and supergiant stars do not display the dust properties we observe.

The Wolf-Rayet star, WR~121a, is a WN7 spectral type \citep{1999AJ....117.1392B} and lies just north ($\sim$3$\arcsec$) of the mid-infrared peak of source 4 (Figure~\ref{fig:mm15b}). \citet{1999AJ....117.1392B} only resolved a single source at this location in the near-infrared, but follow-up near-infrared observations of \citet{2011A&A...532A..92L} find this to also be a double source, which they denote as W43\#1a and W43\#1b. \citet{2011A&A...532A..92L} speculate that this is a physical binary with W43\#1a being the WR star, and W43\#1b is its O-type companion. \citet{2020ApJ...891..104A} also speculate that WR~121a must be a massive binary with an active wind collision between its components to account for the X-ray emission they see via Chandra observations. At shorter and shorter infrared wavelengths, as seen with Spitzer, the brightest infrared peak in the area moves from the dust emission of source 4 to photospheric emission of WR~121a (Figure~\ref{fig:mm15a}). The stellar ages of these massive stars are estimated to be 1--6 Myr \citep{2003ApJ...582..277M,2010A&A...518L..90B}. Rounding out the known massive stars of the cluster is W43\#2, which lies even further from the mid-infrared/radio peak of source 4 and, being a mid-to-early O supergiant \citep{1999AJ....117.1392B}, displays no detectable infrared or radio continuum emission (Figure~\ref{fig:mm15b}). This WR/OB stellar cluster is the most prominent feature of this sub-region, and the massive stars there are likely a key contributor to the overall heating and ionization of W43~Main \citep{2003ApJ...582..277M}.

Source 3 (a.k.a. G30.7660-0.0332 from \citealt{2017ApJ...839..108S}) lies $\sim$10$\arcsec$ northwest of the peak of 4 (Figure~\ref{fig:mm15a}), and at 8, 20, and 37\,$\mu$m clearly appears as a narrow and elongated ($\gtrsim$8$\arcsec$) source of emission with a central peak. The near- and mid-infrared peak location of this elongated source is not emitting at 70\,$\mu$m, but the southwestern-most end of the elongation is the location of millimeter fragment MM15 which does have some modest 70\,$\mu$m emission (as well as cm radio continuum emission). The peak in the mid-infrared of source 3 is is coincident with a tongue of emission seen at 6\,cm at lower resolution, and a faint peak at higher resolution (Figure~\ref{fig:mm15b}). Given its projected close proximity to WR121a, this structure may be an externally heated and ionized ridge, and source 13 may be a YSO (traced by the infrared peak) embedded within that ridge. Under the assumption that it is indeed internally heated, we can confirm that its photometry can be fit by the MYSO models of our SED fitter, and we get a best fit for a MYSO with $8M_{\sun}$.  

Mid-infrared source 6 (a.k.a. G30.7652-0.0375 from \citealt{2017ApJ...839..108S}), which lies $\sim$10$\arcsec$ southeast of the peak of source 4 (Figure~\ref{fig:mm15a}), is clearly visible at all Spitzer wavelengths and is coincident with a cm radio continuum peak (Figure~\ref{fig:mm15b}). It does not have a prominent peak at 70\,$\mu$m, but there is a tongue of emission extending from the location of the far-infrared peak of MM20 toward this source indicating it does emit modestly at this wavelength. Our SED model fitting indicates that this source is a MYSO, with a best fit mass of $24M_{\sun}$. 

When interpreting the locations of sources in W43~Main, it is important to remember that our view is inherently two-dimensional. Although infrared sources 3, 4, and 6 (as well as MM14, MM15, and MM20) appear in projection to lie near the core of the WR/OB stellar cluster, the caustic environment of such a cluster makes it more likely that these infrared and millimeter sources are physically offset along the line of sight and located farther from the cluster than their projected separations suggest. Nevertheless, there is evidence -- particularly for MM14, MM15, and MM20 -- that they remain close enough to be strongly influenced by the cluster while still sufficiently distant to have avoided complete disruption. Indeed, \citet{2003ApJ...582..277M} determines that MM14, MM15, and MM20 all appear to be gravitationally unbound, and are thus probably clumps of material governed by turbulence, and may presently be in the process of being dispersed.  

\subsubsection{The MM6 \& MM8 Ridge: Sources 1 \& 8}

\begin{figure*}[tb!]
\epsscale{1.05}
\plotone{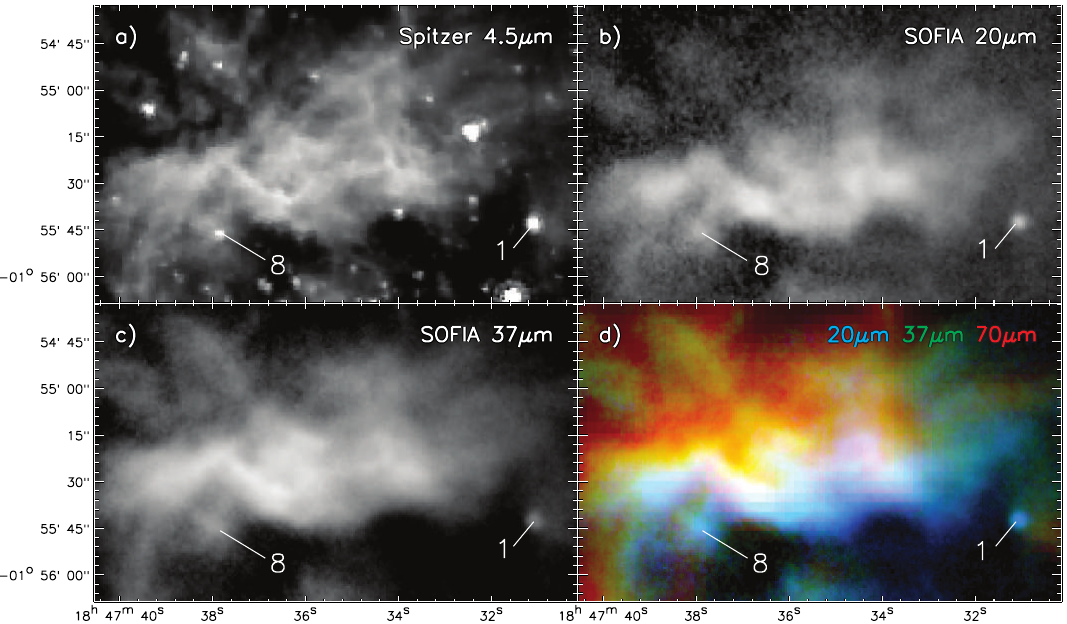}
\caption{Images of the MM6 \& MM8 ridge containing infrared sources 1 and 8, with (a) Spitzer 4.5\,$\mu$m, (b) SOFIA 20\,$\mu$m, (c) SOFIA 37\,$\mu$m, and (d) a three-color RGB image made from Herschel 70\,$\mu$m (red), SOFIA 37\,$\mu$m (green), and SOFIA 20\,$\mu$m (blue) data. In all panels the infrared sources are labeled numerically.\label{fig:mm6a}}
\end{figure*}

\begin{figure*}[tb!]
\epsscale{1.0}
\plotone{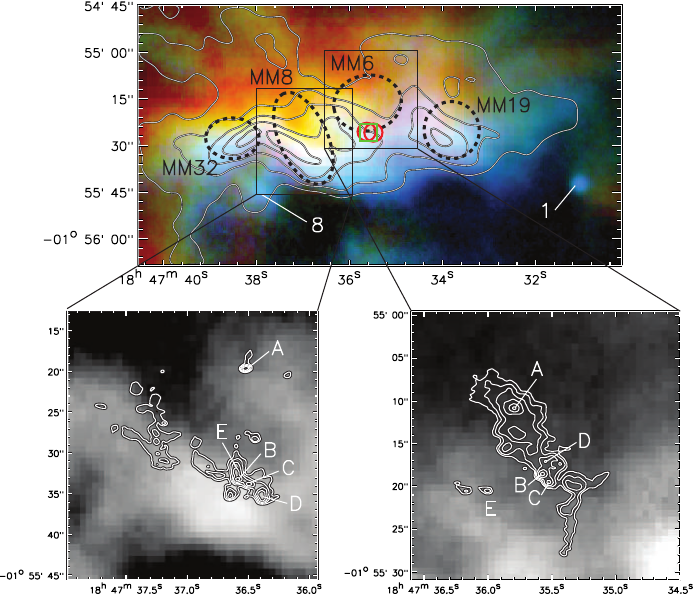}
\caption{(Top) Three-color RGB image of the MM6 \& MM8 ridge from Figure~\ref{fig:mm6a} overlaid with contours from the 6\,cm data from the VLA archive with 4.5$\arcsec$ resolution. White numerical labels indicate the location and name of the mid-infrared sources. The dashed black ellipses denote the 2D Gaussian sizes and positions of the mm fragments MM6, MM8, MM19, and MM32 defined by \citet{2003ApJ...582..277M}. Small red circles mark the locations of the water masers from \citet{2001A&A...368..845V} and \citet{1998A&AS..127..181H}, though they have $\sim$20$\arcsec$ positional uncertainty. The green square marks the location of the 5\,cm OH maser from \citet{1997A&A...325..255B} which has 65$\arcsec$ uncertainty. (Bottom-Left) Closeup 20\,$\mu$m image of the region around MM8, with the 1.3\,mm map of MM8 from \citet{2019ApJ...884...48C} overlaid as contours, with individual mm cores labeled as letters. (Bottom-Right) Closeup 20\,$\mu$m image of the region around MM6, with the 1.3\,mm contours of MM6 from \citet{2019ApJ...884...48C}.
\label{fig:mm6b}}
\end{figure*}

Located $\sim$1$\arcmin$ north of the WR/OB cluster lies an extremely prominent and large (1.5$\arcmin$ long) east-west dust ridge (Figure~\ref{fig:mm6a}). This ridge contains a compact mid-infrared source we label as source 8 (a.k.a. G30.7808-0.0333 from \citealt{2017ApJ...839..108S}), which has a 6\,cm radio continuum peak, and is seen at all Spitzer-IRAC wavelengths. A tongue of emission extends here at 70\,$\mu$m, and thus it does likely emit some modest levels of emission at this wavelength. This compact infrared source is best fit with an SED model of a $8M_{\sun}$ MYSO. 

The infrared-bright ridge of material here was first resolved in both the infrared (12.6\,$\mu$m) and radio continuum (11\,cm) by \citet{1974ApJ...193..283P}. Within this ridge are a string of molecular fragments designated as (going east to west) MM32, MM8, MM6, and MM19 (Figure~\ref{fig:mm6b}), of which MM6 and MM8 are the largest, and this region is thus referred to as the ``MM6 \& MM8 ridge'' \citep{2003ApJ...582..277M}. The SOFIA data reveal no additional compact sources or sharp peaks; instead, the broad maxima along the ridge align with gentle undulations in the dust morphology observed at Spitzer–IRAC wavelengths. Importantly, there does not seem to be any compact sources in the better resolution near-infrared images nor narrow near-infrared peaks associated the broad mid-infrared peaks, and thus the ridge does not display obvious embedded star formation activity in the infrared. The ridge appears to have a similar morphology at all wavelengths from near-infrared to radio, but offset in position. While the Spitzer-IRAC data shows a near-infrared ridge of material that appears more or less coincident with the mid-infrared ridge, the Herschel 70\,$\mu$m image shows a ridge of emission offset slightly more northward than the mid-infrared ridge (i.e., farther way from the WR/OB cluster). The JCMT 450\,$\mu$m data shows a ridge offset even farther northward and away from the WR/OB cluster. Since we do not see any clear evidence of near/mid-infrared peaks indicating the presence of internal stellar/proto-stellar heating, the flux from this ridge may be dominated in the near- to mid-infrared by external heating from the WR/OB cluster. The free-free emission seen at cm radio wavelengths is also likely due to external ionization of the ridge by this cluster. Consistent with this idea, \citet{2003ApJ...582..277M} estimate that free-free emission from external ionization accounts for 10, 40, 40, and 70\% of the flux they measure at 1.3\,mm for fragments MM6, MM8, MM19, and MM32, respectively. The brightest area of the ridge in the mid-infrared is closest to the peak of the MM8 fragment, and lies within the brightest region at 3.6\,cm \citep{2001AJ....121..371B} and 6\,cm (Figure~\ref{fig:mm6b}). \citet{2001AJ....121..371B} show that this ridge of emission is the brightest of all radio-emitting regions in W43~Main at 3.6\,cm. The ridge likely constitutes the northern periphery of the ionization front from the expanding \ion{H}{2} region of the WR/OB stellar cluster.

Though likely dominated by external heating and ionization, there does appear to be some form of star-formation activity present at the ridge location on the sky. Somewhere in the ridge (or in projection to the ridge) OH and water maser emission is present, with nominal coordinates grouped just south of the peak of MM6 (see Figure~\ref{fig:mm6b}). However, the astrometric uncertainties of the maser positions are considerable. \citet{2001A&A...368..845V} find water maser emission here but with 25$\arcsec$ positional uncertainty, and the water maser emission found by \citet{1998A&AS..127..181H} has a 20$\arcsec$ positional uncertainty. \citet{1997A&A...325..255B} find 5\,cm OH maser emission here but with with 65$\arcsec$ uncertainty. All of these positional uncertainties are large, but place the maser emission as coming from somewhere in the ridge, and likely indicate the presence of some star-formation activity there. We speculate further on the nature of this ridge in Section~\ref{sec:trunks}.

\citet{2005A&A...440..121B} find a near-infrared (JHK) source in the ridge which they spectral type to be a O5V-O6V star. \citet{2005A&A...440..121B} do not give positions for the sources in their paper, but state that this source (which they call 18449nr319) is associated with IRAS 18449-0158. Looking at JHK data associated with this IRAS source from the 2MASS archive shows that the closest bright NIR source to the IRAS 18449-0158 position is 2MASS J18473502-0155237. However, there is no associated 3-8\,$\mu$m Spitzer-IRAC source at this location nor at SOFIA or the longer wavelengths we checked. Given this, it could be a foreground object, and not embedded in the ridge. In fact, \citet{2005A&A...440..121B} derive a spectrophotometric distance of $3.8-4.3$\,kpc for the source, which would place it in the foreground of W43. This near-infrared-only source is very close to the nominal water an OH maser positions (Figure~\ref{fig:mm6b}), however given the positional uncertainties of the masers it is not clear if they are related. 

\citet{2019ApJ...884...48C} present high resolution 1.3\,mm ALMA observations of the MM6 and MM8 fragments. Both resolve into a filament-like structure, or string of mm cores. For MM6, most of the cores, and especially the brightest, lie in an areas of the SOFIA map that has a dearth of emission (Figure~\ref{fig:mm6b}). For MM8, we see that the brightest parts of the string of mm cores (except for the absolute brightest, A) seem to lie near (but not coincident with) the 37\,$\mu$m emitting region (Figure~\ref{fig:mm6b}). \citet{2019ApJ...884...48C} find that both MM6 and MM8 have more lower-mass cores and are not dominated by a single bright core as is the case for MM2, MM3, and MM4.

Source 1 (a.k.a. G30.7688-0.0078 from \citealt{2017ApJ...839..108S}) is bright and point-like at all Spitzer-IRAC wavelengths (i.e., Figure~\ref{fig:mm6a}). In the mid-infrared with SOFIA, the source peak appears to be located at the northeast end of a bar or ridge of emission seen at 70\,$\mu$m. Radio continuum emission at cm wavelengths is located along this ridge, and while there is extended radio and 70\,$\mu$m emission present at the location of source 1, there are no peaks at either wavelength coincident with the peak seen in the mid-infrared (Figure~\ref{fig:mm6b}). Nonetheless, our SED modeling does indeed point to this source as being a MYSO (with best mass fit of $8M_{\sun}$).  

\subsubsection{The MM13 Region: Source 2}

\begin{figure*}[tb!]
\epsscale{1.05}
\plotone{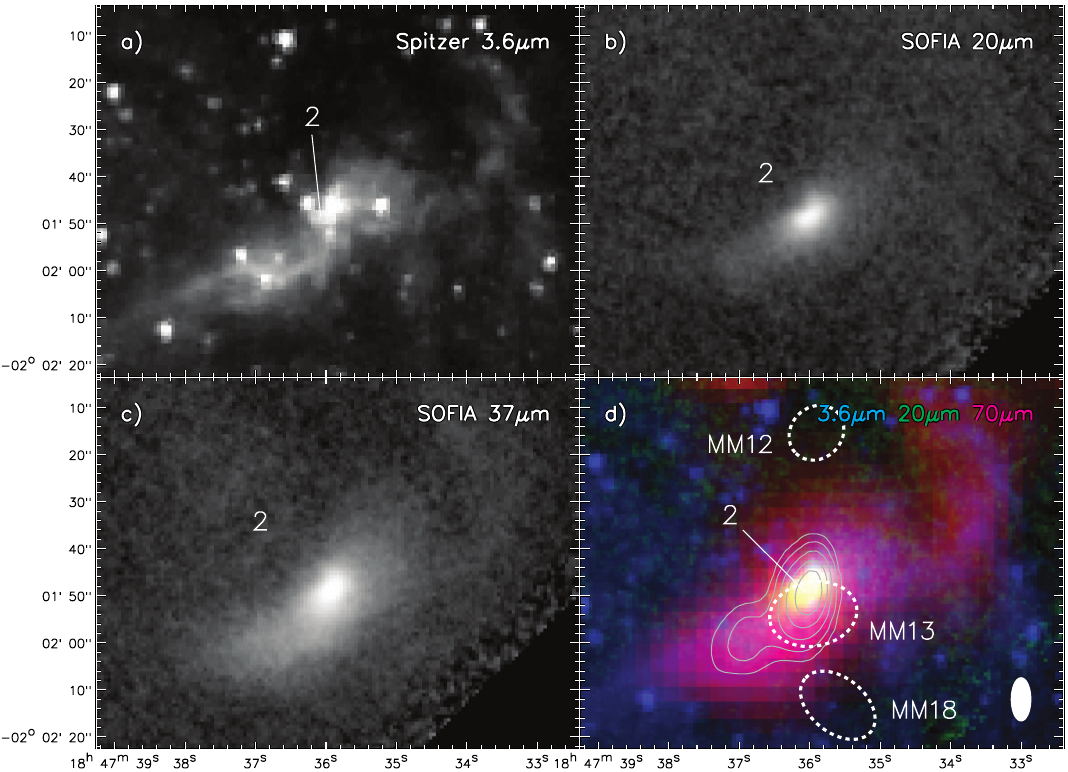}
\caption{Images of the MM13 region containing infrared source 2, with (a) Spitzer 3.6\,$\mu$m, (b) SOFIA 20\,$\mu$m, (c) SOFIA 37\,$\mu$m, and (d) a three-color RGB image made from Herschel 70\,$\mu$m (red), SOFIA 20\,$\mu$m (green), and Spitzer 3.6\,$\mu$m (blue). In panel (d), the dashed white ellipses denote the 2D Gaussian size and position of the mm fragment MM13, as well as nearby MM12 and MM18 \citep[from][]{2003ApJ...582..277M}. Contours are 6\,cm radio continuum emission from \citet{2009A&A...501..539U}, and the filled ellipse on the bottom right corner shows the radio beam size and shape. \label{fig:mm13}}
\end{figure*}

In addition to being a prominent fragment at millimeter wavelengths \citep[e.g.,][]{2003ApJ...582..277M, 2009A&A...507.1467L}, MM13 (Figure~\ref{fig:mm13}) is also associated with a bright free-free emission source (e.g., at 21\,cm by \citealt{1993AJ....106.2349L}, at 6\,cm by \citealt{2009A&A...501..539U}, and H42$\alpha$ by \citealt{2017ApJ...844L..25N}). 

Looking at the location of the MM13 fragment at 20\,$\mu$m with SOFIA, our source 2 appears as a bright peak with a diffuse tail of emission extending towards the southeast for $>$15$\arcsec$ (Figure~\ref{fig:mm13}). At 37\,$\mu$m, the peak is seen to be embedded in a region of broader and the diffuse emission that extends to the northwest as well as southeast (though still more extended towards the southeast). At 70\,$\mu$m the source appears very similar in size and morphology to 37\,$\mu$m, however in the Spitzer-IRAC data, the source is better resolved and found to have an interesting near-infrared ridge-like morphology, visually similar to a photodissociation region. At 8\,$\mu$m the source appears as a brighter knot in a almost 2$\arcmin$-long dust ridge or filament (Figure~\ref{fig:mm13}), which must be hot or contain considerable PAH emission, because it is not very prominent at Herschel wavelengths. At the shortest IRAC wavelengths, the brighter region of the filament, which is coincident with source 2, can be seen peppered with stars (at least in projection). \citet{2015MNRAS.450.4364N} state that this is the location of a potential star cluster, as demonstrated by a stellar over-density around MM13 in the near infrared, spanning an angular diameter of 25$\arcsec$. 

\citet{2003ApJ...582..277M} claim the radio continuum source located here is a full-blown \ion{H}{2} region heavily extinguished by foreground cloud material ($A_{V}\sim80$\,mag). However, \citet{2009A&A...501..539U} find a radio continuum source here (442.5 mJy at 6\,cm) with similar size and extent to our MIR source (Figure~\ref{fig:mm13}). Like in the mid-infrared, the radio source detected by \citet{2009A&A...501..539U} appears as a main peak plus a extension of emission towards the southeast. Indeed, the measured FWHM through the largest dimension is only 10$\arcsec$ (or 0.26\,pc) at 37$\mu$m, confirming the idea that it is a compact mid-infrared source in a larger diffuse region. From the photometry of the compact emission component, our SED models indicate that our source 2 is a $\sim$16$M_{\sun}$ MYSO.

\subsubsection{Sources 7 \& 11}

\begin{figure*}[tb!]
\epsscale{0.90}
\plotone{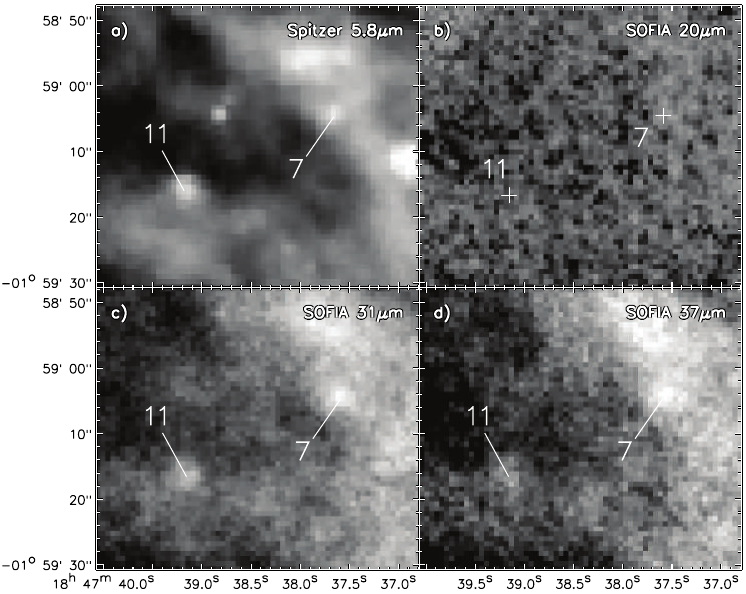}
\caption{Images of the region containing infrared sources 7 and 11, with (a) Spitzer 5.8\,$\mu$m, (b) SOFIA 20\,$\mu$m, (c) SOFIA 31\,$\mu$m, and (d) SOFIA 31\,$\mu$m. In all panels the infrared sources are labeled numerically, but source positions marked by a cross are undetected at that wavelength.\label{fig:source7}}
\end{figure*}

Sources 7 and 11 are located in a position on the sky between the brighter ionized regions of W43~Main and the highly extinguish star forming MM2 \& MM3 ridge (Figures \ref{fig:fig2}-\ref{fig:fig5}). They are located in the southwest arc of the wind-blown cavity (best seen in Figure~\ref{fig:fig1}) caused by the WR/OB stellar cluster. Sources 7 and 11 are visible at both 31 and 37\,$\mu$m, but most clearly seen at 31\,$\mu$m (Figure~\ref{fig:source7}). Both sources are relatively faint with signal-to-noise of $\sim$6 at 31\,$\mu$m. These sources are not detected in the SOFIA 20\,$\mu$m (Figure~\ref{fig:source7}b) or 11\,$\mu$m images, nor at 70\,$\mu$m or in cm radio continuum. However, both have a near-infrared component that can be seen at all Spitzer wavelengths (Figure~\ref{fig:source7}a) but not at shorter 2MASS JHK wavelengths. In order to detect these sources in Spitzer-IRAC filters as well as 31 and 37\,$\mu$m, but not at 11 and 20\,$\mu$m, would require high extinction and high PAH contributions in the 3, 5, and 8\,$\mu$m Spitzer bands. Consistent with this, \citet{2017ApJ...839..108S} classify source 11 (which they named G030.7315-0.0647) a PAH-dominated source based upon its Spitzer-IRAC fluxes. They also cataloged source 7 as G030.7314-0.0576 as a near-infrared source, but were unable to classify it.  

Our analysis does show both sources appear to be a PAH-dominated (see Figure~\ref{fig:ccd}), however our best fits for both sources are for models of intermediate mass YSOs (with best fit masses of $2M_{\sun}$), though both do have MYSOs in their groups of best fits. In the Herschel-SPIRE images (i.e., 250-500\,$\mu$m), both sources appear to be located in a cold dust ridge running from sources 9/10 to sources 12/15/16, with source 7 in a stronger far-infrared emitting area than source 11. Consistent with this, our fits indicate that both sources lie in a medium of high extinction, with modeled higher extinctions for source 7 ($A_V=164-285$) than source 11 ($A_V=12-156$).  Based upon these results we conclude that sources 7 and 11 are likely intermediate mass YSOs in a highly extinguished region.

\subsubsection{Non-Compact Objects of Interest}\label{sec:trunks}

\begin{figure*}[tb!]
\epsscale{1.15}
\plotone{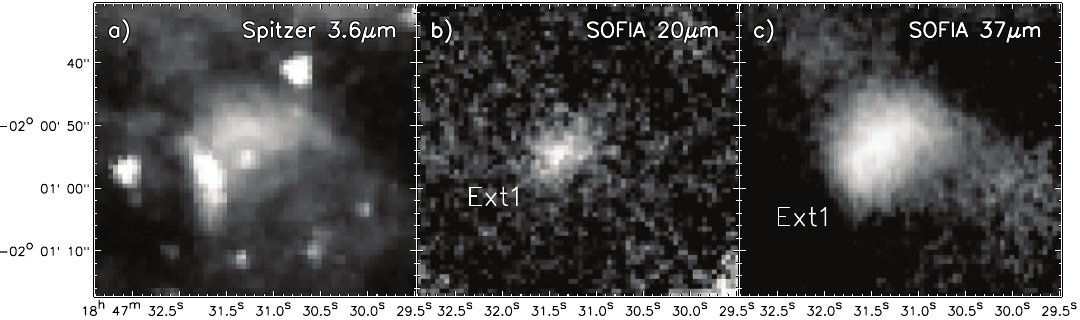}
\caption{Images of infrared source Ext1 at (a) Spitzer 3.6\,$\mu$m, (b) SOFIA 20\,$\mu$m, and (c) SOFIA 37\,$\mu$m. \label{fig:source13}}
\end{figure*}

In this section we will discuss some points of interest regarding infrared sources or features that do not fit into the category of being a compact infrared source.  

\textbf{Source Ext 1:} Mid-infrared source Ext1 is over the size limit for being considered a compact source, having a measured FWHM of 17$\arcsec$ along its widest dimension and 11$\arcsec$ along its narrowest (i.e., $0.5\times0.3$\,pc) at 37\,$\mu$m. It also has no compact peak in the mid-infrared, but instead appears as a large and diffuse region with no discernible peak (Figure~\ref{fig:source13}). However, in the Spitzer-IRAC data, there is a compact source at 3.6\,$\mu$m in the center of the mid-infrared emitting region, and this compact source appears to be surrounded by a partial shell of dust emission. At longer Spitzer-IRAC wavelengths (e.g., 8.0\,$\mu$m) only the shell is seen. A dust shell with a central compact object that is not visible at longer infrared wavelengths may mean we are seeing an evolved star (i.e., a planetary or protoplanetary nebula), but such an object should have low extinction. Ext1 is also fairly bright at 70 and 160\,$\mu$m,  atypical for evolved stars (see \citealt{2025ApJD}), and no optical star is found in the SDSS Photometric Catalog at the center of Ext1, nor within a 20$\arcsec$ radius region centered on the central coordinates of Ext1. 

One alternate possibility for the observed morphology is if we are seeing the periphery of an expanding compact \ion{H}{2} region, embedded in a cold and dusty environment. However, Ext1 does not appear to be emitting free-free radio continuum emission. Though it is located in a ridge of 21\,cm emission, the mid-infrared emission of Ext1 is not coincident with any radio peak \citep{1993AJ....106.2349L}, and there is no obvious detection of Ext1 in the H42$\alpha$ by maps of \citet{2017ApJ...844L..25N}. A second alternative is that the emission from the northeastern edge of the source, which preferentially emits near-infrared (i.e., hotter) radiation, may be signaling that Ext1 is externally heated by one or more non-ionizing stars located to the northeast. We see not evidence for such stars in the optical/near-infrared, but this may be due to high extinction.

Source Ext1 is located on the sky on the periphery of a millimeter fragment MM30 of \citet{2003ApJ...582..277M}, with the central coordinates of its mid-infrared located $\sim$18$\arcsec$ southeast from the millimeter peak of MM30. Though we do not believe Ext1 is a compact source, it does exhibit a trend in integrated flux vs. wavelength of a typical MYSO. We therefore did try fitting the object with the SED fitter, and got reasonable fits for a $12-32M_{\sun}$ source, with high extinctions required to make the fits ($A_V=160-240$). However, we cannot conclude the true nature of this source with the data at hand.

\begin{figure*}[t!]
\epsscale{0.5}
\plotone{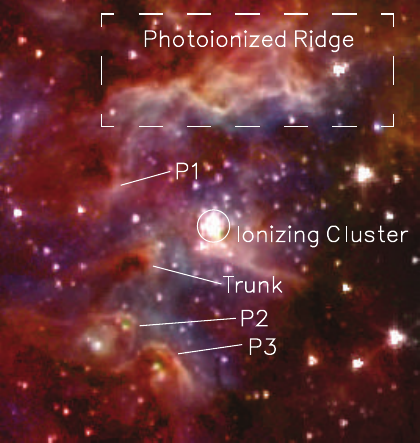}
\caption{Large scale infrared features of W43~Main. Labeled are the locations of the photoionized ridge north of (and heated/ionized) by the Wolf-Rayet ionizing stellar cluster, which is also labeled. The photoablation of natal cloud material near the ionizing cluster has lead to the appearance of several protuberances (labeled P1, P2, and P3) as well as a trunk-shaped structure. The trunk and protuberances all point back to the ionizing cluster and it may be that the MYSOs seen in projection coincident with P2 and P3 are actually contained within the protuberances. This image is the same 4-color image as in Figure~\ref{fig:fig1}, but with higher contrast to show detail. \label{fig:trunks}}
\end{figure*}

\textbf{Photoablated Trunk:} East of millimeter fragment MM20 and infrared sources 3, 4, and 6 lies a region that appears to be infrared-dark at Spitzer and SOFIA wavelengths. As seen in the bottom left corner of Figure~\ref{fig:mm15a}, this dark area looks like a trunk pointing towards the location of the ionizing and heating stellar cluster containing WR121a (a.k.a. W43\#1). The tip of this dark trunk is capped with a region of fairly bright infrared emission seen at all Spitzer and IRAC wavelengths (e.g., Figure~\ref{fig:mm15a}a-c), but at longer wavelengths, like Herschel 70\,$\mu$m (Figure~\ref{fig:mm15a}d), this trunk itself is bright, indicating the trunk is composed of dense cold dust. Figure~\ref{fig:mm15b} shows that the 1.3\,mm fragment MM14 is coincident with the Herschel emission here, and that the cap of hot dust seen in the Spitzer/SOFIA data is also the location of cm radio emission, likely from ionized surface at the tip of the trunk. 

The peak of the emission at the tip of this trunk changes of position as a function of wavelength at far-infrared, millimeter, and cm wavelengths consistent with external heating by the WR/OB cluster. In fact, the multi-wavelength images of MM14 (see Figure~\ref{fig:mm15b}, and especially Figure~\ref{fig:trunks}) give the impression of a photo-ablated trunk or pillar like those seen in the Hubble images of the Eagle Nebula (i.e., M16; \citealt{1996AJ....111.2349H}) or like those seen in other G\ion{H}{2} regions in this survey (e.g., NGC~3603 from \citetalias{2024ApJ...963...55D}, or M17 from \citetalias{2020ApJ...888...98L}).

\textbf{Photoablated Protuberances:} There are several infrared features protruding from the ridges of dust around the location of the WR/OB stellar cluster that appear similar to the photoablated trunk, but are not as elongated. We call these short trunk-like features protuberances, and label three of them (P1, P2, and P3) that are the most pronounced in the mid-infrared.

\textit{P1:} Mid-infrared protuberance P1 appears as a diffuse and extended object (measuring FWHM$\sim$14$\arcsec$ along its narrowest dimension 37\,$\mu$m). It is not detected at 11\,$\mu$m (Figure~\ref{fig:fig2}), but clearly visible at 20, 31, and 37\,$\mu$m (Figure~\ref{fig:mm15a}). Like the photoablated trunk, the center of this extended source moves as a function of wavelength, with the 70\,$\mu$m emission concentrated farthest from the WR/OB star cluster, and 37\,$\mu$m closer, and 20\,$\mu$m even closer still to the cluster (Figure~\ref{fig:mm15b}), indicative of external heating by the cluster. Also like the trunk, the side of the protuberance closest to the stellar cluster displays cm radio continuum emission suggesting it is being ionized by the cluster (Figure~\ref{fig:mm15b}). The 4-color image of Figure~\ref{fig:trunks} also shows this source to appear as a short trunk pointing towards the location of the ionizing WR/OB cluster. P1 can be readily seen in the Spitzer-IRAC bands, and has been cataloged as near-infrared source G30.7780-0.0448 by \citet{2017ApJ...839..108S}.

\textit{P2:} Mid-infrared protuberance P2 is associated with millimeter fragment MM11, but has a more complicated structure than P1. The Herschel 70\,$\mu$m emission extends from source 13 and away from the direction of the WR/OB cluster to just beyond source 14 (Figure~\ref{fig:mm4a}d). At Spitzer-IRAC wavelengths this protuberance is capped by a bright and thin layer of emission, likely coming from the hottest surface layer of dust (Figure~\ref{fig:mm4a}a). Just interior to this apex is source 13, which is likely an embedded protostar housed there. Just exterior (i.e. located closer to the WR/OB cluster) there is a peak in the extended cm radio continuum (Figure~\ref{fig:mm4b}), which likely is tracing the ionized tip of the protuberance. Given its location in projection on the sky, Source 14 may be a YSO embedded deeper within the protuberance. Again, the 4-color image of Figure~\ref{fig:trunks} gives the clearest picture of this protuberance, showing that the tip is pointing towards the location of the ionizing cluster, which is the cause of its ionization and photoablation.

\textit{P3:} This protuberance can be seen as a infrared-dark region surrounding sources 9 and 10 at 3.6 and 20\,$\mu$m, which like P2, is capped by a bright and thin layer of emission, likely coming from the hottest surface layer of dust (Figure~\ref{fig:mm4a}a) facing the ionizing WR/OB cluster. This infrared-dark area is bright at 70\,$\mu$m (Figure~\ref{fig:mm4a}) and longer wavelengths, signifying an area of denser, cold dust. The cm radio continuum in this immediate area is dominated by the UC\ion{H}{2} region, and it is not as clear how much of the extended cm radio continuum seen in Figure\ref{fig:mm4b} is related to the protuberance itself. The appearance of P3 in the 4-color image including Herschel 70\,$\mu$m data (Figure~\ref{fig:trunks}) gives the impression of a short photo-ablated pillar or trunk, brightest at the shorter infrared wavelengths tracing the hot dust on the side facing the WR/OB cluster. Since  they lie in projection coincident with the center of P3, it could be that sources 9/10 (and the UC\ion{H}{2} region) are embedded in the protuberance. 

\textbf{Photoablated Ridge:} First seen as a bright and extended mid-infrared component of W43~Main by \citet{1974ApJ...193..283P}, the infrared ridge north of the WR/OB ionizing cluster has a series of undulations pointing back at the cluster (Figure~\ref{fig:trunks}). Home to millimeter fragments MM6, MM8, MM19, and MM32 (Figure\,\ref{fig:mm6b}), it is clear from the emission as a function of wavelength (i.e., 20\,$\mu$m closest to, 37\,$\mu$m farther from, and 70\,$\mu$m farthest from the ionizing cluster; see Figure~\ref{fig:mm6a}d) that this entire ridge is being photoablated by the ionizing cluster. The whole region is likewise being ionized by the cluster, as evident by the wide-spread presence of cm radio continuum emission (Figure~\ref{fig:mm6b}). Again, the nature of this ridge as described is best evidenced by its morphology as seen in Figure~\ref{fig:trunks}.

\subsection{Discussion of the MYSO Population of W43~Main}\label{sec:mysos}

In this study, we classify 15 sources satisfying our criteria of housing a MYSO out of the 20 SOFIA-FORCAST defined compact sources. Only source 17 is designated a possible MYSO (pMYSO). Therefore, overall, we have determined 16 sources to be either a MYSO or pMSYO ($\sim$80\%) in W43~Main. 

Looking further at the results in Table\,\ref{tb:seds}, the absolute best model fits for the mid-infrared detected YSO candidates in the all of W43~Main yield protostellar masses in the range  $m_*$\,=\,1--64\,$M_{\sun}$, which is approximately equivalent to a range of ZAMS spectral type G5--O4 stars. Note that the ZT MYSO SED models have sampled protostellar mass at at 0.5, 1, 2, 4, 8, 12, 16, 24, 32, 48, 64, 96, 128, 160\,$M_{\sun}$, thus there is a minimum mass granularity that can be explored with the models \citep{2018ApJ...853...18Z}. The most massive source in W43~Main is source 14, having best fit stellar mass of $64M_{\sun}$. This source is also the brightest source in the 11, 31, and 37\,$\mu$m SOFIA images (as well as the 5.8 and 8.0\,$\mu$m Spitzer images). Source 4 is the brightest in the 20\,$\mu$m image (as well as the 3.6 and 4.5\,$\mu$m Spitzer data), and weighs in at $24\,M_{\sun}$. 

The brightest 6\,cm radio continuum sources are associated with source 16 and 10, which we measure to have SED-derived masses of 16 and $12M_{\sun}$, respectively. In fact, only 7 of the 16 sources we believe are MYSOs (Table~\ref{tb:seds}) are associated with a compact cm-wavelength continuum source or peak (44\%). For the MYSOs with no detectable cm radio continuum emission, it is possible that they are in a very young state prior to the onset of a hypercompact \ion{H}{2} region \citep{2010ApJ...721..478H} and not observable via radio continuum emission. Sources 7, 11, 12, and 18 can all be fit with the ZT MYSO SED model fitter, but have best model fits below $8M_{\sun}$, so we assume they are intermediate (or even low) mass YSOs.   

As we have mentioned previously, W43~Main is positioned in the Galaxy near where a spiral arm and the Galactic Bar interact, which is thought to be subject to increased turbulence and complexity due to converging flows of material \citep{2011A&A...529A..41N}, perhaps making their observational properties more akin to the Central Molecular Zone (CMZ) G\ion{H}{2} regions we have studied. In \citetalias{2025ApJD}, we conclude that the CMZ G\ion{H}{2} regions appear to have lower MYSO densities ($\sim$0.06\,MYSOs/pc$^2$) compared to the median we derive for non-CMZ G\ion{H}{2} regions ($\sim$0.16\,MYSOs/pc$^2$). Furthermore, Sgr\,B2 and Sgr\,C appear to have much lower percentages of infrared compact sources that are MYSOs (46\% and 32\%, respectively) than the median for non-CMZ G\ion{H}{2} regions ($\sim$80\%). However, our compact source statistics for W43~Main do not mimic those lower values of the CMZ G\ion{H}{2} regions. Given the number of MYSOs and pMYSOs in W43~Main (16), we derive a MYSO density of 0.12 MYSOs/pc$^2$, which is close to the median MYSO density from all non-CMZ G\ion{H}{2} regions studied to date (0.16\,MYSOs/pc$^2$). W43~Main appears very typical when compared to non-CMZ GHII regions in other indicators as well, including having 80\% of its compact infrared objects being MYSOs (median = 80\%), and having a most massive MYSO of 64\,$M_{\sun}$ (median = 64\,$M_{\sun}$). These numbers seem to imply that the location of W43 at the meeting point of a spiral arm and Galactic Bar does not appear to be adversely affecting the present star formation activity in W43~Main, and at least in the indicators mentioned above, it is behaving like an average G\ion{H}{2} region.

\begin{figure*}[tb!]
\epsscale{0.65}
\plotone{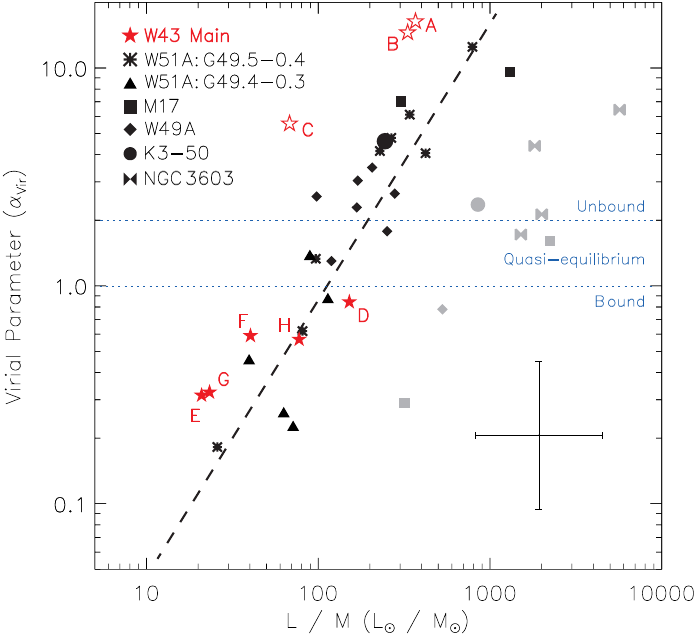}
\caption{\footnotesize Virial parameter ($\alpha_{\rm vir}$) vs. $L/M$ of all infrared sub-regions in all G\ion{H}{2} regions studied so far, with values for the sub-regions of W43~Main given in Table~\ref{tab:w43_nro} plotted as red stars with red labels (see also Figure~\ref{fig:fig6}). The open star symbols mean that we are less confident in the results due to unresolved spectroscopic feature blending. Asterisks are values for the sub-regions in W51A:G49.5-0.4 with triangles for the sub-regions in W51A:G49.4-0.3. The dashed line indicates the best line fit from \citetalias{2019ApJ...873...51L} to both W51A G\ion{H}{2} regions only (slope$\sim$1.28 in log-space). Squares show the sub-regions of M17, diamonds show the data for the sub-regions in W49A, the filled circles are for K3-50, and the bowtie symbols are for NGC~3603. Gray symbols indicate sources that have significantly higher $L/M$ values that place them to the right of the main trend, and may have significant contamination from external heating/ionization. Objects to the lower left are evolutionarily younger, and sources to the upper right, older. Sources above the blue dotted line at $\alpha_{\rm vir} = 2$, are gravitationally unbound and expanding, and all source below the blue dotted line at $\alpha_{\rm vir} = 1$, are self-gravitationally bound and collapsing. Sources in the region between the blue dotted lines ($1 < \alpha_{\rm vir} \leq 2$) are in quasivirial equilibrium, meaning the clump is slightly expanding but still gravitationally bound. The error bar at the bottom left shows the typical uncertainty (a factor of $\sim$2) in both $L/M$ and $\alpha_{\rm vir}$. } \label{fig:almplot}
\end{figure*}

\subsection{Extended Sub-Regions: Evolutionary Analysis Methodology}\label{sec:esm}

As was done for our previous G\ion{H}{2} region studies, we utilized SOFIA observations in conjunction with gas kinematics data to assess the evolutionary status of extended sub-regions, assuming they represent star-forming molecular clumps. Our goal was to explore whether these structures offer insight into the formation history and development of their G\ion{H}{2} regions.  Previous publications in this project involved comparing two independent indicators of molecular clump evolution -- namely, the virial parameter ($\alpha_{\rm vir}$) and the luminosity-to-mass ratio ($L/M$) -- focusing on extended sources within the G\ion{H}{2} regions. That analysis uncovered a clear positive correlation between $\alpha_{\rm vir}$ and $L/M$, with elevated values of both suggesting more evolved sub-regions, thus providing two independent ways of measuring relative ages. 

We applied a similar methodology to the extended sub-regions in W43~Main. Masses were derived through pixel-by-pixel graybody fits using Herschel images from 160 to 500\,$\mu$m, all convolved to a 36$\arcsec$ resolution to generate a template temperature map following the method of \citet{2016ApJ...829L..19L}. This temperature map was then applied to APEX 870\,$\mu$m data for improved mass estimation at 18$\arcsec$ resolution. Bolometric luminosities were determined using a two-temperature graybody model that incorporated integrated fluxes across Spitzer, SOFIA, and Herschel bands. We employed the small–median–filter background subtraction technique described by \citet{2012ApJ...754....5B} for Spitzer 8~$\micron$ maps and implemented in \citet{2016ApJ...829L..19L} across Spitzer, WISE, and Herschel band images. In this method, bright emission from entire W43 region was first masked in each wavelength map, and the median filter size was iteratively adjusted to capture the smoothly varying diffuse emission surrounding the sources. The residual large–scale background was then reconstructed using a distance–weighted interpolation of the filtered map toward the masked regions. 

To compute the virial parameters, we adopted the approach outlined by \citet{2019ApJ...873...51L}, which compares the gravitational potential and kinetic energies through the relationship $\alpha_{\rm vir} = M_{\rm vir}/M$. The kinetic term was derived from the FWHM of the integrated $^{13}$CO\,(1--0) line profiles obtained with the Nobeyama~45\,m telescope, following Equation~2 of \citet{2019ApJ...873...51L}. These data, originally presented by \citet{2021PASJ...73S.129K}, have an angular resolution of $\sim$20$\arcsec$ and a spectral resolution of 0.65\,km\,s$^{-1}$. To assess the impact of line blending and multiple velocity components on the derived linewidths, we examined the higher--resolution $^{13}$CO\,(2--1) data from the IRAM~30\,m observations of \citet{2013A&A...560A..24C}, which provide $\sim$11$\arcsec$ spatial and 0.2\,km\,s$^{-1}$ spectral resolution. Although the IRAM data reveal narrow subcomponents and complex velocity structures, the integrated profiles agree well with those from the Nobeyama data when compared at a common effective resolution.

As was mentioned previously, the galactic location of W43~Main is thought to be subject to high turbulence. We are also viewing the G\ion{H}{2} region through the tangent of the spiral arm, and thus there is unrelated cold molecular material at different distances (and thus gas velocities) along the line of sight. 

For W43~Main, the $^{13}$CO data are far more complicated than the data for previously studied Galactic plane G\ion{H}{2} regions, but not as bad as the situation for the CMZ G\ion{H}{2} regions studied in \citetalias{2025ApJD}, which had such complicated CO velocities as to be impossible tell which gas was associated with which G\ion{H}{2} substructure. \citet{2013A&A...560A..24C} claim that the large range in $^{13}$CO velocities ($\Delta$v=$\sim$30 km/s) appear to be a gradient from $\sim$80 km/s at the southwestern end of W43~Main to $\sim$115 km/s in the northeast end which may indicate a single cloud undergoing rotation. However, \citet{2021PASJ...73S.129K} state that W43~Main has four main gas components centered at 82, 94, 103, and 115\,km/s, which are separate gas flows all physically associated with the region and driving the on-going star formation there. We independently compared the $^{13}$CO channel maps to the mid-infrared to sub-mm dust maps, allowing us to match CO clumps at the velocities where they appear morphologically similar to the dust in the star-forming clumps. This provided further confidence that we are analyzing CO emission related to the denser and warmer material within the W43~Main star-forming clumps themselves. For each clump, we adopt the velocity dispersion derived from the CO velocity component with the strongest emission (as measured from the higher spectral resolution data of \citealt{2021PASJ...73S.129K}), and treat this dominant component as the primary indicator of the clump’s kinematics, following standard practice in molecular cloud studies \citep[e.g.,][]{2006ApJS..163..145J}. Nevertheless, given the complex structure of W43~Main described above, multiple velocity components frequently overlap along the line of sight, and individual clumps may therefore encompass a broader range of velocities than is captured by this single representative value. In this context, the velocity dispersions quoted here reflect the dominant CO component within a broader and continuous velocity field. Figure~\ref{fig:fig6} shows the locations of our CO clumps (red dashed circles labeled A-H), and it can be seen that they are centered on the sub-region structures seen in the mid-infrared and/or far-infrared. Our clumps at the periphery of the W43~Main cavity (i.e., D-H) show up best at 82\,km/s. These results are consistent with those of \citet{2021PASJ...73S.129K} who found that the 82\,km/s gas traces the brightest dust morphology seen in Spitzer 8$\mu$m data. However, the clumps we identified closest to the WR/OB cluster (A-C), were best seen at 115\,km/s, again consistent with the results of \citet{2021PASJ...73S.129K} who find the highest intensity ratio at this velocity is located near the stellar cluster region. 

Table~\ref{tab:w43_nro} provides the measurements and physical properties we determined for each molecular clump from the $^{13}$CO data, including the CO line full-width-at-half-maximum (FWHM) and velocity dispersion ($\sigma_v$). Also included are the total clump mass ($M$), luminosity-to-mass ratio ($L/M$), virial parameter ($\alpha_{\rm vir}$), and warm and cold dust temperatures ($T_{\rm cold}$ and $T_{\rm warm}$) derived from the multi-wavelength infrared continuum data. To further assess evolutionary states, we plotted $\alpha_{\rm vir}$ against $L/M$ for the sub-regions of W43~Main and compared them to those in previous studies (Figure~\ref{fig:almplot}). 

Most of our velocity dispersion measurements are of uncontaminated CO lines, however the three clumps at 115\,km/s (A-C) do have shoulders or marginally-resolved secondary peaks, indicating confusion with perhaps unrelated gas components along the line of sight. Therefore, the FWHM and dispersion measurements (Table~\ref{tab:w43_nro}) for these clumps (and thus $\alpha_{\rm vir}$) must be considered as upper limits. These three clumps are also the closest clumps in projection to the WR/OB stellar cluster which may be responsible for their more complex kinematic properties. Given the enhanced uncertainty of these measurements, sources A, B, and C, are shown with open symbols in Figure~\ref{fig:almplot}. All three lie at higher $\alpha_{\rm vir}$ values than the general trend. If we were to assume that the sources should have 70\% narrower dispersion due to line confusion, these clumps would lie closer to the trend, but most importantly to our main analysis (discussed in the next section), they still would be well within the $\alpha_{\rm vir}$ range of being gravitationally unbound structures.

\subsection{Discussion of Results for the Extended Sub-regions}\label{sec:es}

In terms of clump masses, our analyses indicate that the W43~Main clumps range from 90 to 2224 $M_{\sun}$, with a median of approximately 548 $M_{\sun}$. This is close to twice the median clump mass of M17 ($\sim$265$M_{\sun}$), but comparable to that of W51A:G49.5-0.4 ($\sim$409$M_{\sun}$), which are the two sources in our observed sample with the closest Lyman continuum photon rates (log$N_{LyC}=51.01$ and 51.03, respectively) to that of W43~Main (50.76). However, the clump mass ranges for M17 ($20<M<4340 M_{\sun}$) and W51A:G49.5-0.4 ($107<M<9930 M_{\sun}$) extended to much larger masses than W43~Main. In fact, only NGC~3603 tops out at a lower clump mass ($1560 M_{\sun}$). 

It can be seen from Figure~\ref{fig:almplot}, that the star-forming clumps of W43~Main have a large spread in relative ages, as they are distributed from the lower-left to upper right of the plot. Clumps A, B, and C appear towards the upper left in the plot (indicating that they are the most evolved) and spatially likely lie closest to the WR-OB cluster. Consistent with this, \citet{2003ApJ...582..277M} determine that all three of their mm fragments nearest in projection to the WR/OB cluster, MM14, MM15, and MM20, are gravitationally unbound. They further postulate that they are fragments in the process of being dispersed by the cluster. Indeed, MM15 and MM20 are unresolved in our far-infrared and CO data and are thus both contained within our A clump, and MM14 is coincident with our B clump, both of which we confirm appear to be unbound (i.e., they have $\alpha_{\rm vir}>2$).

The clumps that appear to be the least evolved from our analyses (i.e. they are located closest to the lower-left corner of the plot in Figure~\ref{fig:almplot}) are E and G. Clump E contains the MM1 region, and G contains the the MM2 region, both of which contain protoclusters that are highly embedded and appear at our shorter wavelengths as infrared dark regions. All of the clumps labeled D through H lie more than 2\,pc in projection from the WR/OB cluster and appear to be gravitationally bound structures ($\alpha_{\rm vir}<2$).

The large ranges in L/M and $\alpha_{\rm vir}$ (and hence relative ages) seen in Figure~\ref{fig:almplot} for W43~Main is similar to that of W51A:G49.5-0.4. The reason given for this broad range of ages in the case of W51A:G49.5-0.4 is that the G\ion{H}{2} region contains sequential star formation proceeding from one side of the host molecular cloud to the other \citep{2019ApJ...873...51L}. In the case of W43~Main, the sources that appear more evolved (A, B, and C) also appear to lie closer to the WR-OB cluster whereas the younger regions (D-H) lie along or just beyond the expanding shell of the large-scale feedback driven cavity. We conclude that this not only implies sequential star formation is occurring in W43~Main but it also hints that the expansion of the cavity may be triggering present and future star formation. We will discuss this possibility in the context of the results from previous studies of the evolutionary history of W43~Main in the next section.

\begin{deluxetable*}{lccccccc}
\tabletypesize{\scriptsize}
\tablecolumns{8}
\tablewidth{0pt}
\tablecaption{Derived Parameters of Subregions in W43 Main \label{tab:w43_nro}}
\tablehead{\colhead{Source} & \colhead{FWHM} & \colhead{$\sigma_v$} & \colhead{$M$} & \colhead{$L/M$} & \colhead{$\alpha_{\rm vir}$} & \colhead{$T_{\rm warm}$} & \colhead{$T_{\rm cold}$} \\ 
\colhead{} & \colhead{(km s$^{-1}$)} & \colhead{(km s$^{-1}$)} & \colhead{($M_{\odot}$)} & \colhead{($L_{\odot}/M_{\odot}$)} & \colhead{} & \colhead{(K)} & \colhead{(K)} }
\startdata
A & 3.14 & 1.33 & 90.3 & 368.4 & 16.43 & 264.0 & 23.8\\
B & 3.08 & 1.31 & 95.9 & 333.0 & 14.73 & 274.0 & 24.0\\
C & 3.62 & 1.54 & 539.7 & 68.2 & 5.53 & 389.0 & 25.0\\
D & 1.30 & 0.55 & 693.3 & 151.9 & 0.84 & 305.0 & 23.8\\
E & 1.78 & 0.75 & 2,224 & 21.0 & 0.31 & 303.0 & 24.0\\
F & 1.48 & 0.63 & 555.3 & 40.3 & 0.59 & 305.0 & 24.2\\
G & 1.63 & 0.69 & 1,689 & 23.3 & 0.32 & 305.0 & 24.0\\
H & 1.00 & 0.43 & 285.3 & 77.6 & 0.57 & 305.0 & 23.8\\
\enddata
\tablecomments{Velocity dispersion ($\sigma_v$) and FWHM are derived from NRO $^{13}$CO spectra. Temperatures are from the SED fits.}
\end{deluxetable*}

\subsection{The Evolutionary History of W43~Main}\label{sec:alm}

The expansive W43 complex lies at the intersection of the Galactic Bar and the Scutum-Centaurus Arm in a region where it is believed that there are large-scale gas flows converging which should trigger highly turbulent conditions \citep[e.g.,][]{2011A&A...529A..41N}. Our analyses of the sub-regions of W43~Main combined with observations of many others reveal it to be a complex and dynamic G\ion{H}{2} region affected by both feedback and gas collisions.  

In terms of feedback, on the largest scales, \citet{2010A&A...518L..90B} show that the W43 complex lies at the top of a long cavity, that is at least 70~pc long. Most of this cavity is filled with faint free-free emission. W43~Main G\ion{H}{2} region powers the upper portion of the chimney, and the southern portion is powered by a older group of massive stars at lower galactic latitudes which may have compressed the clouds to the galactic north, perhaps triggered the initial star formation in W43. At the smaller scale -- local to the W43~Main -- we see that feedback in the form of radiation pressure, ionization, and/or stellar winds from the WR/OB cluster has created a $\sim$3.5\,pc bubble toward the west, as well as north and south. The W43-MM1 region and W43-MM2 \& 3 ridge may lie deep enough within their host IRDCs and far enough from the cluster to be beyond any major influences caused by this feedback. However, the clumps B, C, and D, which lie $\leq$2.5\,pc in projection from the WR-OB cluster, show indications of recent star formation (e.g., MYSOs: Sources 8, 9, 13, and 14; a UC\ion{H}{2} region: Source 10; as well as masers) that are likely influenced (or were maybe even triggered to form) by the feedback from the WR/OB cluster. Apart from this being consistent with the broader conclusion from our evolutionary analyses that there appears to be sequential star formation in W43~Main (Section~\ref{sec:es}), further circumstantial evidence for this in our data comes from the numerous trunks and protuberances seen in these clumps pointing toward the WR/OB cluster (see Section~\ref{sec:trunks} and Figure~\ref{fig:trunks}) as well as the fact that these clumps appear to be highly ionized preferentially on the surfaces facing the WR/OB cluster (see Figures~\ref{fig:mm4b}, \ref{fig:mm15b}, and \ref{fig:mm6b}). \citet{2017ApJ...844L..25N} come to a similar conclusion, stating they see direct evidence that the expanding ionized bubble from the WR/OB cluster is interacting with these molecular clumps in their radio recombination line and HCO$^+$ observations.  

In terms of gas collisions, \citet{2021PASJ...73S.129K} claim that there are broad bridging features between the four separate velocity components separated by $\sim$20 km/s (as seen in $^{13}$CO data), suggesting cloud–cloud interactions within the region. They (as well as \citealt{2024A&A...685A.101L}) hypothesize that these cloud-cloud collisions may be responsible for producing the local “mini-starbursts” in the W43-MM1 region and W43-MM2 \& 3 ridge. Though cloud–cloud interactions should be powerful sources of turbulence on the scales of W43~Main, we measure relatively narrow line widths for the major star-forming sub-regions themselves (see Table~\ref{tab:w43_nro}). Consistent with \citet{2013A&A...560A..24C}, the largest line widths we measure in the $^{13}$CO data are the regions located closest in projection to the WR/OB cluster, which is not likely due to cloud-cloud collisions, but rather the turbulence driven by the local feedback from the stellar cluster itself. This implies that even if there are colliding gas flows in and around the W43~Main G\ion{H}{2} region, potential turbulence driven by these flows does not appear to be disrupting the star formation. On the contrary, these flows are likely also funneling material into the region allowing for long-term sustained star formation. Indeed, the WR/OB cluster here is estimated to be 1–6 Myr old \citep{2003ApJ...582..277M,2010A&A...518L..90B}, while our mid-infrared observations are tracing the present epoch of star formation, and the starless cores seen in mm observations \citep[e.g.,][]{2023A&A...674A..75N} indicate there will likely continue to be star formation into the future. This all indicates at least 3 epochs of sustained star formation activity in W43~Main so far. 

Despite potentially intense dynamical interplay of bar–arm gas flows, it appears the effects of any potential cloud-cloud collisions may be minor and isolated to smaller-scale triggered star formation (i.e., MM1 and MM2 \& MM3 regions within W43~Main). This is different than what we saw in the Galactic Center G\ion{H}{2} regions where we postulated that the combination of high turbulence and galactic shear may be causing suppressed MYSO formation and disruption of the natal molecular environment (preventing sustained star formation). W43~Main has an average MYSO population for its Lyman continuum photo rate and still maintains a large local reservoir of material (exhibited by its far-infrared and molecular gas maps). 

By using the Galactocentric radius of a cloud, along with its spatial size, rotational velocity, and mass, \citet{2012ApJ...758..125D} define a shear parameter $S_g$, where values higher than 1 mean that the shearing force would be strong enough to destroy a cloud. Employing this, \citet{2013A&A...560A..24C} measure the shear parameter of the entire W43 complex to be $S_g=0.77$, thus showing that the molecular clouds here are not being disrupted by shear forces of Galactic motion. This may be why, despite the complex local gas dynamics, the large scale properties and compact-source statistics of W43~Main are similar to typical Galactic \ion{H}{2} regions rather than the more extreme conditions seen in the Central Molecular Zone (which may further be implying that it is predominantly Galactic shear, not turbulence, that suppresses continued star formation in the CMZ G\ion{H}{2} regions). Along these lines, \citet{2012MNRAS.422.3178E} find the clump formation efficiency in W43~Main does not appear to be different than star forming regions elsewhere in the plane of the Galaxy and conclude that environment does not factor into the high clump formation efficiency. Indeed, numerical simulations by \citet{2020ApJ...900...82P} seem to indicate that massive star-forming environments may actually require such converging flows as those seen in W43~Main to drive and sustain star formation. Therefore, maybe W43~Main exists where it does and has the properties it does because it is at the convergence of the Galactic bar and spiral arm, instead of in spite of that fact. 

\section{Summary}\label{sec:sum} 

In this eighth paper of the SOFIA-FORCAST series on Milky Way G\ion{H}{2} regions, we present an analysis of the massive star-forming complex W43~Main. We produced 11, 20, 31, and 37\,$\mu$m maps of the region, capturing the brightest central infrared-emitting area ($9\farcm2\times10\farcm3$, or 14.7$\times$16.5\,pc) at a spatial resolution of $\lesssim$3$\arcsec$. These data provide the highest-resolution infrared images to date of the full extent of this G\ion{H}{2} region at these wavelengths. To investigate the morphological and physical characteristics of both the compact sources and the extended dust substructures within W43~Main, we compared the SOFIA-FORCAST maps with previous multi-wavelength observations, ranging from the near-infrared to radio, obtained with various ground- and space-based facilities. We further employed a MYSO SED fitting algorithm to constrain the properties of the compact infrared sources under the assumption they are internally heated by protostars, and applied evolutionary analyses to the extended subregions under the assumption that they trace star-forming molecular clumps.

W43~Main is believed to be located at the meeting point of the Scutum spiral arm and the Galactic Bar, representing a Galactic environment distinct from our previously studied G\ion{H}{2} regions. It is believed that this Galactic location should have higher turbulence due to the converging flows of material, and that this could affect the star formation activities in W43~Main. In contrast, our analyses indicate that W43~Main is largely representative of the G\ion{H}{2} regions studied to date, with most of its properties, including its Lyman continuum photon production rate ($logN_{LyC}=50.76$ photons/s), closely aligned with the sample average ($log\overline{N_{LyC}}=50.79$ photons/s). Below we itemize further details supporting this claim, as well as the other major conclusions from this work. 

1) The dominant large-scale shape of W43~Main in the SOFIA data is that of a cavity, blowing out to the west, fulfilling our definition of a ``cavity-type'' G\ion{H}{2} region. There are several large ridges and extended dust structures seen in the mid-infrared which are coincident with cm radio continuum emission, and are thus likely tracing ionization fronts. The hottest dust, as evidenced by the prevalence of emission at shorter infrared wavelengths, exists in and around the location of the WR/OB stellar cluster. This cluster has sculpted trunks and protrusions in the surrounding dust ridges through photoablation, which are most clearly visible in the SOFIA images.

2) The W43-MM1 ridge is a well-studied part of W43~Main that contains millimeter fragments MM1 and MM5. This region appears as a infrared-dark region at Spitzer and shorter SOFIA wavelengths, but contains 3 highly embedded sources seen at 31 and 37\,$\mu$m, which we label 18, 19, and 20. Our infrared source 20 lies within MM1 and source 18 lies within MM5. All three infrared sources are believed to be YSOs since they are located at the center of outflows (to within our astrometric error), and sources 20 and 19 are coincident with methanol masers, which are signposts of massive YSOs. Further evidence of the YSO nature of these three infrared sources comes from our SED modeling, which implies that sources 19 and 20 are massive YSOs, and source 18 is a low-to-intermediate mass YSO. 

3) The W43-MM2 \& MM3 region is another well-studied sub-region of W43~Main. This region also appears to have no extended emission in the Spitzer or SOFIA images, but is not empty space, as the entire an area of widespread far-infrared emission, indicating the presence of an extensive cold dust ridge. This ridge contains millimeter fragments MM2, MM3, and MM10. We find all three millimeter fragments to be coincident with compact SOFIA sources (source 5, 16, and 12, respectively). Source 16 is found to be a MYSO and is coincident with the UC\ion{H}{2} region within MM3. Source 5 is also found to be a MYSO, whereas source 12 is found to be a low-to-intermediate mass YSO.

4) The subregions of W43~Main containing the MM4, MM11, and MM14 millimeter fragments all appear in the SOFIA data to be photoablated trunks or protuberances, carved by the WR/OB cluster. In the cases of MM4 and MM11,  infrared compact sources are found to perhaps be embedded within these photoablated features. 

5) We find 20 compact infrared objects in W43~Main, of which 16 (80\%) are determined to be MYSOs or possible MYSOs from our analyses. The most massive MYSO is source 14, having a best fit mass of $64M_{\sun}$, and this source is the brightest object on our field at 11, 31, and 37\,$\mu$m. Only 7 (44\%) of the sources believed to be MYSOs are associated with a cm radio continuum source or peak, signifying the rest may be in a very young state prior to the onset of a UC\ion{H}{2} region. 

6) W43~Main has an MYSO density typical for a G\ion{H}{2} region of its size and power (as measured by its Lyman continuum photon rate) with a derived value (0.12\,MYSOs/pc$^2$) close to the median value of all non-CMZ G\ion{H}{2} regions studied thus far (0.16\,MYSOs/pc$^2$). W43~Main's highest mass MYSO ($64M_{\sun}$) is the same value as the median for the highest mass MYSOs across all G\ion{H}{2} regions studied in this survey. 

7) Using L/M and $\alpha_{\rm vir}$ as evolutionary proxies, we find that the W43~Main G\ion{H}{2} region contains star-forming sub-regions over a rather broad relative age range, demonstrating sequential star formation is occurring there. Furthermore, some star-forming sub-regions show photo-ablated structures pointing back toward the WR/OB cluster, and/or have ionized surfaces facing the cluster. This perhaps demonstrates that feedback from a prior epoch of star formation (i.e., the WR/OB cluster) may be triggering the present epoch of star formation we find occurring within these clumps.

8) Prior studies have found four velocity components in the molecular gas associated with W43~Main and have posited that they have undergone cloud-cloud collisions responsible for producing the local “mini-starbursts” in the W43-MM1 region and W43-MM2 \& 3 ridge. Such large scale flows should be powerful sources of turbulence, injecting kinetic energy into the host cloud of W43~Main, however we measure relatively narrow line widths for the major star-forming sub-regions and do not see any unusual effects in our results due to the complex dynamics of W43~Main's unique Galactic location.

\begin{acknowledgments}
This research is based on archival data from the NASA/DLR Stratospheric Observatory for Infrared Astronomy (SOFIA). SOFIA was jointly operated by the Universities Space Research Association, Inc. (USRA), under a contract with NASA, and the Deutsches SOFIA Institut (DSI), under a contract from DLR to the University of Stuttgart. This work is also based in part on archival data obtained with the Spitzer Space Telescope, which was operated by the Jet Propulsion Laboratory, California Institute of Technology under a contract with NASA. This work is also based in part on archival data obtained with Herschel, an European Space Agency (ESA) space observatory with science instruments provided by European-led Principal Investigator consortia and with important participation from NASA. This research has made use of \textit{Aladin Sky Atlas}, CDS, Strasbourg Astronomical Observatory, France.  

The lead author wishes to acknowledge NASA funding via an ADAP Award (80NSSC24K0640) which made this work possible.

\facilities{SOFIA(FORCAST), Spitzer, Herschel}
\software{sofia\_redux \citep{https://doi.org/10.5281/zenodo.8219569}}
\end{acknowledgments}

\vspace{5mm}

\bibliography{references}{}
\bibliographystyle{aasjournal}

\clearpage

\appendix
\section{Data Release}

The fits images used in this study are publicly available at the following URL: {\it https://dataverse.harvard.edu/ dataverse/SOFIA-GHII}. The data include the \textit{SOFIA} FORCAST 11, 20, 31 and 37\,$\mu$m final image mosaics of W43~Main their exposure maps.   

\section{Additional Photometry of Sources in W43~Main}\label{sec:appendixB}

As stated in Section~\ref{sec:cps}, in addition to the flux densities derived from the SOFIA-FORCAST data, we performed additional aperture photometry for all compact sources using archival Spitzer-IRAC data at 3.6, 4.5, 5.8, and 8.0\,$\mu$m, deriving flux densities at each wavelength. Furthermore, we used Herschel-PACS archival data to derive flux densities for all compact sources at 70 and 160\,$\mu$m. As was done to the FORCAST data, we applied the same optimal extraction and background subtraction techniques to the Spitzer and Herschel data, and measured both non-background subtracted and background subtracted flux densities for all sources at all wavelengths. Table~\ref{tb:CIRAC} lists the photometry values we derived from the Spitzer-IRAC data for all sources within W43~Main. Table~\ref{tb:BPACS} gives the measured photometry values for all sources from the Herschel-PACS data.

\begin{deluxetable*}{lrrrrrrrrrrrr}
\tabletypesize{\scriptsize}
\tablecaption{Spitzer-IRAC Photometry of Sources in W43~Main \label{tb:CIRAC}}
\tablehead{\colhead{  }&
           \multicolumn{3}{c}{${\rm 3.6\mu{m}}$}&
           \multicolumn{3}{c}{${\rm 4.5\mu{m}}$}&
           \multicolumn{3}{c}{${\rm 5.8\mu{m}}$}&
           \multicolumn{3}{c}{${\rm 8.0\mu{m}}$}\\[-4pt]
           \cmidrule(lr){2-4} \cmidrule(lr){5-7} \cmidrule(lr){8-10} \cmidrule(lr){11-13}\\[-12pt]
           \colhead{ Source }&
           \colhead{ $R_{\rm int}$ } &
           \colhead{ $F_{\rm int}$ } &
           \colhead{ $F_{\rm int-bg}$ } &
                      \colhead{ $R_{\rm int}$ } &
           \colhead{ $F_{\rm int}$ } &
           \colhead{ $F_{\rm int-bg}$ } &
                      \colhead{ $R_{\rm int}$ } &
           \colhead{ $F_{\rm int}$ } &
           \colhead{ $F_{\rm int-bg}$ } &
                      \colhead{ $R_{\rm int}$ } &
           \colhead{ $F_{\rm int}$ } &
           \colhead{ $F_{\rm int-bg}$ } \\[-6pt]
	   \colhead{  } &
	   \colhead{ ($\arcsec$) } &
	   \colhead{ (mJy) } &
	   \colhead{ (mJy) } &
	   \colhead{ ($\arcsec$) } &
	   \colhead{ (mJy) } &
	   \colhead{ (mJy) } &
	   \colhead{ ($\arcsec$) } &
	   \colhead{ (mJy) } &
	   \colhead{ (mJy) } &
	   \colhead{ ($\arcsec$) } &
	   \colhead{ (mJy) } &
	   \colhead{ (mJy) } \\[-12pt]
}
\startdata
Ext1	&	13.8	&	222	&	128	&	13.8	&	270	&	155	&	13.8	&	2580	&	1350	&	13.8	&	6770	&	3110	\\
1	&	5.4	&	79.2	&	39.6	&	5.4	&	128	&	75.7	&	5.4	&	240	&	175	&	5.4	&	1520	&	474	\\
2	&	8.4	&	176	&	129	&	8.4	&	321	&	266	&	8.4	&	1440	&	981	&	8.4	&	4470	&	3100	\\
3	&	5.4	&	162	&	UR	&	5.4	&	266	&	UR	&	5.4	&	1180	&	307	&	5.4	&	3940	&	1390	\\
4	&	5.4	&	840	&	UR	&	5.4	&	1110	&	UR	&	5.4	&	2450	&	UR	&	5.4	&	5490	&	UR	\\
5	&	9.0	&	29.2	&	UR	&	9.0	&	26.5	&	UR	&	9.0	&	257	&	UR	&	9.0	&	517	&	UR	\\
6	&	4.2	&	261	&	193	&	4.2	&	486	&	377	&	4.2	&	1400	&	849	&	4.2	&	3120	&	1380	\\
7	&	3.6	&	11.8	&	2.58	&	3.6	&	18.6	&	4.12	&	4.8	&	298	&	54.1	&	4.8	&	917	&	104	\\
8	&	9.0	&	253	&	56.4	&	9.0	&	354	&	104	&	7.8	&	1580	&	389	&	7.8	&	4670	&	711	\\
9	&	7.8	&	179	&	67.4	&	7.8	&	260	&	106	&	7.8	&	2110	&	905	&	7.8	&	5860	&	2470	\\
10	&	4.8	&	65.5	&	20.7	&	4.8	&	122	&	27	&	4.8	&	908	&	354	&	4.8	&	2480	&	495	\\
11	&	3.6	&	9.5	&	2.23	&	4.2	&	18.2	&	4.99	&	4.8	&	215	&	39.6	&	4.8	&	702	&	137	\\
12	&	2.4	&	3.5	&	UR	&	3.6	&	9.12	&	UR	&	3.0	&	29.8	&	UR	&	3.0	&	83.4	&	UR	\\
13	&	3.6	&	53.3	&	18.1	&	3.6	&	175	&	116	&	3.6	&	832	&	290	&	3.0	&	1170	&	307	\\
14	&	12.0	&	470	&	277	&	12.0	&	635	&	361	&	12.0	&	4970	&	2660	&	12.0	&	14000	&	6670	\\
15	&	3.0	&	9.8	&	0.898	&	4.8	&	25.5	&	6.69	&	6.0	&	249	&	55.8	&	6.0	&	667	&	162	\\
16	&	3.6	&	29.6	&	9.91	&	6.0	&	98.8	&	51.1	&	6.0	&	482	&	220	&	7.2	&	1470	&	727	\\
17	&	3.0	&	9.05	&	4.57	&	4.2	&	65	&	53.7	&	4.2	&	186	&	95.3	&	6.0	&	674	&	65.8	\\
18	&	2.4	&	2.14	&	UD	&	2.4	&	2.34	&	UD	&	2.4	&	22	&	UD	&	2.4	&	77.8	&	UD	\\
19	&	2.4	&	2.29	&	UR	&	2.4	&	4.39	&	1.1	&	2.4	&	24.9	&	1.57	&	\nodata	&	\nodata	&	\nodata	\\
20	&	2.4	&	1.91	&	UD	&	2.4	&	1.2	&	UD	&	2.4	&	19.5	&	UD	&	2.4	&	69.7	&	UD	
\enddata
\tablecomments{Entries with `UD' means the sources were undetected in that band. 'UR' means that the flux could not be resolved from nearby compact or extended emission sources. For sources marked UD or UR, the $F_{\rm int}$ value is used as the upper limit in the SED modeling for that filter. 
For Source 19, the 8\,$\mu$m data has severe negative flux artifacts in the area that do not permit photometry, hence no values are given here or used in the SED modeling.}
\label{tb:IRAC}
\end{deluxetable*}

\begin{deluxetable*}{lrrrrrr}
\tabletypesize{\scriptsize}
\tablecolumns{7}
\tablewidth{0pt}
\tablecaption{Herschel-PACS Photometry of Sources in W43~Main \label{tb:BPACS}}
\tablehead{\colhead{  }&
           \multicolumn{3}{c}{${\rm 70\mu{m}}$}&
           \multicolumn{3}{c}{${\rm 160\mu{m}}$}\\[-4pt]
           \cmidrule(lr){2-4} \cmidrule(lr){5-7}\\[-12pt]
           \colhead{ Source }&
           \colhead{ $R_{\rm int}$ } &
           \colhead{ $F_{\rm int}$ } &
           \colhead{ $F_{\rm int-bg}$ } &
           \colhead{ $R_{\rm int}$ } &
           \colhead{ $F_{\rm int}$ } &
           \colhead{ $F_{\rm int-bg}$ } \\[-6pt]
	   \colhead{  } &
	   \colhead{ ($\arcsec$) } &
	   \colhead{ (Jy) } &
	  \colhead{ (Jy) } &
	   \colhead{ ($\arcsec$) } &
	   \colhead{ (Jy) } &
	   \colhead{ (Jy) } \\[-12pt]
}
\startdata
Ext1	&	22.4	&	663	&	UR	&	22.4	&	1010	&	UR	\\
1	&	16.0	&	362	&	UR	&	22.4	&	490	&	UR	\\
2	&	22.4	&	875	&	UR	&	22.4	&	1040	&	UR	\\
3	&	16.0	&	1140	&	UD	&	22.4	&	1090	&	UD	\\
4	&	16.0	&	1340	&	UR	&	22.4	&	1150	&	UR	\\
5	&	16.0	&	454	&	UR	&	22.4	&	2430	&	UR	\\
6	&	16.0	&	1230	&	UR	&	22.4	&	1090	&	UR	\\
7	&	16.0	&	465	&	UD	&	22.4	&	957	&	UD	\\
8	&	16.0	&	668	&	UR	&	22.4	&	1020	&	UR	\\
9	&	16.0	&	2210	&	UR	&	22.4	&	2300	&	UR	\\
10	&	22.4	&	2080	&	UR	&	22.4	&	2330	&	UR	\\
11	&	16.0	&	429	&	UD	&	22.4	&	1160	&	UD	\\
12	&	16.0	&	150	&	UR	&	22.4	&	1040	&	UR	\\
13	&	16.0	&	1600	&	UR	&	22.4	&	1790	&	UR	\\
14	&	16.0	&	1520	&	UR	&	22.4	&	1650	&	UR	\\
15	&	16.0	&	492	&	UR	&	22.4	&	1570	&	UR	\\
16	&	22.4	&	981	&	UR	&	22.4	&	1790	&	UR	\\
17	&	16.0	&	386	&	UD	&	22.4	&	1550	&	UR	\\
18	&	16.0	&	609	&	UR	&	22.4	&	3070	&	UR	\\
19	&	16.0	&	662	&	UR	&	22.4	&	3060	&	UR	\\
20	&	22.4	&	976	&	UR	&	22.4	&	2590	&	UR	
\enddata
\tablecomments{UR and UD have same meaning as discussion in caption of Table \ref{tb:IRAC}. For these sources, the $F_{\rm int}$ value is used as the upper limit in the SED modeling.}
\end{deluxetable*}

\end{document}